\documentclass[twocolumn,pre,superscriptaddress,amsmath,amssymb]{revtex4}

\usepackage{graphicx}
\usepackage{dcolumn}
\usepackage{bm}
\usepackage{float}
\usepackage{color}

\begin{document}


\title{Computational Design of Anisotropic Stealthy Hyperuniform Composites with Engineered Directional Scattering Properties}

\author{Wenlong Shi}
\affiliation{Materials Science and Engineering, Arizona State
University, Tempe, AZ 85287}
\author{David Keeney}
\affiliation{Materials Science and Engineering, Arizona State
University, Tempe, AZ 85287}
\author{Duyu Chen}
\affiliation{Materials Research Laboratory, University of California, Santa Barbara, California 93106, United States}
\author{Yang Jiao}
\email[correspondence sent to: ]{yang.jiao.2@asu.edu}
\affiliation{Materials Science and Engineering, Arizona State
University, Tempe, AZ 85287} \affiliation{Department of Physics,
Arizona State University, Tempe, AZ 85287}
\author{Salvatore Torquato}
\email[correspondence sent to: ]{torquato@electron.princeton.edu}
\affiliation{
Department of Chemistry, Princeton University, Princeton, New Jersey 08544, USA}
\affiliation{
Department of Physics,  Princeton University, Princeton, New Jersey 08544, USA}
\affiliation{
Princeton Institute  of Materials, Princeton University, Princeton, New Jersey 08544, USA}
\affiliation{
Program in Applied and Computational Mathematics, Princeton University, Princeton, New Jersey 08544, USA}



\date{\today}

\begin{abstract}
Disordered hyperuniform materials are an emerging class of exotic amorphous states of matter that endow them with singular physical properties, including large isotropic photonic band gaps, superior resistance to fracture, and nearly optimal electrical and thermal transport properties, to name but a few. Here, we generalize the Fourier-space based numerical construction procedure for designing and generating digital realizations of {\it isotropic} disordered hyperuniform two-phase heterogeneous materials (i.e., composites) developed by Chen and Torquato [Acta Mater. {\bf 142}, 152 (2018)] to {\it anisotropic} microstructures with targeted spectral densities. Our generalized construction procedure explicitly incorporates the {\it vector-dependent} spectral density function ${\tilde \chi}_{_V}({\bf k})$ of {\it arbitrary form} that is realizable. We demonstrate the utility of the procedure by generating a wide spectrum of {\it anisotropic} stealthy hyperuniform (SHU) microstructures with ${\tilde \chi}_{_V}({\bf k}) = 0$ for ${\bf k} \in \Omega$, i.e., complete suppression of scattering in an ``exclusion'' region $\Omega$ around the origin in the Fourier space. We show how different exclusion-region shapes with various discrete symmetries, including circular-disk, elliptical-disk, square, rectangular, butterfly-shaped and lemniscate-shaped regions of varying size, affect the resulting statistically anisotropic microstructures as a function of the and phase volume fraction. The latter two cases of $\Omega$ lead to directionally hyperuniform composites, which are stealthy hyperuniform only along certain directions, and are non-hyperuniform along others. We find that, while the circular-disk exclusion regions give rise to isotropic hyperuniform composite microstructures, the directional hyperuniform behaviors imposed by the shape asymmetry (or anisotropy) of certain exclusion regions give rise to distinct anisotropic structures and degree of uniformity in the distribution of the phases on intermediate and large length scales along different directions. Moreover, while the anisotropic exclusion regions impose strong constraints on the {\it global} symmetry of the resulting media, they can still possess almost isotropic {\it local} structures. Both the isotropic and anisotropic hyperuniform microstructures associated with the elliptical-disk, square and rectangular $\Omega$ possess phase-inversion symmetry over certain range of volume fractions and a percolation threshold $\phi_c \approx 0.5$. On the other hand, the directionally hyperuniform microstructures associated with the butterfly-shaped and lemniscate-shaped $\Omega$ do not possess phase-inversion symmetry and percolate along certain directions at much lower volume fractions. We also apply our general procedure to construct stealthy non-hyperuniform systems. Our construction algorithm enables one to control the statistical anisotropy of composite microstructures via the shape, size and symmetries of $\Omega$, which is crucial to engineering directional optical, transport and mechanical properties of two-phase composite media.


\end{abstract}







\maketitle


\newpage
\section{Introduction}

Disordered hyperuniform materials are exotic states of matter \cite{To03, To18a} that lie between a perfect crystal and liquid. These systems are similar to liquids or glasses in that they are statistically isotropic and generally possess no Bragg peaks, and yet they completely suppress large-scale normalized density fluctuations like crystals and in this sense possess a hidden long-range order \cite{To03, Za09, To18a}. A hyperuniform many-particle system, disordered or ordered, is one in which the static structure factor $S({\bf k})$ vanishes in the infinite-wavelength (or zero-wavenumber) limit, i.e., $\lim_{|{\bf k}|\rightarrow 0}S({\bf k}) = 0$, where ${\bf k}$ is the wavevector \cite{To03}. Here $S({\bf k})$ is defined as  $S({\bf k}) \equiv 1 + \rho\Tilde{h}({\bf k})$, where $\Tilde{h}({\bf k})$ is the Fourier transform of the total correlation function $h({\bf r}) = g_2({\bf r}) - 1$, and $g_2({\bf r})$ is the pair correlation function, and $\rho$ is the number density of the system. Note that this definition implies that the forward scattering contribution to the diffraction pattern is omitted. Equivalently, a hyperuniform point pattern is one which the local number variance $\sigma_N^2(R)$ associated with a spherical observation window of radius $R$ grows more slowly than the window volume in the large-$R$ limit \cite{To03, To18a}.

The concept of hyperuniformity was subsequently generalized by Torquato and co-workers to two-phase heterogeneous materials \cite{Za09} and random scalar, vector and tensor fields \cite{To16a}. For two-phase heterogeneous materials (i.e., composites), the focus of this work, the quantity of interest is the spectral density function $\tilde{\chi}_{_V}({\bf k})$,  which is Fourier transform of the auto-covariance function $\chi_{_V}({\bf r}) = S_2({\bf r}) - \phi^2$, where $S_2({\bf r})$ and $\phi$ are respectively the two-point correlation function and volume fraction for the phase of interest in the composite (see Sec. II for detailed discussions) \cite{To02a, Za09, chen2018designing}. A hyperuniform heterogeneous two-phase material possesses a vanishing spectral density function in the zero-wavenumber limit, i.e., $\lim_{|{\bf k}|\rightarrow 0}\Tilde{\chi}_{_V}({\bf k}) = 0$. Since $\Tilde{\chi}_{_V}({\bf k})$ is trivially proportional to the scattering intensity \cite{To02a}, which indicates that the scattering of a disordered hyperuniform composite is completely suppressed at the infinite-wavelength limit. Equivalently, a hyperuniform heterogeneous medium, disordered or not, possesses a local volume fraction variance
$\sigma_{_V}^2(R)$ that decreases {\it more rapidly} than $R^d$ for large $R$, i.e., $\lim_{R\rightarrow\infty}\sigma_{_V}^2(R) \cdot R^d = 0$, where $R$ is the radius of spherical observation windows used to compute $\sigma_{_V}^2(R)$. This behavior is to be contrasted with those of  typical disordered two-phase media for which the variance decays as $R^{-d}$.


A variety of exotic correlated disordered systems, which can be in both equilibrium and non-equilibrium settings, and come in both quantum-mechanical and classical varieties, are known to be hyperuniform. Examples include the density fluctuations in early
universe \cite{ref3}, disordered jammed packing of hard particles
\cite{ref4, ref5, ref6, ref7}, certain exotic classical ground
states of many-particle systems \cite{ref8, ref9, ref10, ref11,
ref12, ref13, ref14, ref15}, jammed colloidal systems \cite{ref16,
ref17, ref18, ref19}, driven non-equilibrium systems \cite{ref20,
ref21, ref22, ref23, salvalaglio2020hyperuniform}, certain quantum ground states \cite{ref24, ref25}, avian photoreceptor patterns \cite{ref26}, organization of
adapted immune systems \cite{ref27}, amorphous silicon
\cite{ref28, ref29}, a wide class of disordered cellular materials
\cite{ref30}, dynamic random organizing systems
\cite{hexner2017noise, hexner2017enhanced, weijs2017mixing,
lei2019nonequilibrium, lei2019random}, electron density distributions \cite{Ge19, sakai2022quantum}, vortex distribution in superconductors \cite{Ru19, Sa19}, certain medium/high-entropy alloys \cite{chen2021multihyperuniform}, disordered 2D materials \cite{Zh20, Ch21, PhysRevB.103.224102, Zh21}, amorphous carbon nano-tubes \cite{nanotube}, and certain metallic glasses \cite{zhang2023approach}. The readers are referred to the review article by Torquato \cite{To18a} for further details about disordered hyperuniform states of matter.

The unique characteristics of disordered hyperuniform materials, i.e., the combination of both disordered liquid-like structures on small-scales and crystal-like hidden order on large scales, endow them with many superior physical properties, including wave propagation characteristics \cite{ref31, ref32, ref33, scattering, granchi2022near, park2021hearing, klatt2022wave, tavakoli2022over, cheron2022wave}, diffusion and electrical properties \cite{ref34, torquato2021diffusion, maher2022characterization}, mechanical properties
\cite{ref35, puig2022anisotropic} as well as optimal multifunctional characteristics \cite{ref36, kim2020multifunctional, torquato2022extraordinary}. A particularly important discovery was the demonstration that disordered ``stealthy'' hyperuniform materials possess large, complete and isotropic photonic band gaps, blocking all directions and polarizations \cite{ref31, ref32}, which had been thought not to be possible. Such exotic disordered photonic band gap materials enable waveguide  geometries that have advantages over their periodic counterparts \cite{Fl13}, which have also important ramifications for electronic and phononic device applications \cite{tang2022soft}. Subsequently, it was shown that disordered stealthy hyperuniform materials can be made fully transparent to electromagnetic waves \cite{new1, torquato2021nonlocal, new2}. These discoveries set a new paradigm for engineered disorder in photonic metamaterials \cite{wu2017effective, zhang2019metasurface} and optical applications \cite{zhang2019experimental, zhang2021hyperuniform, zhang2022reconfigurable}. Designer disordered hyperuniform materials have also been successfully fabricated or synthesized using different techniques \cite{ref37, ref38}. A recent review article \cite{yu2021engineered} describes engineered metamaterials for photonic applications with an emphasis on disordered hyperuniform materials.

An inverse problem of great importance is the generation of realizations of heterogeneous two-phase materials with a prescribed set of statistical descriptors, which is usually referred to as the material {\it construction} problem \cite{To02a, Ye98a, Ye98b}. To this end, a variety of numerical construction methods have been developed, including Gaussian random field method \cite{roberts1997statistical, dhu_rand_field}, phase recovery method \cite{niezgoda2008delineation, fullwood2008microstructure, cherkasov2021adaptive}, multi-point statistics \cite{okabe2005pore, hajizadeh2011multiple, tahmasebi2013cross, tahmasebi2012multiple}, and machine-learning based methods \cite{xu2014descriptor, cang2017microstructure, cang2018improving, yang2018microstructural,li2018transfer, cheng2022data} to name but a few.


\begin{figure}[ht]
\begin{center}
$\begin{array}{c}\\
\includegraphics[width=0.45\textwidth]{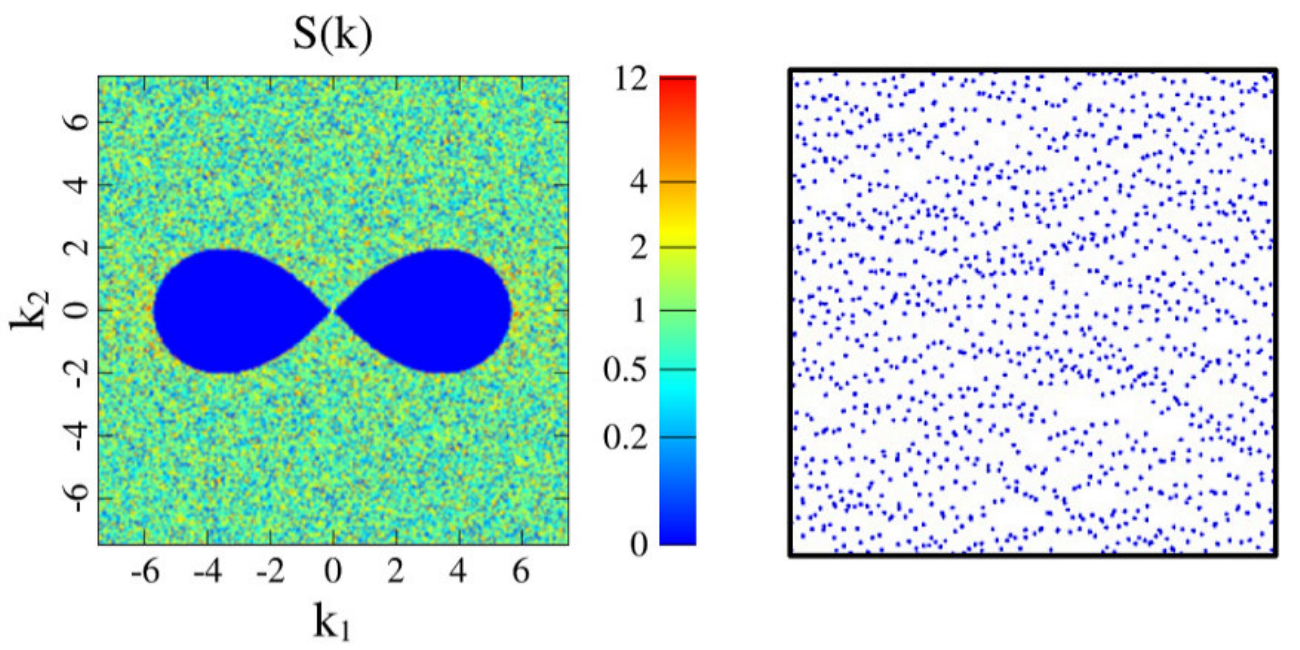}
\end{array}$
\end{center}
\caption{An example of directionally hyperuniform system \cite{To16a}. Left panel: A directionally hyperuniform scattering pattern in which the exclusion region $\Omega$ is a lemniscate shape around the origin where the scattering intensity is exactly zero (darkest shade). This ``stealthy'' pattern clearly shows that hyperuniformity depends on the direction in which the origin ${\bf k} = {\bf 0}$
is approached. Right panel: a statistically anisotropic point configuration associated with the scattering pattern.} \label{fig_anisotropy}
\end{figure}

\begin{figure*}[ht]
\begin{center}
$\begin{array}{c}\\
\includegraphics[width=0.85\textwidth]{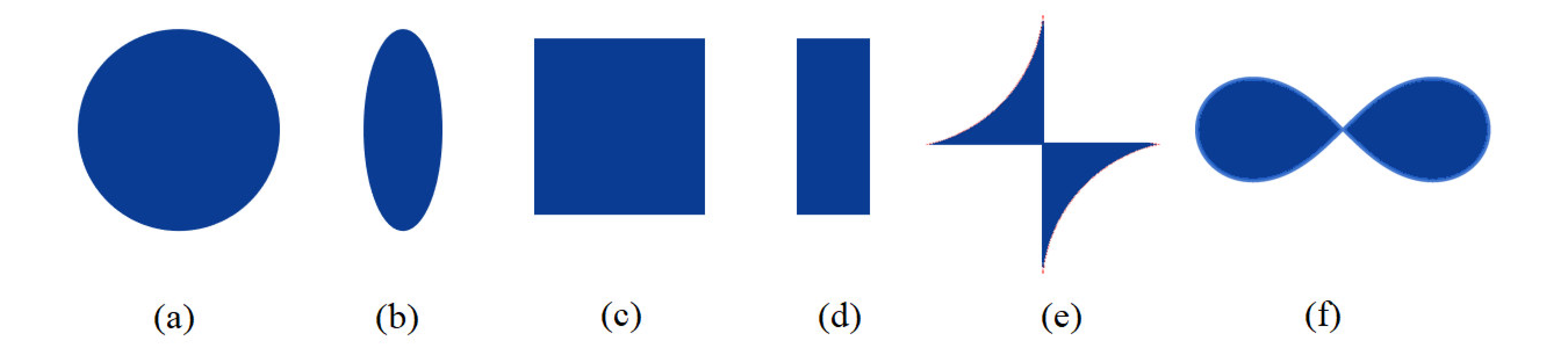}
\end{array}$
\end{center}
\caption{Illustration of the Fourier-space exclusion regions $\Omega$ investigated in this work. From left to right: (a) circular-disk, (b) elliptical-disk, (c) square, (d) rectangular, (e) butterfly-shaped and (f) lemniscate-shaped regions. } \label{fig_shape}
\end{figure*}

One of the most widely used construction methods is a procedure devised by Yeong and Torquato, who formulated the construction as an energy minimization problem and solved it using stochastic optimization \cite{Ye98a, Ye98b}. In particular, an energy is defined as the sum of the squared differences between a prescribed set of statistical microstructural descriptors and the corresponding descriptors computed from a trial microstructure. The simulated annealing method \cite{kirkpatrick1983optimization} is subsequently employed to evolve the trial microstructure in order to minimize the energy. The statistical descriptors such as various correlation functions \cite{To02a}, can be obtained either by sampling from microstructure images or theoretical considerations. In the former case, the problem is typically referred to as reconstruction. In the latter instance, the target functions need to satisfy the necessary {\it realizability} conditions in order to achieve a successful construction \cite{torquato2006necessary}. The Yeong-Torquato procedure has been employed to incorporate a variety of statistical descriptors \cite{jiao2007modeling, jiao2008modeling, jiao2009superior, chen2019hierarchical, chen2020probing} and applied to a wide spectrum of heterogeneous two-phase material systems \cite{gerke2015improving, karsanina2018hierarchical, feng2018accelerating, jiao2013modeling, chen2015dynamic, jiao2014modeling, guo2014accurate, chen2016stochastic, gerke2019calculation, karsanina2022stochastic, chen2022quantifying}.


Very recently, Chen and Torquato \cite{Ch18a} further generalized the Yeong-Torquato procedure to generate microstructures of {\it isotropic} disordered hyperuniform heterogeneous two-phase materials, from realizable {\it angular-averaged} spectral density functions ${\tilde \chi}_{_V}({|{\bf k}|})$, including both stealthy and non-stealthy hyperuniform ones. Importantly, this allowed them to design disordered stealthy hyperuniform dispersion that possesses nearly optimal effective conductivity, which are also transparent to electromagnetic radiation for certain wavelengths and can be readily experimentally fabricated using 3D printing and lithographic technologies.

Hyperuniformity has recently been generalized to treat structurally anisotropic systems \cite{To16a}, which necessitates the introduction of the concept of ``directional hyperuniformity'', i.e., hyperuniformity along certain directions in Fourier space and non-hyperuniformity along other directions. Figure \ref{fig_anisotropy} shows a directionally (anisotropic) hyperuniform scattering pattern (left panel) possessing a lemniscate-shaped region around the origin in which the scattering intensity is exactly zero (darkest shade), which was originally studied in Ref. \cite{To16a}. In this example, hyperuniformity clearly depends on the direction in which the origin ${\bf k} = {\bf 0}$ is approached. Specifically, the pattern is stealthy hyperuniform along the horizontal direction (i.e., $S({\bf k}) = 0$ for $k_x<K$ and $k_y = 0$, see definition below), and nonhyperuniform along the vertical direction \cite{To18a}. The right panel shows a statistically anisotropic point configuration that corresponds to the scattering pattern \cite{To16a}. It is seen that the points are arranged in ``wavy'' chains along the horizontal direction, and no such ``wavy'' patterns are observed along the vertical direction. Other examples of directionally hyperuniform point patterns include exotic ground states of directional pair potentials whose Fourier representations are non-zero on compact sets around the origin and zero everywhere else \cite{martis2013exotic}. Such anisotropic hyperuniform systems have important implications for the design of waveguide with engineered direction-dependent performance \cite{waveguide}. To the best of our knowledge, no systematic investigations on directionally hyperuniform heterogeneous two-phase materials have been carried out.


Here, we generalize the Fourier-space based numerical construction procedure developed by Chen and Torquato \cite{Ch18a} to generate digital realizations of {\it anisotropic} disordered hyperuniform heterogeneous two-phase material microstructures with designed spectral densities. Controlling the statistical anisotropy of composite microstructures via
the shape, size and discrete symmetries of $\Omega$ is crucial to engineering directional optical, transport and mechanical properties of the two-phase media for a wide spectrum of applications \cite{waveguide, ref33}, including achieving exotic anisotropic dispersion relations for electromagnetic and acoustic wave propagation \cite{damaskos1982dispersion, itin2010dispersion}. Our generalized construction procedure explicitly incorporates the vector-dependent spectral density function ${\tilde \chi}_{_V}({\bf k})$. We demonstrate the utility of the procedure by applying it to render a wide spectrum of anisotropic {\it stealthy hyperuniform} (SHU) composites \cite{To16b}, which possess a spectral density
\begin{equation}
{\tilde \chi}_{_V}({\bf k}) = 0, \quad\text{for}\quad {\bf k} \in \Omega,
\label{eq_1}
\end{equation}
where $\Omega$ is an ``exclusion'' region \cite{To15} around the origin in the $\bf k$-space in which scattering is completely suppressed, excluding forward scattering at ${\bf k=0}$. Thus, SHU composites anomalously suppress local volume fraction fluctuations from intermediate to infinite wavelengths. We systematically investigate the effects of shape anisotropy and size of the exclusion region $\Omega$ with certain discrete symmetries  on the resulting composite microstructure by constructing corresponding statistically anisotropic realizations as a function of
phase volume fraction. We begin by investigating SHU media with circular-disk regions and then consider anisotropic shapes, namely, elliptical-disk, square, and rectangular regions. Moreover, we study directionally hyperuniform composites associated with butterfly-shaped and lemniscate-shaped $\Omega$ regions, which are stealthy and hyperuniform only along certain directions, and are non-hyperuniform along others. These selected representative exclusion-region shapes are schematically illustrated in Fig. \ref{fig_shape}.



We find that while the circular-disk exclusion regions give rise to isotropic hyperuniform structures on both global and local scales, distinct anisotropic structures and degree of uniformity in the distribution of the phases on intermediate and large length scales along different directions, leading to directional hyperuniform behaviors, can result from the shape asymmetry (or anisotropy) of certain exclusion regions. Moreover, while the anisotropic exclusion regions impose strong constraints on the {\it global} symmetry of the resulting media, they can still possess almost isotropic {\it local} structures. Both the isotropic and anisotropic hyperuniform microstructures associated with the elliptical-disk, square and rectangular $\Omega$ possess phase-inversion symmetry over certain range of volume fractions and a percolation threshold $\phi_c \approx 0.5$. On the other hand, the directionally hyperuniform microstructures associated with the butterfly-shaped and lemniscate-shaped $\Omega$ do not possess phase-inversion symmetry and percolate along certain directions at much lower volume fractions. 


It is noteworthy that our present results together with theoretical predictions for effective properties that depend on the spectral density enable the inverse design of heterogeneous two-phase materials with desirable properties. Such predictive formulations include not only nonlocal theories for the effective dynamic dielectric constant \cite{kim2020effective, torquato2021nonlocal} that accurately accounts  for multiple scattering but also the spreadability for time-dependent diffusive transport behaviors \cite{torquato2021diffusion, wang2022dynamic}. Such theories allow one to achieve desirable anisotropic composite properties by tuning the microstructure to have a targeted spectral density ${\tilde \chi}_{_V}({\bf k})$ \cite{xu2022correlation, iyer2020designing, farooq2018spectral}. Once an optimized ${\tilde \chi}_{_V}({\bf k})$ is obtained, our procedure enables one to render realizations of the designed microstructure, which can subsequently experimental manufactured using, e.g., 3D printing techniques.

The rest of the paper is organized as follows: In Sec. II, we provide definition of correlation functions, spectral density and hyperuniformity in heterogeneous two-phase material systems. In Sec. III, we discuss the numerical construction procedure in detail, including its mathematical formulation as a constrained optimization problem and its solution via stochastic simulated annealing method. In Sec. IV, we present constructions of realizations of a variety of disordered SHU composites with prescribed exclusion-region shapes,
sizes and symmetries as a function of phase volume fraction. In Sec. VI, we provide concluding remarks and outlook of future work. The effects of large exclusion regions and application of our general method to non-hyperuniform stealthy composites, along with other supporting information and results, are presented in the Appendix.

\section{Definitions}
\label{definition}



\subsection{Correlation Functions}


Consider a two-phase random heterogeneous material (i.e., medium or a composite), which is a sub-domain of $d$-dimensional Euclidean space, i.e., $\mathcal{V} \subseteq \mathbb{R}^d$ of volume $V \leq +\infty$, composed of two regions $\mathcal{V} = \mathcal{V}_1 \cup \mathcal{V}_2$. $\mathcal{V}_1$ is the phase 1 region of volume $V_1$ and fraction $\phi_1 = V_1/V$; and $\mathcal{V}_2$ is the phase 2 region of volume $V_2$ and volume fraction $\phi_2 = V_2/V$. In the infinite-volume limit $V\rightarrow \infty$, $V_i$ ($i=1, 2$) also increases proportionally such that the ratio $V_i/V$ (i.e., the volume fraction $\phi_i$) tends to a well-defined constant. The statistical properties of each phase $i$ of the system are specified by the countably infinite set of {\it $n$-point correlation functions} $S_n^{(i)}$, which are defined by
\cite{To02a}:
\begin{equation}\label{Sndef}
S_n^{(i)}(\mathbf{x}_1, \ldots, \mathbf{x}_n) = \left\langle\prod_{i=1}^n I^{(i)}(\mathbf{x}_i)\right\rangle,
\end{equation}
where $I^{(i)}({\bf x})$ is the indicator function for phase $i$, i.e.,
\begin{equation}
I^{(i)}({\bf x}) =
\begin{cases}
1, \quad &\text{if } {\bf x} \in  \mathcal{V}_1\\
0, \quad &\text{otherwise.}
\end{cases}
\end{equation}
The function $S_n^{(i)}(\mathbf{x}_1, \ldots, \mathbf{x}_n)$ can also be interpreted to be the probability of randomly throwing down $n$ points at
positions $\mathbf{x}_1, \ldots, \mathbf{x}_n$ and having all of the points fall into the same phase $i$. Henceforth, we consider statistically homogeneous, i.e., there is no preferred origin in the system and thus $S_n^{(i)}$ only depends on the relative displacement between the points \cite{To02a}. The one-point function is simply the volume fraction of phase $i$, $\phi_i$, which is
a constant, i.e.,
\begin{equation}
S_1^{(i)}(\mathbf{x_1}) =  \phi_i,
\end{equation}
and the two-point $S_2^{(i)}({\bf r})$ depends only the relative displacement ${\bf r}={\bf x}_2-{\bf x}_1$. For media
without long-range order, $S_2({\bf r})$ possesses the following asymptotic behavior:
\begin{equation}
\lim_{|{\bf r}|\rightarrow \infty} S_2(\mathbf{r}) = \phi^2,
\end{equation}
where we henceforth drop indicating phase $i$ 
and simply refer to the phase of interest.


Upon subtracting the long-range
behavior from $S_2$, one obtains the {\it autocovariance function}, i.e.,
\begin{equation}
\chi_{_V}({\bf r})= S_2(\mathbf{r})-\phi^2
\end{equation}
which is generally an $L^2$-function. The associated spectral density
function $\Tilde{\chi}_{_V}({\bf k})$ is given by
\begin{equation}
\Tilde{\chi}_{_V}({\bf k}) = \int_{\mathbb{R}^d} \chi_{_V}({\bf r})e^{-i{\bf
k}\cdot{\bf r}}d{\bf r},
\end{equation}
which is the Fourier transform of $\chi_{_V}(r)$ and is obtainable from scattering
intensity measurements \cite{debye}.

The autocovariance function obeys the bounds \cite{torquato2006necessary}:
\begin{equation}
-\text{min}\{(1-\phi)^2, \phi^2\} \leq \chi_{_V}({\bf r})\leq (1-\phi)\phi,
\end{equation}
where $\phi$ is the volume fraction of the reference phase.
We remark that it is an open problem to identify additional necessary and sufficient conditions that the autocovariance function must satisfy in order to correspond to a binary stochastic process.


\subsection{Hyperuniform Random Heterogeneous Two-Phase Materials}


For a two-phase heterogeneous material, the quantity of interest is the local volume fraction variance $\sigma_{_V}^2(R)$, which was first introduced in Ref. \cite{lu1990local}:
\begin{equation}
\sigma_{_V}^2(R) = \frac{1}{v_1(R)}\int_{\mathbb{R}^d} I({\bf r})\alpha_2(r; R)d{\bf r},
\end{equation}
where $\alpha_2(r; R)$ is the scaled intersection volume, i.e., the intersection volume of two spherical windows of radius $R$ whose centers are separated by a distance $r$, divided by the volume $v_1(R)$ of the window, i.e.,
\begin{equation}
v_1(R)=\frac{\pi^{d/2} R^d}{\Gamma(1+d/2)}.
\label{v1}
\end{equation}



A hyperuniform random medium is one whose
$\sigma_{_V}^2(R)$ decreases more rapidly than $R^d$ for large $R$, i.e.,
\begin{equation}
\lim_{R\rightarrow\infty}\sigma_{_V}^2(R) \cdot R^d = 0.
\end{equation}
This behavior is to be contrasted with those of  typical disordered two-phase media for which the variance decays as $R^{-d}$, i.e., as the inverse of the window volume $v_1(R)$.


The hyperuniform condition is equivalently given by
\begin{equation}
\label{eq_hyper} \lim_{|{\bf k}|\rightarrow 0}\Tilde{\chi}_{_V}({\bf k})
= 0,
\end{equation}
which implies that  the  direct-space autocovariance function $\chi_{_V}({\bf r})$  exhibits both positive  and  negative  correlations  such  that  its volume  integral  over all space is exactly zero \cite{To16b}, i.e.,
\begin{equation}
\int_{\mathbb{R}^d} \chi_{_V}({\bf r})d{\bf r} = 0,
\end{equation}
which is a direct-space sum rule for hyperuniformity. 


For hyperuniform two-phase media whose spectral density goes to zero as a power-law scaling as $|\bf k|$ tends to zero \cite{To16a}, i.e.,
\begin{equation}
    {\tilde \chi}_{_V}({\bf k})\sim |{\bf k}|^\alpha
\end{equation}
there are three different scaling regimes (classes) that describe the associated large-$R$ behaviors of the local volume fraction variance:
\begin{equation}
\sigma^2_{_V}(R) \sim
\begin{cases}
R^{-(d+1)}, \quad\quad\quad \alpha >1 \qquad &\text{(Class I)}\\
R^{-(d+1)} \ln R, \quad \alpha = 1 \qquad &\text{(Class II)}\\
R^{-(d+\alpha)}, \quad 0 < \alpha < 1\qquad  &\text{(Class III).}
\end{cases}
\label{eq:classes}
\end{equation}
Classes I and III are the strongest and weakest forms of hyperuniformity, respectively. Class I media include all crystal structures \cite{To03}, many quasicrystal structures \cite{Og17} and exotic disordered media \cite{Za09, Ch18a}. Examples of Class II systems include some quasicrystal structures \cite{Og17}, perfect glasses \cite{zhang2017classical}, and maximally random jammed packings \cite{ref4, ref5, ref6, Za11c, Za11d}. Examples of Class III systems include classical disordered ground states \cite{Za11b}, random organization models \cite{ref20}, perfect glasses \cite{zhang2017classical}, and perturbed lattices \cite{Ki18a}; see Ref. \cite{To18a} for a more comprehensive list of systems that fall into the three hyperuniformity classes. SHU media, the focus of this study, are also of class I. Known examples of such media are periodic packings of spheres as well as unusual disordered sphere packings derived from stealthy point patterns for which $\Omega$ is a spherical/circular-disk region \cite{To16b}.

\section{Methods}

\subsection{Mathematical Formulation}

In the {\it construction} problem, one aims to find a digital realization (e.g., represented by a binary-valued matrix with entries equal to 0 or 1) associated with a prescribed set of statistical descriptors. Here, we focus on the construction of a disordered SHU two-phase medium with a prescribed spectral density $\tilde \chi_{_V}({\bf k})$, if realizable. We consider digitized medium in a square domain of length $L$ in $\mathbb{R}^2$ with periodic boundary conditions. We note that our construction formulation and procedure can be readily generalized to three dimensional systems. In this case, the indicator function $I({\bf r})$ also takes a discrete form, i.e., $I({\bf r})$ is only defined on a discrete set of ${\bf r} = n_1 {\bf e}_1 + n_2 {\bf e}_2$, where ${\bf e}_i$ are unit vectors along the orthogonal directions and $n_1, n_2 = 0, 1, \ldots, N$ with $N$ being the system size or ``resolution'' (i.e., number of pixels along each dimension). The Fourier-space wavevectors for the system also take discrete values, i.e., ${\bf k} = (2\pi/L)(n_1 {\bf e}_1 + n_2 {\bf e}_2)$. The corresponding spectral density for the digital realization is given by
\begin{equation}
    \tilde \chi_{_V}({\bf k}) = \frac{1}{L^2}{\tilde m^2}({\bf k})|\tilde J (\bf k)|^2
    \label{eq_chi_k}
\end{equation}
where $\tilde J (\bf k)$ is the {\it generalized collective coordinate} \cite{To18a, Ch18a} defined as
\begin{equation}
 \tilde J (\bf k) = \sum_{{\bf r}}\exp(i{\bf k}\cdot{\bf r}) J({\bf r})
   \label{eq_I_k}
\end{equation}
where the sum is over all pixel centers ${\bf r}$, and
\begin{equation}
J({\bf r}) = I({\bf r}) - \phi.
   \label{eq_J}
\end{equation}
The quantity ${\tilde m}({\bf k})$ is the Fourier transform
of the shape function (or indicator function) $m({\bf r})$ of a square pixel which is given by
\begin{equation}
{\tilde m}({\bf k}) =
\begin{cases}
\frac{sin(k_x/2)}{k_x/2}\frac{sin(k_y/2)}{k_y/2}, \quad k_x\neq 0,  k_y\neq 0 \\
\frac{sin(k_x/2)}{k_x/2}, \quad\quad\quad\quad k_x\neq 0,  k_y = 0 \\
\frac{sin(k_y/2)}{k_y/2}, \quad\quad\quad\quad k_x = 0,  k_y\neq 0\\
1, \quad\quad\quad\quad\quad\quad\quad k_x = 0, k_y = 0.
\end{cases}
\end{equation}
where $k_x$ and $k_y$ are the components of the discrete wavevector ${\bf k}$. 

Here we are interested in disordered SHU system, which is characterized by $\tilde \chi_{_V}({\bf k}) = 0$ for wavevectors in the exclusion region, i.e., ${\bf k}\in \Omega$. This directly imposes a set of constraints on the discrete indicator function $I({\bf r})$ through Eqs. (\ref{eq_chi_k}) to (\ref{eq_J}), i.e.,
\begin{equation}
    {\tilde m^2}({\bf k})|\sum_{\bf r}\exp(i{\bf k}\cdot{\bf r})[I({\bf r})-\phi)]|^2 = 0
    \label{eq_chi_Ir}
\end{equation}
for ${\bf k}\in \Omega$. We note Eq. (\ref{eq_chi_Ir}) represents a set of $N_\Omega$ number of nonlinear equations of $I({\bf r})$ (and equivalently of $J({\bf r})$), where $N_\Omega$ is the number of {\it independent} $\bf k$ points in $\Omega$. We note that due to the symmetry of the spectral density function $\tilde \chi_{_V}({\bf k})$ \cite{Ch18a}, only a half of ${\bf k}$ points in $\Omega$ are independent and thus, $N_\Omega \sim \frac{1}{2}Vol(\Omega)$. For a digital realization of linear size $L$, the total number of pixels is $N = L^2$. The number of unknowns in Eq. (\ref{eq_chi_Ir}), i.e., the value of each pixel, is also $N$. We are interested in the cases $N_\Omega < N$, i.e., the number of constraints (equations) is much smaller than that of unknowns. Therefore, Eq. (\ref{eq_chi_Ir}) does not have unique solutions and we will employ stochastic optimization method to iteratively solve Eq. (\ref{eq_chi_Ir}).

As with stealthy point configurations \cite{To18a, ref9}, it is also convenient to introduce a ratio $\chi$ between the number of constraints (equations) and the number of unknowns (degrees of freedom):
\begin{equation}
    \chi = N_\Omega/(N-2)
    \label{eq_chi_ratio}
\end{equation}
which is the fraction of constrained degree of freedom in the systems. We note 2 degrees of freedom associated with the trivial overall translation of the entire system are subtracted in Eq. (\ref{eq_chi_ratio}). In the case of point configurations, it has been shown that increasing $\chi$ leads to increased degree of order in the SHU many-particle systems \cite{To18a, Ba08}. It is expected that increasing $\chi$ requires suppression of local volume fraction fluctuations on a broader range of length scales, which leads to microstructures with very fine and uniformly dispersed phase morphologies.





\subsection{Simulation Annealing Procedure}


We employ the simulated annealing procedure \cite{kirkpatrick1983optimization} to solve Eq. (\ref{eq_chi_k}), which has been widely used in composite construction problems \cite{Ye98a, Ye98b, jiao2007modeling, jiao2008modeling, jiao2009superior, jiao2013modeling, jiao2014modeling}. In particular, the construction problem is formulated as
an ``energy'' minimization problem, with the energy functional $E$
defined as follows
\begin{equation}
\label{eq208} E =
\sum\limits_{{\bf k}\in \Omega_I}\left[{\tilde \chi_{_V}({\bf k})-\tilde \chi_{_V}^{0}({\bf k})}\right]^2,
\end{equation}
where $\tilde \chi_{_V}^{0}({\bf k})$ is the target spectral density function
and  $\tilde \chi_{_V}({\bf k})$ is the corresponding
function associated with a trial microstructure, $\Omega_I$ is the set of {\it independent} ${\bf k}$ points due to the symmetry of $\tilde \chi_{_V}({\bf k})$, which is defined as
\begin{equation}
\Omega_I = \{ {\bf k}\in \Omega,~ k_x >0, ~k_x = 0 \cap k_y>0 \}.
\end{equation}
For stealthy systems, we have $\tilde \chi_{_V}^{0}({\bf k}) = 0$ for ${\bf k}\in \Omega$. Thus, Eq. (\ref{eq208}) simply reduces to
\begin{equation}
\label{eq209} E =
\sum\limits_{{\bf k}\in \Omega_I}\left[{\tilde \chi_{_V}({\bf k})}\right]^2.
\end{equation}
We note that our procedure can be employed to generate realizations with arbitrary $\tilde \chi_{_V}^{0}({\bf k})$, which allows us to engineer anisotropic scattering properties for these composites.


The simulated annealing procedure is then employed to solve the
aforementioned minimization problem. Specifically, starting from
an initial trial configuration (i.e., old realization) which
contains a fixed number of pixels for each phase consistent with
the volume fraction of that phase with an energy $E_{old}$, two randomly selected pixels
associated with different phases are exchanged to generate a new
trial microstructure. Relevant correlation functions are sampled
from the new trial configuration and the associated energy $E_{new}$ is
evaluated, which determines whether the new trial configuration
should be accepted or not via the probability \cite{kirkpatrick1983optimization}:
\begin{equation}
\label{eq_pacc} p_{acc} = min \{\exp(-\Delta E/T), 1\},
\end{equation}
where $\Delta E = E_{new} - E_{old}$ is the energy difference between the new and old
trial configurations and $T$ is a virtual temperature that is
chosen to be initially high and slowly decreases according to a
cooling schedule. An appropriate cooling
schedule reduces the chances that the system gets stuck in a
shallow local energy minimum. In practice, a power law schedule
$T(n) = \gamma^n T_0$ is usually employed, where $T_0$ is the
initial temperature, $n$ is the cooling stage and $\gamma \in (0,
1)$ is the cooling factor ($\gamma = 0.99$ is used here). The
simulation is terminated when $E$ is smaller than a prescribed
tolerance (e.g., $10^{-6}$ in this case).

Generally, a large number of trial configurations ($\sim 10^7$) need to be searched to generate a
successful construction. Therefore, highly efficient
methods \cite{Ch18a} are used that enable one to
rapidly obtain the spectral density function of a new configuration by updating the corresponding function associated
with the old configuration, instead of completely re-computing
the function from scratch.
In particular, the generalized collective coordinate $\tilde I({\bf k})$ is tracked during the construction process. At the
beginning of the simulation, $\tilde I({\bf k})$ of the initial configuration is computed from scratch and the values for all ${\bf k}$ are stored. After each pixel exchange, since only a single pixel of the phase of interest changes its position from ${\bf r}_{old}$ to ${\bf r}_{new}$, we can then obtain the updated $\tilde I({\bf k})$ values by only explicitly computing the contributions from this changed pixel, i.e.,
\begin{equation}
\tilde I({\bf k}) \leftarrow \tilde I({\bf k}) + \delta \tilde I_{new}({\bf k}) - \delta \tilde I_{old}({\bf k}),
\end{equation}
where
\begin{equation}
\delta \tilde I_{new}({\bf k}) = \exp(i{\bf k}\cdot {\bf r}_{new}), \quad \delta \tilde I_{old}({\bf k}) = \exp(i{\bf k}\cdot {\bf r}_{old})
\end{equation}
Once the updated $\tilde I({\bf k})$ is obtained, the updated $\tilde \chi_{_V}({\bf k})$ can be immediately computed using Eq. (\ref{eq_chi_k}).

To enhance the convergence of the construction, we also employ surface optimization technique \cite{jiao2008modeling}: towards the end of the construction process, instead of randomly selecting pixels throughout the systems with equal probability, the pixels that are isolated or on the surface of connected regions of the phase of interest are assigned  larger probabilities of being selected compared to those within the phase regions. Interestingly, as we will show in Sec. IV, even with surface optimization the construction renders realizations containing dispersed pixels or small clusters of pixels for large $\chi$ values.

In our subsequent constructions, we mainly consider realizations in a square domain with $L = 300$ pixels and periodic boundary conditions. We have also investigated smaller and larger system sizes, including $L = 150$ and $L = 600$ pixels to verify that a resolution of $L = 300$
pixels does not affect the construction results.


\section{Results}

\begin{figure*}[ht]
\begin{center}
$\begin{array}{c}\\
\includegraphics[width=0.7\textwidth]{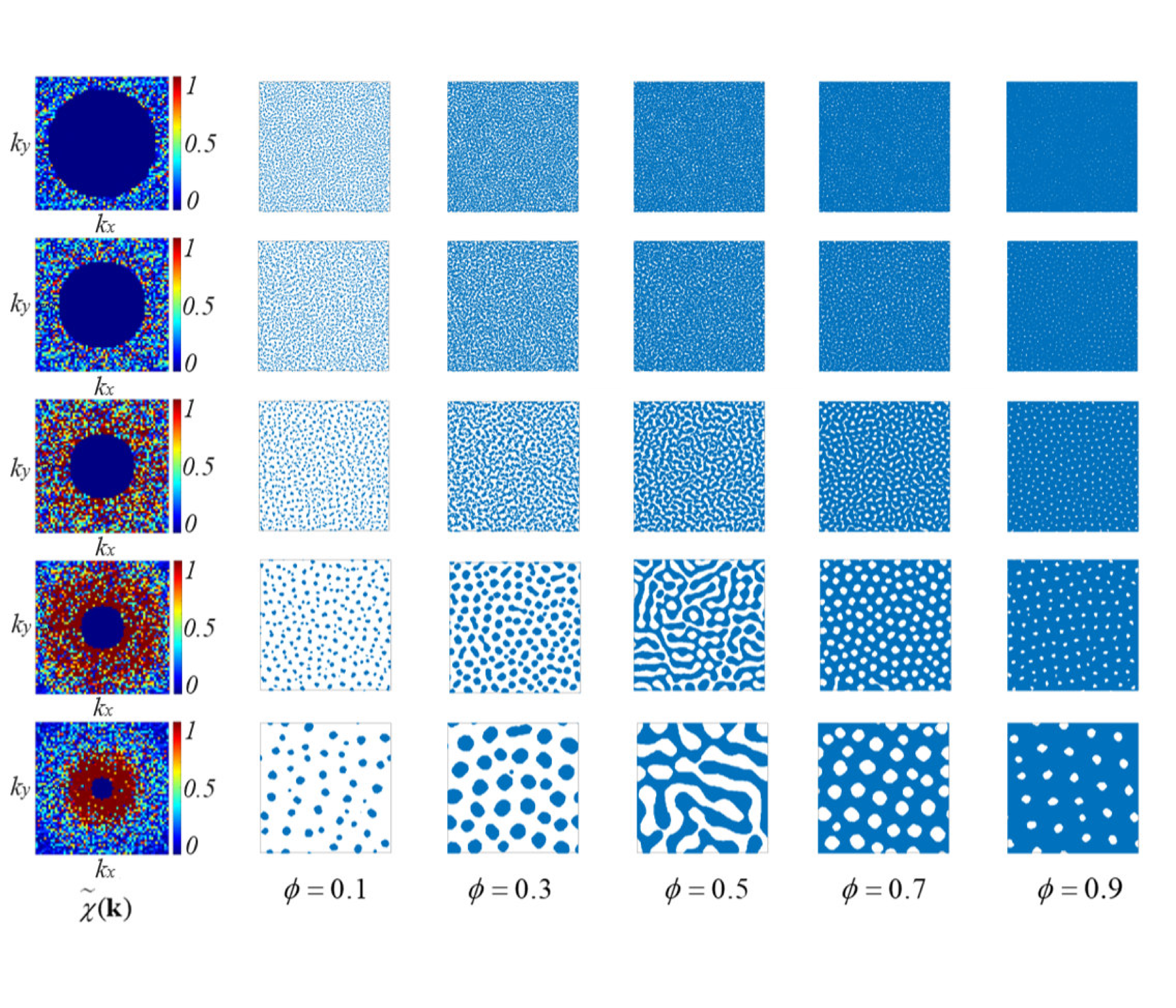}
\end{array}$
\end{center}
\caption{Isotropic hyperuniform composites with circular-disk exclusion regions with varying radius $\ell$ and phase volume fraction $\phi$. The left most column shows representative $\tilde \chi_{_V}({\bf k})$ (associated with $\phi = 0.5$) for each radius $\ell$: from bottom to top $\ell = 5, 10, 15, 20, 25$ (in units of $2\pi/L$), corresponding to $\chi \approx 8.7\times10^{-4}, 3.5\times10^{-3}, 7.9\times10^{-3}, 1.4\times10^{-2}, 2.2\times10^{-2}$, respectively. The phase volume fractions for the realizations from left to right are $\phi = 0.1, 0.2, 0.3, 0.4, 0.5$, respectively. The linear size of the system is $L = 300$ pixels.} \label{fig_circle}
\end{figure*}

\begin{figure*}[ht]
\begin{center}
$\begin{array}{c}\\
\includegraphics[width=0.9\textwidth]{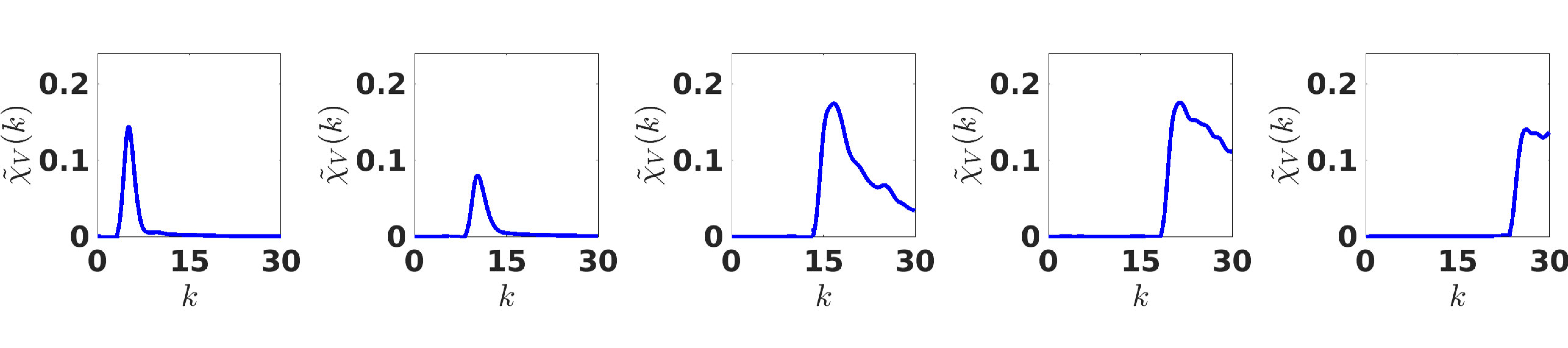}
\end{array}$
\end{center}
\caption{Representative angle-averaged $\tilde \chi_{_V}({k})$ (where $k = |{\bf k}|$) associated with the constructed SHU composites with $\phi = 0.5$ and a circular-disk $\Omega$ region in the Fourier space for different radius $\ell$: from left to right $\ell = 5, 10, 15, 20, 25$ (in units of $2\pi/L$).} \label{fig_circle_chi}
\end{figure*}

In this section, we present constructions of anisotropic SHU composite microstructures across volume fractions associated with exclusion regions with a prescribed shape, size and symmetry. Specifically, we consider circular-disk, elliptical-disk, square, rectangular, butterfly-shaped and lemniscate-shaped exclusion regions.  The size of $\Omega$ for a given exclusion-region shape is controlled by a characteristic length scale $\ell$ of the shape (e.g., radius for circles, semi-axis length for ellipses, edge lengths for squares and rectangles), which will specified in detail below. We note that the parameter $\chi$ that characterizes the ratio of the number of constraints (i.e., number of independent wavevectors ${\bf k}$ in $\Omega$) over the number of total degrees of freedom can be estimated as $\chi \approx Vol(\Omega(\ell))/L^2$ for a given $\ell$, which will also be specified for each case below.

\begin{figure*}[ht]
\begin{center}
$\begin{array}{c}\\
\includegraphics[width=0.7\textwidth]{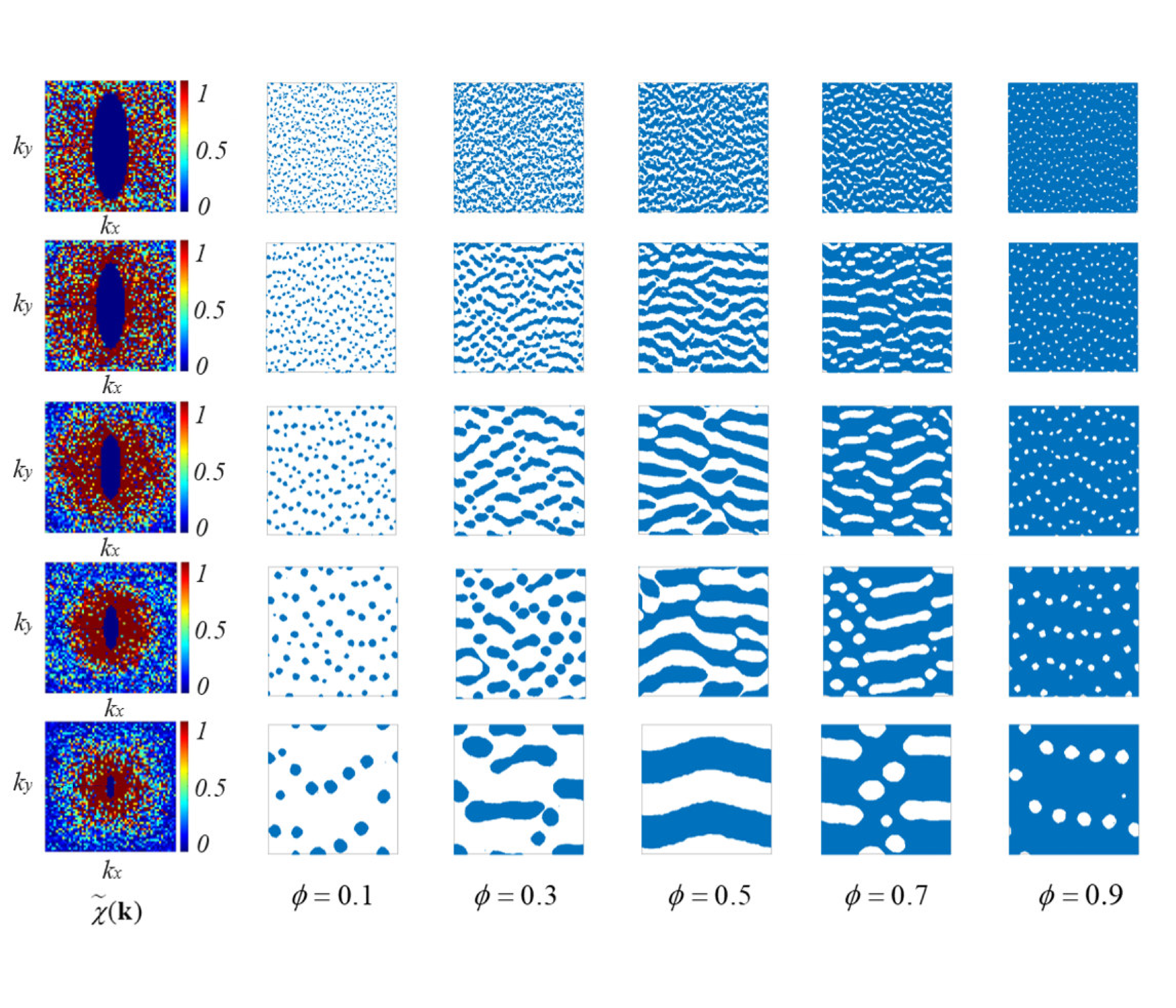}
\end{array}$
\end{center}
\caption{Anisotropic hyperuniform composites with elliptical-disk exclusion regions with varying size $\ell$ (i.e., the length of long semi-axis) and volume fraction $\phi$. The left most column shows representative $\tilde \chi_{_V}({\bf k})$ (associated with $\phi = 0.5$) for each long semi-axis $\ell$: from bottom to top $\ell = 5, 10, 15, 20, 25$ (in units of $2\pi/L$), corresponding to $\chi \approx 2.9\times10^{-4}, 1.2\times10^{-3}, 2.6\times10^{-3}, 4.7\times10^{-3}, 7.3\times10^{-3}$, respectively. The phase volume fractions for the realizations from left to right are $\phi = 0.1, 0.2, 0.3, 0.4, 0.5$, respectively. The linear size of the system is $L = 300$ pixels.} \label{fig_ellipse}
\end{figure*}

\begin{figure*}[ht]
\begin{center}
$\begin{array}{c}\\
\includegraphics[width=0.9\textwidth]{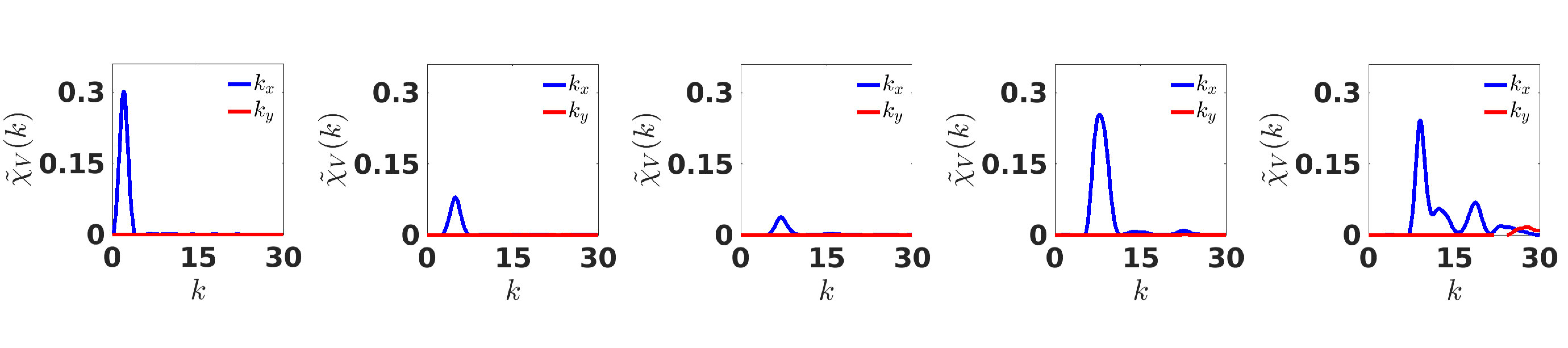}
\end{array}$
\end{center}
\caption{Representative averaged $\tilde \chi_{_V}({k})$ (where $k = |{\bf k}|$) along two orthogonal directions (horizontal and vertical) in the Fourier space associated with the constructed SHU composites with $\phi = 0.5$ and an elliptic exclusion region for different long semi-axis $\ell$: from left to right $\ell = 5, 10, 15, 20, 25$ (in units of $2\pi/L$).} \label{fig_ellipse_chi}
\end{figure*}

\subsection{Isotropic Media with Circular-Disk Exclusion Regions}

To validate the construction procedure based on vector spectral densities $\tilde \chi_{_V}({\bf k})$, we first generate realizations of SHU composites possessing zero spectral density with a circular-disk exclusion region $\Omega(\ell)$ with continuous rotational symmetry and varying radius $\ell$ (see Fig. \ref{fig_shape}a), i.e.,
\begin{equation}
\Omega(\ell) = \{ {\bf k} = \frac{2\pi}{L} {\bf n}: {\bf n}\in \mathbb{Z}^3, (\frac{k_x}{\ell})^2 + (\frac{k_y}{\ell})^2 \le 1 \}.
\end{equation}
This corresponds to the isotropic stealthy system investigated in Ref. \cite{Ch18a} characterized by angularly-averaged spectral density $\tilde \chi_{_V}({k}) = 0$ for $k\le \ell$. The size of $\Omega$ is chosen to be $\ell = 5, 10, 15, 20, 25$, corresponding to $\chi \approx 8.7\times10^{-4}, 3.5\times10^{-3}, 7.9\times10^{-3}, 1.4\times10^{-2}, 2.2\times10^{-2}$, respectively. The full spectrum of phase volume fractions from 0 to 1 was investigated and composite microstructures with $\phi = 0.1, 0.3, 0.5, 0.7, 0.9$ are shown here, which are representative of a wide spectrum of morphologies of the system.


The generated microstructures are shown in Fig. \ref{fig_circle}, along with the representative $\tilde \chi_{_V}({\bf k})$ associated with $\phi = 0.5$ for different $\ell$ (left column). Figure \ref{fig_circle_chi} shows the corresponding angle averaged spectral density function $\tilde \chi_{_V}({k})$ ($k = |{\bf k}|$), which exhibit a ``peak'' immediately beyond the exclusion region and the peak value diminishes as the exclusion region size increases. We note that similar features are observed for $\tilde \chi_{_V}({\bf k})$ associated the same region size for all phase volume fractions, and all different $\Omega$ regions considered here. Therefore, we only show the representative cases with $\phi = 0.5$

The realizations from left to right corresponding to increasing $\phi$; and from bottom to top corresponding to increasing $\ell$. Taking the realization with $\phi = 0.1$ and $\ell = 5$ as a starting point, it can be seen that the phase of interest forms statistically isotropic particles with similar sizes, which overall possess an isotropic distribution. These are expected for a circular-disk $\Omega(\ell)$ and are consistent with the observations reported in Ref. \cite{Ch18a}. As $\phi$ increases while keeping $\ell$ the same, it can be seen that the size of the particles increases with increasing $\phi$, and the ``particles'' merge into ligaments which eventually form a connected morphology. 

Interestingly, the microstructures appear to possess the so-called {\it phase-inversion symmetry} \cite{To03}, i.e., the microstructure associated with a volume fraction $\phi$ is statistical equivalent to that with a volume fraction $1-\phi$ with the two phases exchanged. This can be clearly seen by visually comparing the pairs of microstructures with $\phi = 0.1$ and $0.9$, and $\phi = 0.3$ and $0.7$ shown in Fig. \ref{fig_circle}. Detailed analysis of the rescaled autocovariance functions, which are reported in Sec. IV.G and Appendix A below, indicates the constructed microstructures with a volume fraction $\phi$ approximately in the range [0.4, 0.6] possess a high degree of phase inversion symmetry (i.e., the corresponding microstructures possess virtually identical rescaled autocovariance functions, see Sec. IV.G). As a consequence of the phase inversion symmetry, the resulting {\it isotropic} composite microstructures possess a percolation threshold $\phi_c = 0.5$ \cite{To03}, at which a system-spanning cluster of the phase of interest (i.e., the ``blue'' phase) first emerges.


In addition, we observe an interesting morphology evolution when increasing the size $\ell$ of the circular-disk exclusion region while keeping $\phi$ fixed. Using the realizations with $\phi = 0.1$ as examples, it can be clearly seen that as $\ell$ increases the particles become finer, i.e., they get smaller in size while remain a statistically isotropic morphology (expect for very small particle where the intrinsic anisotropy of the square grid is manifested) and statistically isotropic distributions. Similar trend is also observed for other volume fractions, e.g., for the connected morphology, the width of the ligaments gets smaller as $\ell$ increases. We note that microstructures at the same $\phi$ and different $\ell$ are not related by a simple scaling, i.e., they are not self-similar to one another.



These behaviors can be understood from the formulation of the construction problem. As discussed in Sec. III.A, the construction is formulated as a constrained optimization problem, in which the variables are the discrete values (1 or 0) of individual pixels and constraints are imposed through the zero-value spectral density $\tilde \chi_{_V}({\bf k})$ at certain wavevectors ${\bf k}$. The number of constraints is determined by the number of ${\bf k}$ points in the circular-disk exclusion region. When the region size $\ell$ is small, the pixels are less constrained and can be organized freely to form clustering morphologies. On the other hand, for large $\ell$ the degrees of freedom for each pixel are activated in order to satisfy the large number of constraints. In this case, the pixels behavior like non-overlapping particles moving on the construction grid, and thus are difficult to form clusters. In addition, increasing $\ell$ (i.e., the size of $\Omega$) in Fourier space requires the resulting composites need to be very ``uniform'' on smaller and smaller scales in real space. This forces the realizations to possess finer morphology composed of individual pixels instead of clusters of pixels.

To further verify this, we consider very large exclusion regions (i.e., large $\chi$ values) and present the construction results in Appendix C. It can be seen in Fig. \ref{fig_large_circle} that as $\chi$ increases, individual pixels in the realizations are mutually separated as if they were particles with repulsive interactions, and the over distribution resembles that for a SHU point configuration \cite{To18a},  albeit on a square grid. Although the realizations associated with large $\Omega$ (i.e., large $\ell$) exhibit interesting particle-like behaviors, in the subsequent discussions, we will focus on cases with relatively small $\ell$, corresponding to heterogeneous two-phase material regime.



In summary, we have found that the circular-disk exclusion regions result in isotropic hyperuniform structures on both global and local scales. The isotropic composite microstructures with $\phi$ approximately in the range [0.4, 0.6] possess phase-inversion symmetry. We speculate that the emergence of phase-inversion symmetry is due to the fact that the stealthy spectral density function effectively imposes constraints on the microstructures beyond the exclusion region, especially for microstructures with $\phi$ in the vicinity of 0.5. Increasing phase volume fractions lead to enhanced connectedness of the phase of interest, which percolates at $\phi_c = 0.5$. Moreover, increasing the size of the exclusion region results in a finer phase morphology, which is required to suppress local volume fraction fluctuations over a broader range of length scales associated with the larger exclusion region.





\begin{figure*}[ht]
\begin{center}
$\begin{array}{c}\\
\includegraphics[width=0.7\textwidth]{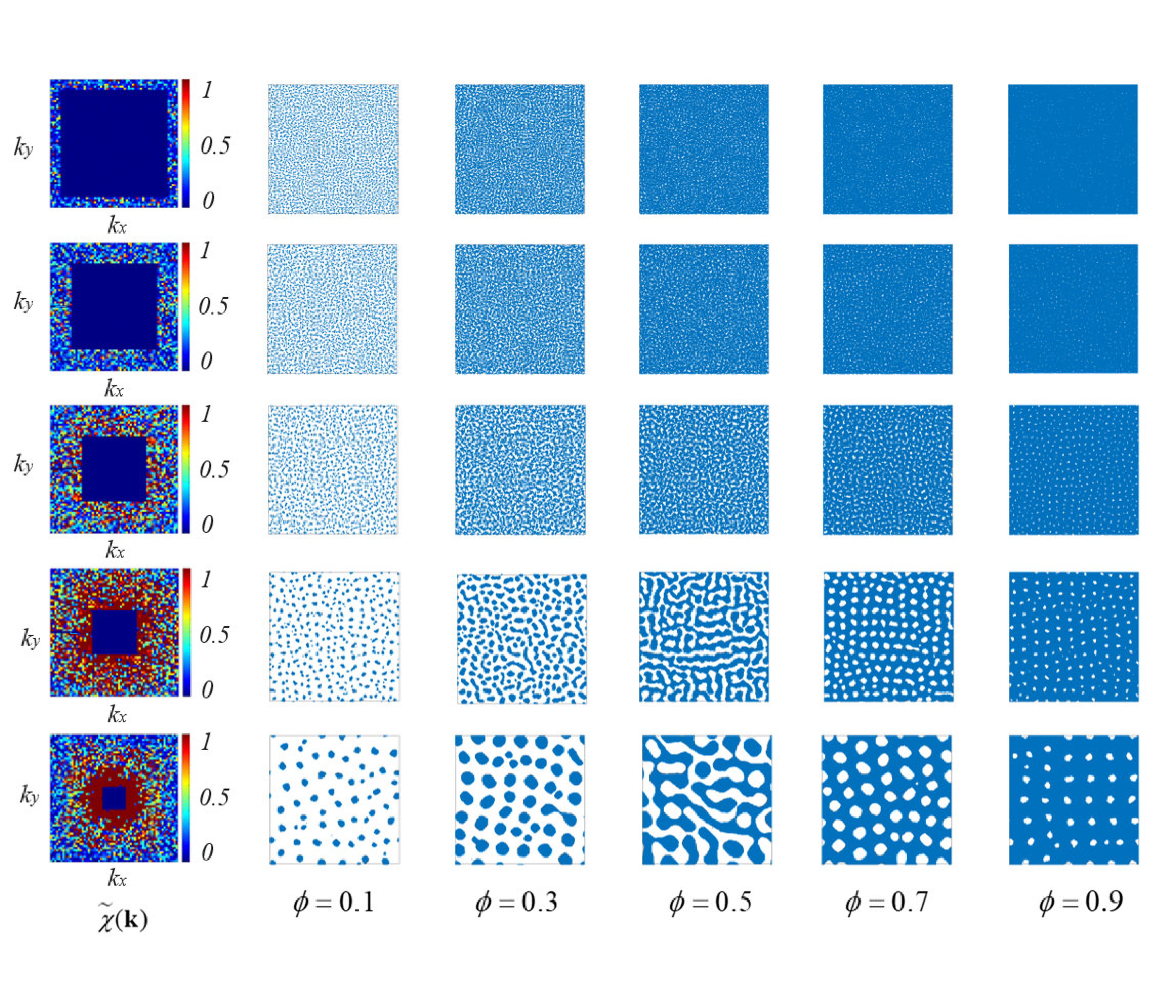}
\end{array}$
\end{center}
\caption{Anisotropic hyperuniform composites with square exclusion regions with varying size $\ell$ (i.e., edge length) and volume fraction $\phi$. The left most column shows representative $\tilde \chi_{_V}({\bf k})$ (associated with $\phi = 0.5$) for each edge length $\ell$: from bottom to top $\ell = 10, 20, 35, 40, 50$ (in units of $2\pi/L$), corresponding to $\chi \approx 1.1\times10^{-3}, 4.4\times10^{-3}, 9.9\times10^{-3}, 1.8\times10^{-2}, 2.8\times10^{-2}$, respectively. The phase volume fractions for the realizations from left to right are $\phi = 0.1, 0.2, 0.3, 0.4, 0.5$, respectively. The linear size of the system is $L = 300$ pixels.} \label{fig_square}
\end{figure*}

\begin{figure*}[ht]
\begin{center}
$\begin{array}{c}\\
\includegraphics[width=0.9\textwidth]{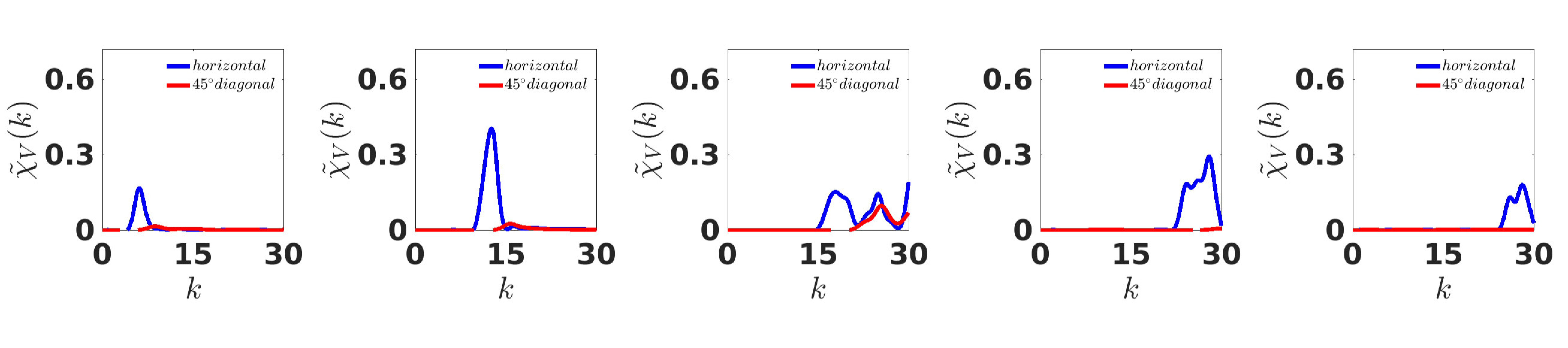}
\end{array}$
\end{center}
\caption{Representative averaged $\tilde \chi_{_V}({k})$ (where $k = |{\bf k}|$) along two directions (horizontal and diagonal) in the Fourier space associated with the constructed SHU composites with $\phi = 0.5$ and a square exclusion region for different edge length $\ell$: from left to right $\ell = 5, 10, 15, 20, 25$ (in units of $2\pi/L$).} \label{fig_square_chi}
\end{figure*}

\begin{figure*}[ht]
\begin{center}
$\begin{array}{c}\\
\includegraphics[width=0.7\textwidth]{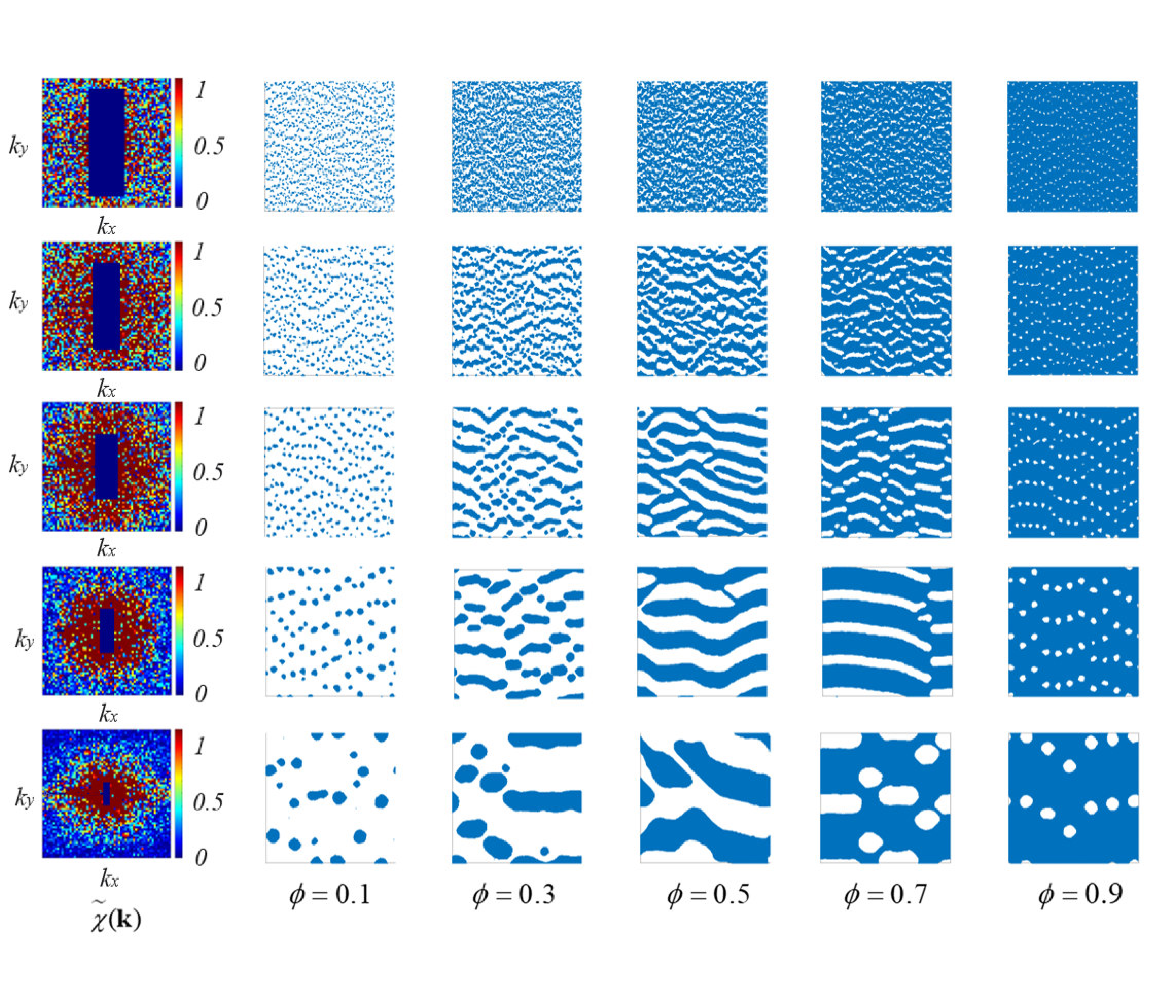}
\end{array}$
\end{center}
\caption{Anisotropic hyperuniform composites with rectangular exclusion regions with varying size $\ell$ (i.e., long-edge length) and volume fraction $\phi$. The left most column shows representative $\tilde \chi_{_V}({\bf k})$ (associated with $\phi = 0.5$) for each long-edge length $\ell$: from bottom to top $\ell = 10, 20, 30, 40, 50$ (in units of $2\pi/L$), corresponding to $\chi \approx 3.7\times10^{-4}, 1.5\times10^{-3}, 3.3\times10^{-3}, 5.9\times10^{-3}, 9.3\times10^{-3}$, respectively. The phase volume fractions for the realizations from left to right are $\phi = 0.1, 0.2, 0.3, 0.4, 0.5$, respectively. The linear size of the system is $L = 300$ pixels.} \label{fig_rectangle}
\end{figure*}

\begin{figure*}[ht]
\begin{center}
$\begin{array}{c}\\
\includegraphics[width=0.9\textwidth]{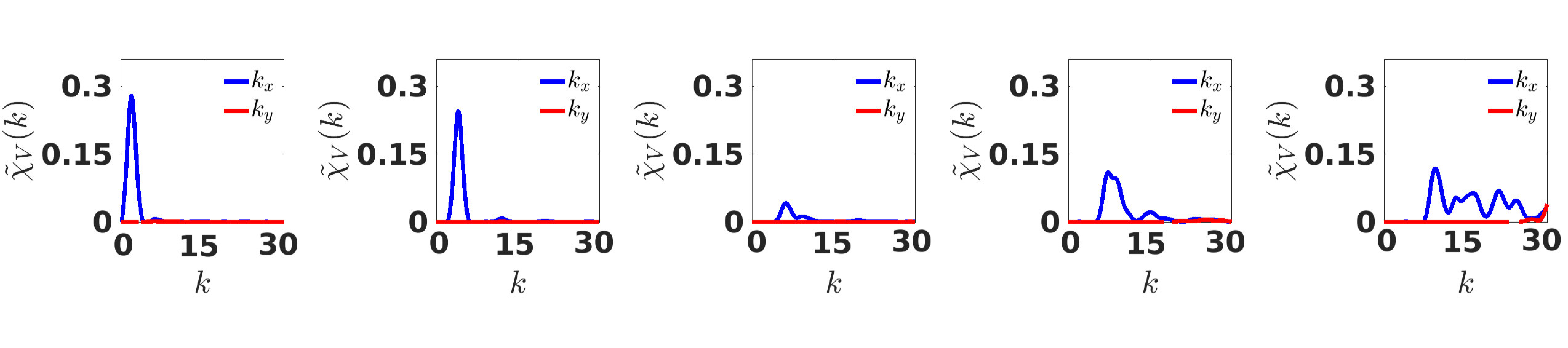}
\end{array}$
\end{center}
\caption{Representative averaged $\tilde \chi_{_V}({k})$ (where $k = |{\bf k}|$) along two orthogonal directions (horizontal and vertical) in the Fourier space associated with the constructed SHU composites with $\phi = 0.5$ and a rectangle exclusion region for different edge length $\ell$: from left to right $\ell = 5, 10, 15, 20, 25$ (in units of $2\pi/L$).} \label{fig_rectangle_chi}
\end{figure*}

\subsection{Anisotropic Media with Elliptical-Disk Exclusion Regions}




We now consider the generation of statistically anisotropic media 
associated with an elliptical-disk exclusion region $\Omega$, which is an affine transformation of a circular disk. Hence, it breaks the continuous rotational symmetry of a circle and possesses two-fold rotational symmetry. Specifically, we consider an elliptic exclusion region with an aspect ratio of 1/3 and a long semi-axis with length $\ell$ (see Fig. \ref{fig_shape}b), defined as
\begin{equation}
\Omega(\ell) = \{ {\bf k} = \frac{2\pi}{L} {\bf n}: {\bf n}\in \mathbb{Z}^3, (\frac{k_x}{\ell/3})^2 + (\frac{k_y}{\ell})^2 \le 1 \}.
\end{equation}
The representative $\tilde \chi_{_V}({\bf k})$ with varying $\ell$ (left column) and the associated realizations are shown in Fig. \ref{fig_ellipse}. Figure \ref{fig_ellipse_chi} shows the corresponding averaged spectral density function $\tilde \chi_{_V}({k})$ ($k = |{\bf k}|$) along the horizontal and vertical directions.

We find that, similar to the circular-disk $\Omega$ cases, increasing $\phi$ leads to increased connectivity and degree of clustering in the constructed microstructures for all $\ell$ values. The microstructures with $\phi$ approximately in the range [0.4, 0.6] also possess a high degree of phase-inversion symmetry (see Sec. IV.G for details) and percolation of the ``blue'' phase along the horizontal direction is observed at $\phi_c \approx 0.5$. Interestingly, the anisotropy in $\tilde \chi_{_V}({\bf k})$ does not result in elongated ``particles'' at low $\phi$ on a local scale, as one might speculate; but rather leads to significantly anisotropic distributions of the particles on a global scale. For example, it can be clearly seen that the particle phase in realizations with $\phi = 0.1$ form ``necklace''-like chains in the horizontal direction. The particles in the chains eventually connect to one another as $\phi$ increases to form connected bands. In addition, increasing $\ell$ leads to finer morphologies, with reduced particle size at lower $\phi$ and reduced ligament width at higher $\phi$, while the anisotropy effects along the horizontal direction persist for all $\phi$ and $\ell$ values. 



\subsection{Anisotropic Media with Square Exclusion Regions}

Next, we consider the effects of a square exclusion region with four-fold rotational symmetry and edge length $\ell$ (see Fig. \ref{fig_shape}c), which is defined as
\begin{equation}
\Omega(\ell) = \{ {\bf k} = \frac{2\pi}{L} {\bf n}: {\bf n}\in \mathbb{Z}^3, |k_x| \le \ell/2, |k_y| \le \ell/2 \}.
\end{equation}
The representative $\tilde \chi_{_V}({\bf k})$ with varying $\ell$ (left column) and the associated realizations are shown in Fig. \ref{fig_square}. Figure \ref{fig_square_chi} shows the corresponding averaged spectral density function $\tilde \chi_{_V}({k})$ ($k = |{\bf k}|$) along the horizontal and diagonal directions. 


We find that again the anisotropy of the exclusion region in the spectral density is not manifested in the local morphology of the phases, i.e., the particles formed by the pixels are still statistically isotropic {\it locally} and no square particles were observed. On the other hand, the {\it global} distributions of the entire particulate microstructure exhibit a high degree of four-fold rotational symmetry. For example, at low $\phi$, the particles are essentially arranged on a distorted square lattice (which is most apparent for the case with $\ell = 5$ and $\phi = 0.3$). At high $\phi$, the connected phase morphology are composed of perpendicularly arranged ligaments, as clearly illustrated in the case with $\ell = 10$ and $\phi = 0.5$. Similar effects for increasing volume fraction (i.e., increasing phase connectivity) and increasing $\ell$ (i.e., leading to finer morphologies) are also observed. In particular, the composite microstructures with $\phi$ approximately in the range [0.4, 0.6] are found to possess phase-inversion symmetry  and the ``blue'' phase percolates along both orthogonal directions at $\phi_c \approx 0.5$ (see Sec. IV.G for details).


\subsection{Anisotropic Media with Rectangular Exclusion Regions}

We also investigate $\tilde \chi_{_V}({\bf k})$ with a rectangular exclusion region, possessing two-fold rotational symmetry and an aspect ratio of $1/3$ and the length of long edge being $\ell$ (see Fig. \ref{fig_shape}d), which is defined as
\begin{equation}
\Omega(\ell) = \{ {\bf k} = \frac{2\pi}{L} {\bf n}: {\bf n}\in \mathbb{Z}^3, |k_x| \le \ell/6, |k_y| \le \ell/2 \}.
\end{equation}
The representative $\tilde \chi_{_V}({\bf k})$ with varying $\ell$ (left column) and the associated realizations are shown in Fig. \ref{fig_square}. Figure \ref{fig_rectangle_chi} shows the corresponding averaged spectral density function $\tilde \chi_{_V}({k})$ ($k = |{\bf k}|$) along the horizontal and vertical directions. 


We find that, as expected, the shape asymmetry of the exclusion region is also manifested in the morphology of the realizations. Specifically, the elongation effect seems to dominate the morphology of the realizations, leading to chain-like arrangements of ``particles'' at low $\phi$ and stripes at high $\phi$. Increasing volume fraction again leads to increasing phase connectivity, resulting in a directional percolation along the horizontal direction at $\phi_c \approx 0.5$, similar to that observed in the microstructures associated with the elliptical-disk exclusion regions. The anisotropic microstructures are found to possess phase-inversion symmetry for $\phi$ approximately in the range [0.4, 0.6] (see Sec. IV.G for details). Finer morphologies due to increasing $\ell$ are also observed.



\begin{figure*}[ht]
\begin{center}
$\begin{array}{c}\\
\includegraphics[width=0.7\textwidth]{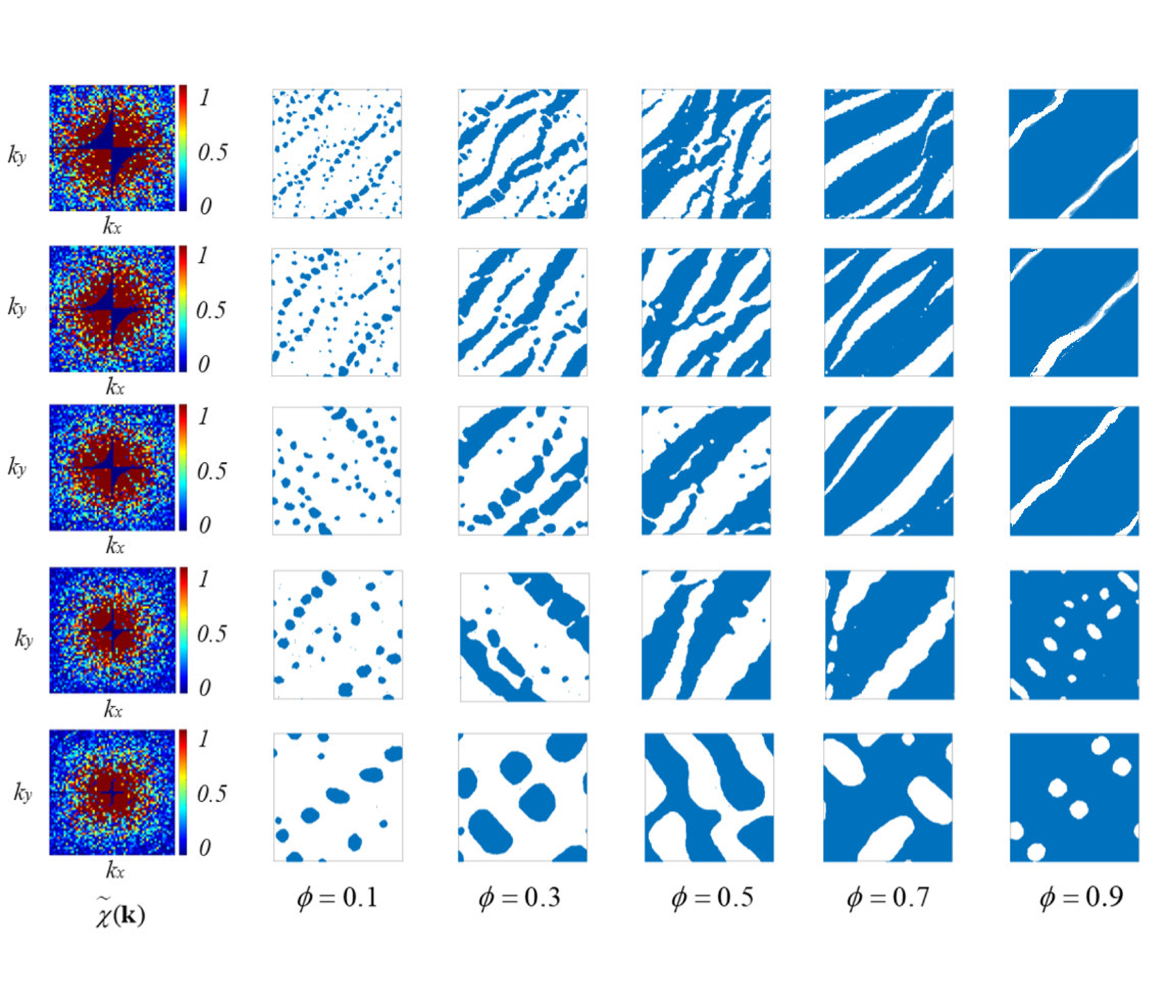}
\end{array}$
\end{center}
\caption{Direally hyperuniform composites with butterfly-shaped exclusion regions with varying size $\ell$ and volume fraction $\phi$. The left most column shows representative $\tilde \chi_{_V}({\bf k})$ (associated with $\phi = 0.5$) for each $\ell$: from bottom to top $\ell = 5, 10, 15, 20, 25$ (in units of $2\pi/L$), corresponding to $\chi \approx 2.7\times10^{-4}, 8.9\times10^{-4}, 1.9\times10^{-3}, 3.2\times10^{-3}, 4.8\times10^{-3}$, respectively. The phase volume fractions for the realizations from left to right are $\phi = 0.1, 0.2, 0.3, 0.4, 0.5$, respectively. The linear size of the system is $L = 300$ pixels.} \label{fig_butterfly}
\end{figure*}

\begin{figure*}[ht]
\begin{center}
$\begin{array}{c}\\
\includegraphics[width=0.9\textwidth]{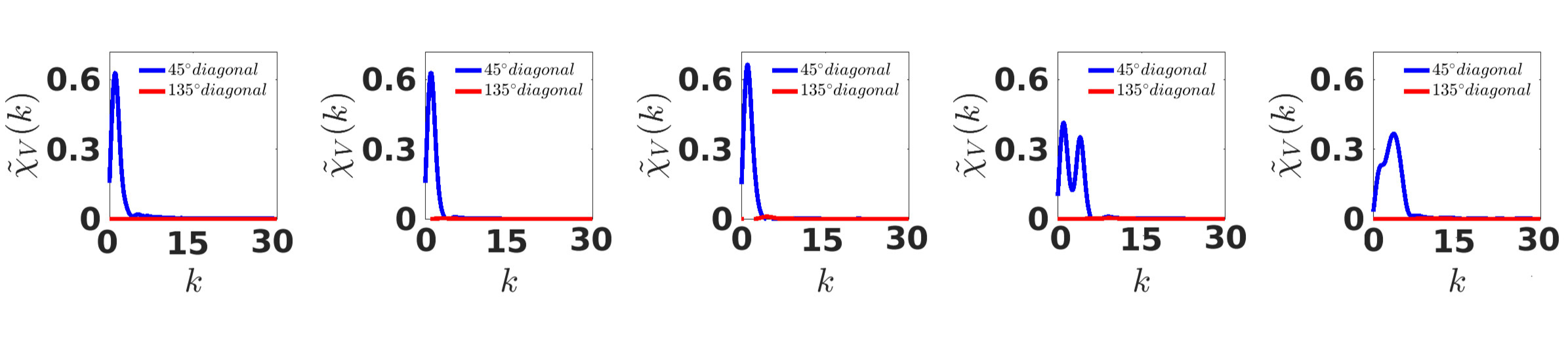}
\end{array}$
\end{center}
\caption{Representative averaged $\tilde \chi_{_V}({k})$ (where $k = |{\bf k}|$) along two orthogonal diagonal directions in the Fourier space associated with the constructed SHU composites with $\phi = 0.5$ and a butterfly-shaped exclusion region for different size $\ell$: from left to right $\ell = 5, 10, 15, 20, 25$ (in units of $2\pi/L$).} \label{fig_butterfly_chi}
\end{figure*}

\begin{figure*}[ht]
\begin{center}
$\begin{array}{c}\\
\includegraphics[width=0.7\textwidth]{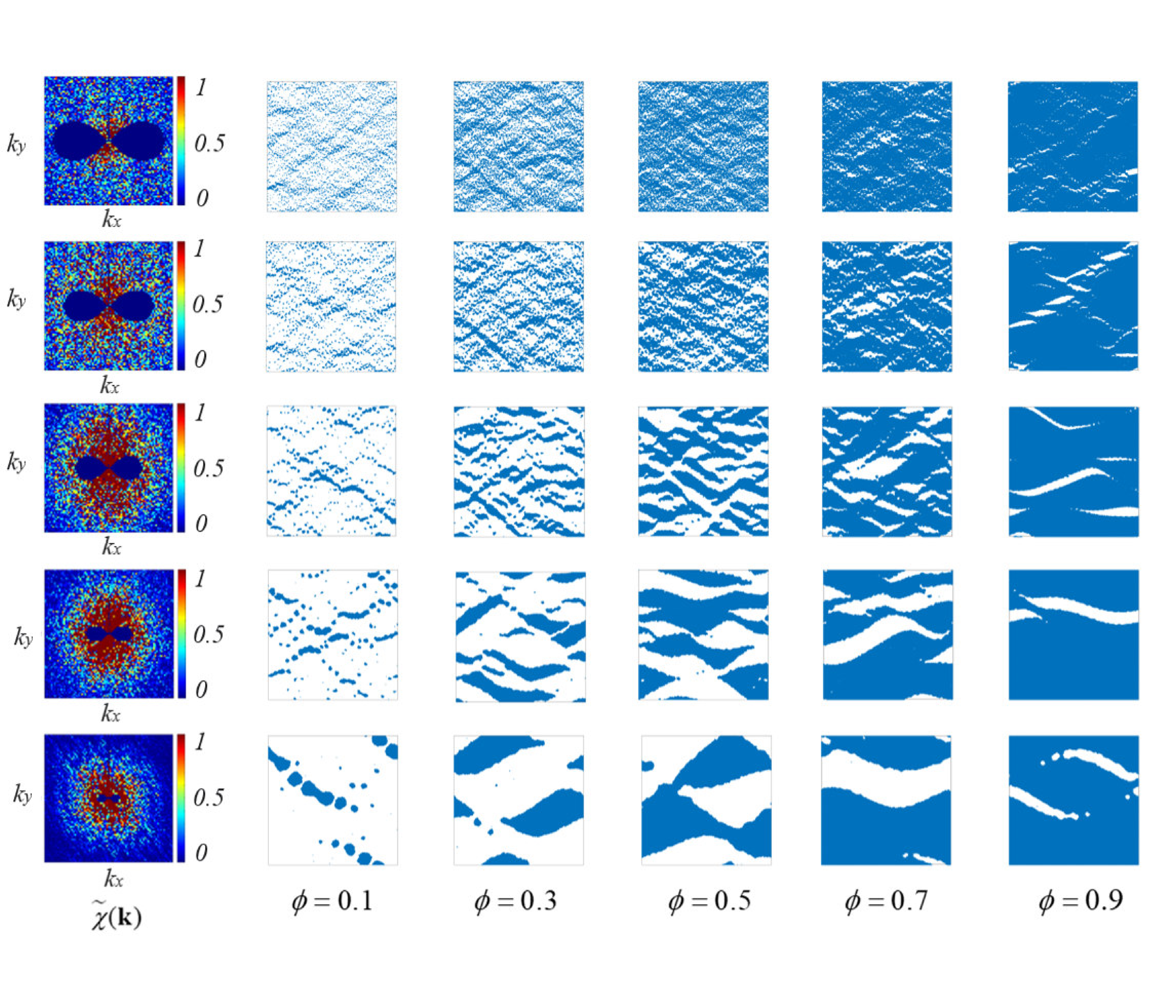}
\end{array}$
\end{center}
\caption{Directionally hyperuniform composites with lemniscate-shaped exclusion regions with varying size $\ell$ and volume fraction $\phi$. The left most column shows representative $\tilde \chi_{_V}({\bf k})$ (associated with $\phi = 0.5$) for each $\ell$: from bottom to top $\ell = 5, 10, 15, 20, 25$ (in units of $2\pi/L$), corresponding to $\chi \approx 5.6\times10^{-4}, 2.2\times10^{-3}, 5.0\times10^{-3}, 8.8\times10^{-3}, 1.4\times10^{-2}$, respectively. The phase volume fractions for the realizations from left to right are $\phi = 0.1, 0.2, 0.3, 0.4, 0.5$, respectively. The linear size of the system is $L = 300$ pixels.} \label{fig_lemniscate}
\end{figure*}

\begin{figure*}[ht]
\begin{center}
$\begin{array}{c}\\
\includegraphics[width=0.9\textwidth]{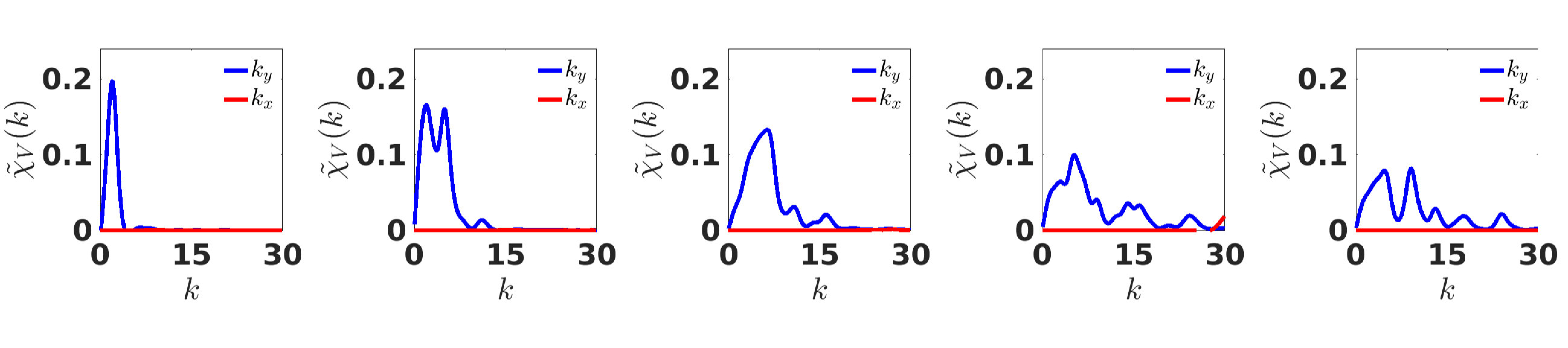}
\end{array}$
\end{center}
\caption{Representative averaged $\tilde \chi_{_V}({k})$ (where $k = |{\bf k}|$) along two orthogonal directions (horizontal and vertical) in the Fourier space associated with the constructed SHU composites with $\phi = 0.5$ and a lemniscate-shaped exclusion region for different size $\ell$: from left to right $\ell = 5, 10, 15, 20, 25$ (in units of $2\pi/L$).} \label{fig_lemniscate_chi}
\end{figure*}

\begin{figure*}[ht]
\begin{center}
$\begin{array}{c}\\
\includegraphics[width=0.85\textwidth]{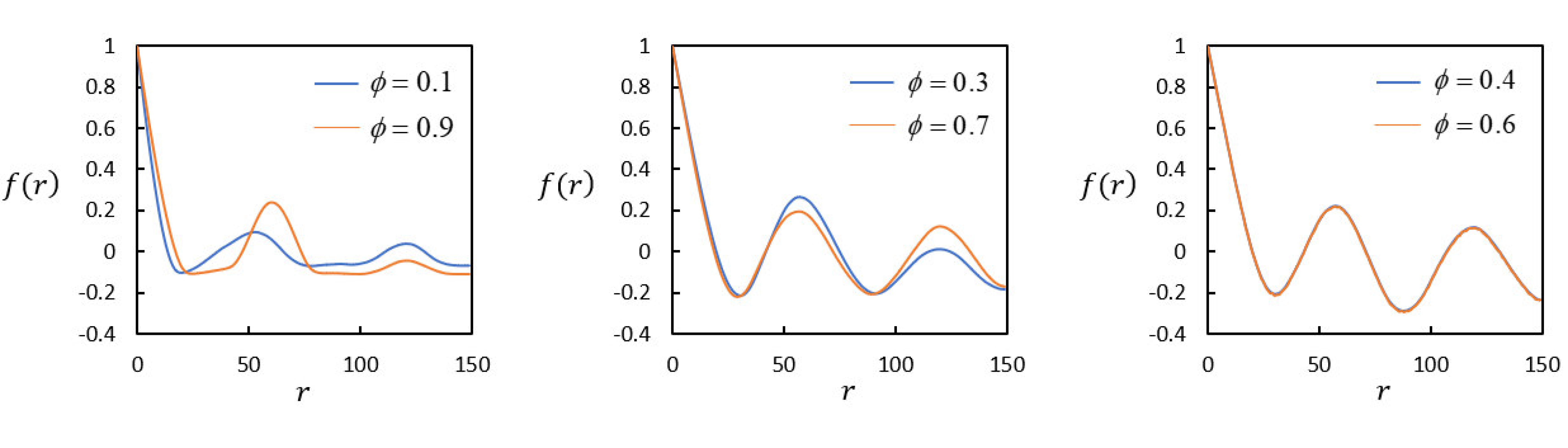}
\end{array}$
\end{center}
\caption{Rescaled autocovariance functions $f(r)$ for isotropic stealthy hyperuniform microstructures associated with the circular-disk exclusion region with $\ell = 5$. The microstructures possess phase inversion symmetry for $\phi$ approximately in the range [0.4, 0.6].} \label{fig_fr_circle}
\end{figure*}

\subsection{Directionally Hyperuniform Media with Butterfly-Shaped Exclusion Regions}



In the previous cases, the composite systems are stealthy hyperuniform along all directions, although the characteristic length scales on which the scattering in the system is completely suppressed (corresponding to $\tilde \chi_{_V}({\bf k}) = 0$) are different along different directions. In this section, we investigate systems which are directionally hyperuniform, i.e., the complete suppression of scattering only occurs in certain directions in the Fourier space. In particular, we consider two representative spectral density functions respectively with a butterfly-shaped exclusion region and lemniscate exclusion region.

The butterfly-shaped $\Omega$ is defined as the region in the 2nd and 4th quadrants enclosed by the concave superdisk curve \cite{jiao2008optimal, atkinson2012maximally} (see Fig. \ref{fig_shape}e), which possesses two-fold rotational symmetry, i.e.,
\begin{equation}
\Omega(\ell) = \{ {\bf k} = \frac{2\pi}{L} {\bf n}: {\bf n}\in \mathbb{Z}^3, |\frac{k_x}{\ell}|^p+|\frac{k_y}{\ell}|^p \le 1, k_x\cdot k_y \le 0 \}
\end{equation}
where $p = 1.5$ in this study. The associated composites thus only show hyperuniform behaviors when ${\bf k}$ approaches the origin in 2nd and 4th quadrants. Figure \ref{fig_butterfly} show the realizations associated the butterfly-shaped $\Omega$, as well as representative $\tilde \chi_{_V}({\bf k})$ for different $\ell$ values. The corresponding averaged spectral density function $\tilde \chi_{_V}({k})$ ($k = |{\bf k}|$) along the two diagonal directions are shown in Fig. \ref{fig_butterfly_chi}. 


We find that, consistent with previous examples, increasing connectivity with increasing $\phi$ and finer morphologies with increasing $\ell$. However, unlike the previous cases where the constructed anisotropic media are hyperuniform along all directions, these directionally hyperuniform microstructures do not possess phase-inversion symmetry anymore, as can be clearly seen in Fig. \ref{fig_shape} and the associated with autocovariance functions in Appendix A. Increasing $\phi$ leads to percolation of the ``blue'' phase along the 45-degree diagonal direction at $\phi_c \approx 0.45$, which is lower than that for the microstructures possessing phase-inversion symmetry.  

The realizations for all $\ell$ and $\phi$ values show strong anisotropy along the diagonal directions of the construction domain, i.e., chain-like arrangements of particles at low $\phi$ and stripe-like bands at high $\phi$, which is consistent with the anisotropy directions of the $\tilde \chi_{_V}({\bf k})$. Specifically, the systems are hyperuniform along the directions of these chain-like or stripe-like structures, and non-hyperuniform along the perpendicular directions. This can be intuitively understood by imagining move observation windows along the chains/stripes. Once the window size is larger than the average distance between a pair of chains/stripes, moving the window along the chain/stripe direction would not result in large volume fraction fluctuations. On the other hand, moving the window along the directions perpendicular to that of chains/stripes is expected to lead to large fluctuations as the window would alternatively contains the white or dark phases.

\subsection{Directionally Hyperuniform Media with Lemniscate-Shaped Exclusion Regions}

The lemniscate $\Omega$ (see Fig. \ref{fig_shape}f) which possesses two-fold rotational symmetry is defined by \cite{cox2014galois}
\begin{equation}
\Omega(\ell) = \{ {\bf k} = \frac{2\pi}{L} {\bf n}: {\bf n}\in \mathbb{Z}^3, \rho^2 \le \ell^2 \cos(2\theta) \}
\end{equation}
where is $\rho$ the Fourier space polar coordinates $\rho^2 = k_x^2 + k_y^2$, $\theta = \tan^{-1}(k_x/k_y)$, and $\ell$ is the length scale parameter. Similar to the butter-fly shaped case, the associated composites only exhibit hyperuniform behaviors along certain directions in the Fourier spaces, i.e., those are enclosed in the lemniscate exclusion region.



Figure \ref{fig_lemniscate} show the realizations associated with the lemniscate-shaped $\Omega$, as well as representative $\tilde \chi_{_V}({\bf k})$ for different $\ell$ values. The corresponding averaged spectral density function $\tilde \chi_{_V}({k})$ ($k = |{\bf k}|$) along the horizontal and vertical directions are shown in Fig. \ref{fig_lemniscate_chi}. Similar to the media associated with the butterfly-shaped exclusion regions, the resulting directionally hyperuniform microstructures do not possess phase-inversion symmetry. Percolation is found to occur at a much lower volume fraction $\phi_c \approx 0.4$ along the anisotropy (i.e., horizontal) direction, compared to that for the microstructures possessing phase-inversion symmetry. It can been seen that the realizations contain chain-like (for low $\phi$) and stripe-like (for high), along the horizontal directions, leading to strong anisotropy in this direction. Overall, these structural elements are arranged in a wavy manner along the horizontal direction. A similar pattern in directionally hyperuniform point configurations with a lemniscate-shaped structure factor was also observed \cite{To18a}, see Fig. \ref{fig_anisotropy}. The emergence of the wavy structures might result from the unique direction dependence scattering behavior imposed by the lemniscate pattern in $\tilde \chi_{_V}({\bf k})$.


In summary, we have found that the directional hyperuniformity imposed by the butterfly-shaped and lemniscate-shaped exclusion regions are realized by the formation of anisotropic chain-like or stripe-like structures in the construction media. These two distinct exclusion-region shapes lead to two distinct sets of anisotropic directions in the constructed composite microstructures (i.e., diagonal directions in the case of butterfly-shaped $\Omega$ and horizontal directions in the case of lemniscate-shaped $\Omega$). Among these directions, the anisotropic chain/stripe structures possess a much more uniform distribution of the phases, giving rise to hyperuniformity along these directions. On the other hand, large gaps and voids are present between the chains/stripes, which result in large local volume fraction fluctuations along the directions perpendicular to these chains/stripes, leading to non-hyperuniform behaviors along these directions. 



\subsection{Phase Inversion Symmetry and Percolation Threshold}

In this section, we provide a detailed analysis of the phase-inversion symmetry of the constructed composite microstructures. A random medium possesses phase-inversion symmetry if the morphology of phase 1 at volume fraction $\phi_1$ is statistically identical to that of phase 2 in the system where the volume fraction is $\phi_2 = 1- \phi_1$, and hence \cite{To03}
\begin{equation}
 S_2^{(1)}(r; \phi_1, \phi_2) = S_2^{(2)}(r; \phi_2, \phi_1).    
\end{equation}
Equivalently, the corresponding microstructures with phase-inversion symmetry should possess identical rescaled autocovariance function
\begin{equation}
 f(r) = \frac{ S_2^{(1)}(r)-\phi_1^2}{\phi_1(1-\phi_1)} = \frac{ S_2^{(2)}(r)-\phi_2^2}{\phi_2(1-\phi_2)}.
 \label{eq_fr}
\end{equation}

Figure \ref{fig_fr_circle} shows the comparison of the rescaled autocovariance function $f(r)$ for the isotropic microstructures associated with the circular-disk exclusion ($\ell = 5$) and varying volume fractions $\phi$ of the ``blue'' phase, i.e., phase ``1'' (see Fig. \ref{fig_circle}). It can be seen as the phase volume fraction increases towards 0.5, the corresponding $f(r)$ functions match better with one another. For $\phi = \phi_1 \ge 0.4$, the corresponding $f(r)$ derived from the two phases are virtual identical to one another, indicating the corresponding microstructures possess a degree of phase-inversion symmetry. We subsequently analyze the rescaled autocovariance functions for the microstructures associated with the elliptical-disk, square and rectangular exclusion regions (see Appendix A) and find that all of these anisotropic hyperuniform microstructures possess phase-inversion symmetry for $\phi$ approximately in the range [0.4, 0.6].  

It is well known that a statistically isotropic medium with phase-inversion symmetry must possess a percolation threshold $\phi_c = 0.5$ \cite{To03}. Indeed, we observed that the isotropic microstructures associated with the circular-disk exclusions all percolate at $\phi_c = 0.5$. Interestingly, our results suggest this statement can be possibly generalized to the anisotropic hyperuniform microstructures possessing phase-inversion symmetry as well. In particular, for the microstructures associated with the elliptical-disk and rectangular exclusion regions, we observe that percolation first occurs along the anisotropic horizontal direction. For the microstructures associated with the square exclusion regions, percolation simultaneously occurs along the both the orthogonal directions.

On the other hand, the directionally hyperuniform microstructures associated with the butterfly-shaped and lemniscate-shaped $\Omega$ do not possess phase inversion symmetry. This can be seen by visually comparing the corresponding microstructures and quantitatively comparing the associated $f(r)$ functions (see Appendix A). We speculate that this is because the very asymmetric spectral density function $\tilde \chi_{_V}({k})$ associated with $\Omega$ regions can only be realized by distinctly different topology and morphology of the phases at different $\phi$, and thus breaks the phase-inversion symmetry. As a consequence, the percolation threshold for these microstructures (along certain directions) are significantly lower than that for microstructures with phase-inversion symmetry, i.e., $\phi_c \approx 0.45$ for the butterfly-shaped $\Omega$ and $\phi \approx 0.4$ for the lemniscate-shaped $\Omega$. We note that the percolation threshold values reported here are only estimate based on the constructed microstructures. A more detailed study is required in order to precisely determine the threshold values \cite{perco}. 




\begin{figure*}[ht]
\begin{center}
$\begin{array}{c}\\
\includegraphics[width=0.85\textwidth]{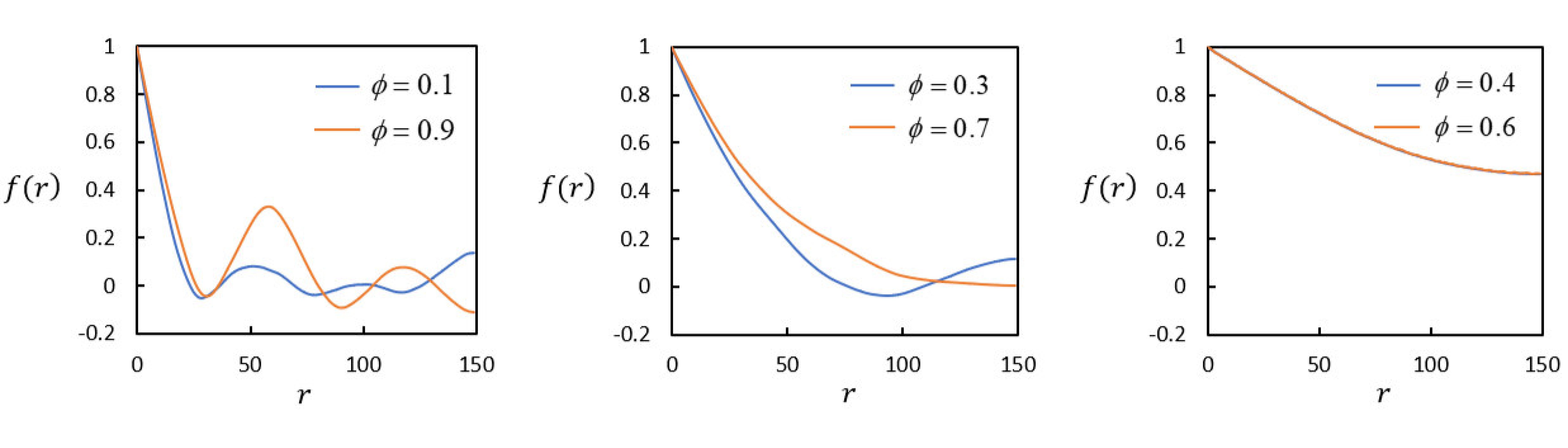}
\end{array}$
\end{center}
\caption{Rescaled autocovariance functions $f(r)$ along the horizontal direction for anisotropic stealthy hyperuniform microstructures associated with the elliptical-disk exclusion region with $\ell = 5$. } \label{fig_fr_ellipse}
\end{figure*}

\begin{figure*}[ht]
\begin{center}
$\begin{array}{c}\\
\includegraphics[width=0.85\textwidth]{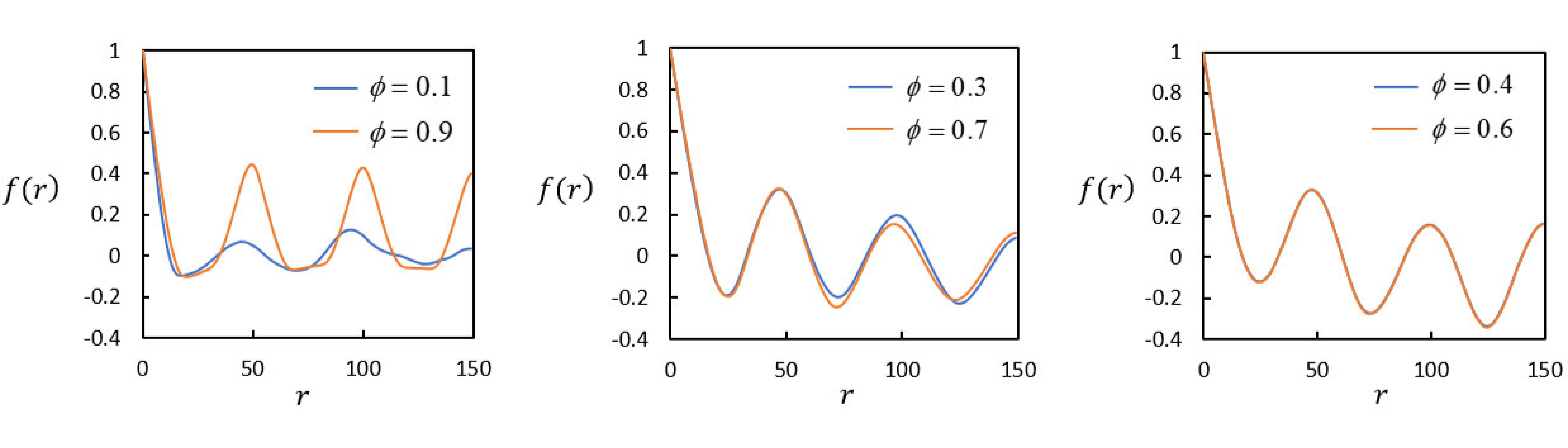}
\end{array}$
\end{center}
\caption{Rescaled autocovariance functions $f(r)$ along the two orthogonal directions for anisotropic stealthy hyperuniform microstructures associated with the square exclusion region with $\ell = 5$. } \label{fig_fr_square}
\end{figure*}

\begin{figure*}[ht]
\begin{center}
$\begin{array}{c}\\
\includegraphics[width=0.85\textwidth]{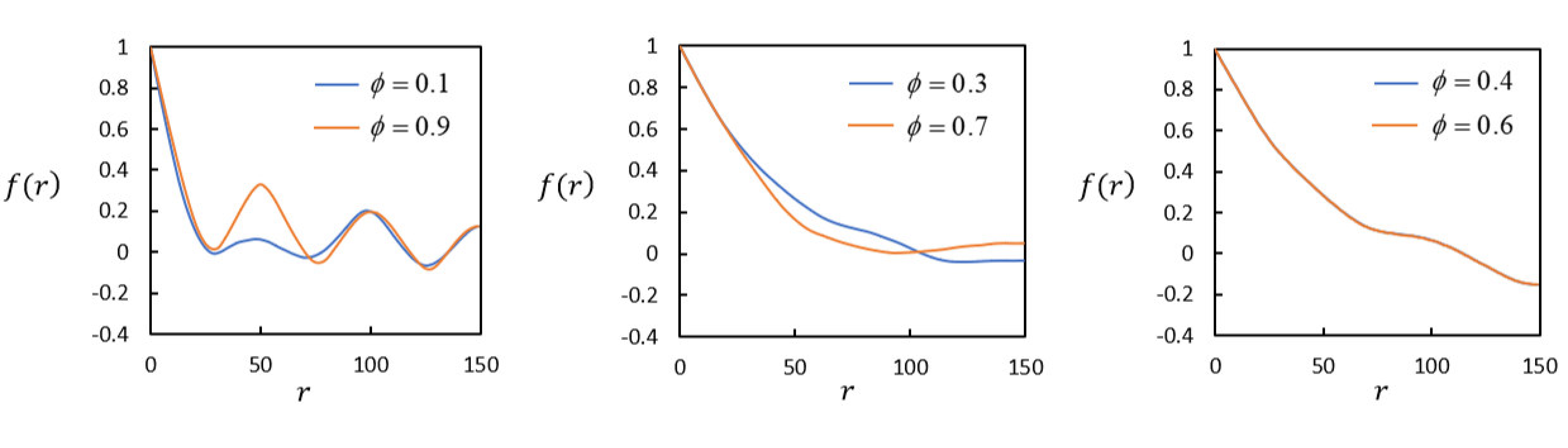}
\end{array}$
\end{center}
\caption{Rescaled autocovariance functions $f(r)$ along the horizontal direction for anisotropic stealthy hyperuniform microstructures associated with the rectangular exclusion region with $\ell = 5$.} \label{fig_fr_rectangle}
\end{figure*}

\begin{figure*}[ht]
\begin{center}
$\begin{array}{c}\\
\includegraphics[width=0.85\textwidth]{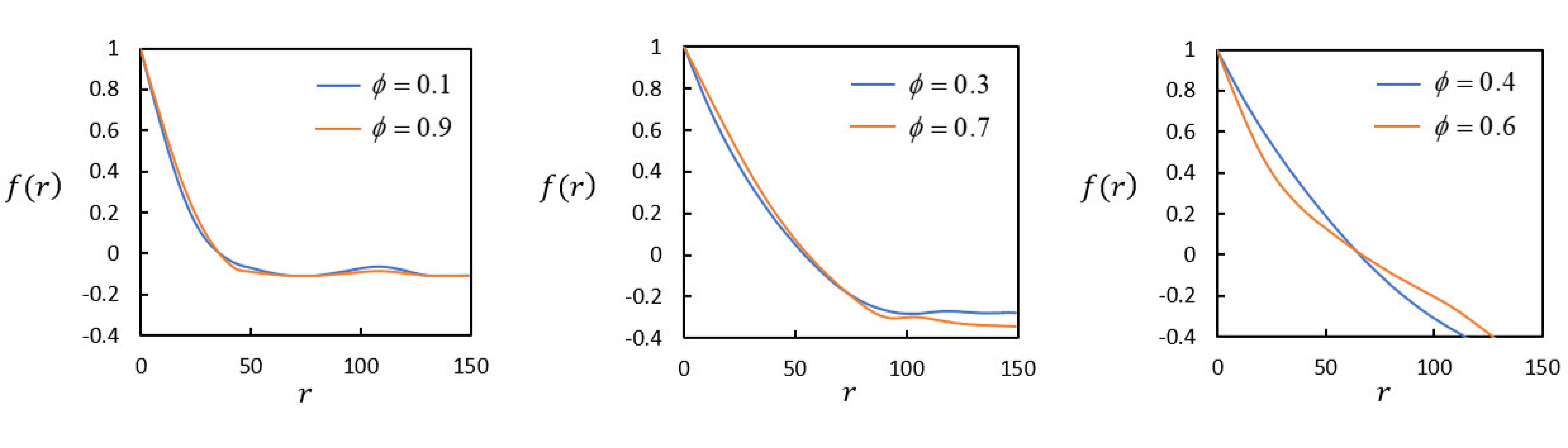}
\end{array}$
\end{center}
\caption{Rescaled autocovariance functions $f(r)$ along the 45-degree diagonal direction for directionally hyperuniform microstructures associated with the butterfly-shaped exclusion region with $\ell = 5$.} \label{fig_fr_butterfly}
\end{figure*}

\begin{figure*}[ht]
\begin{center}
$\begin{array}{c}\\
\includegraphics[width=0.85\textwidth]{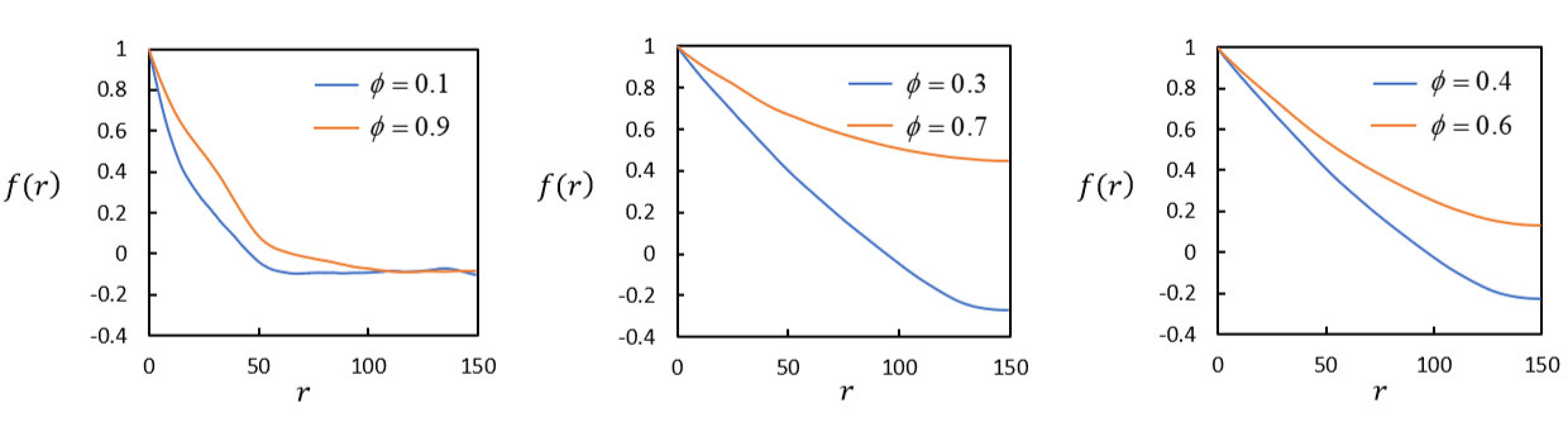}
\end{array}$
\end{center}
\caption{Rescaled autocovariance functions $f(r)$ along the horizontal direction for directionally hyperuniform microstructures associated with the lemniscate-shaped exclusion region with $\ell = 5$.} \label{fig_fr_lemniscate}
\end{figure*}

\begin{figure*}[ht]
\begin{center}
$\begin{array}{c}\\
\includegraphics[width=0.75\textwidth]{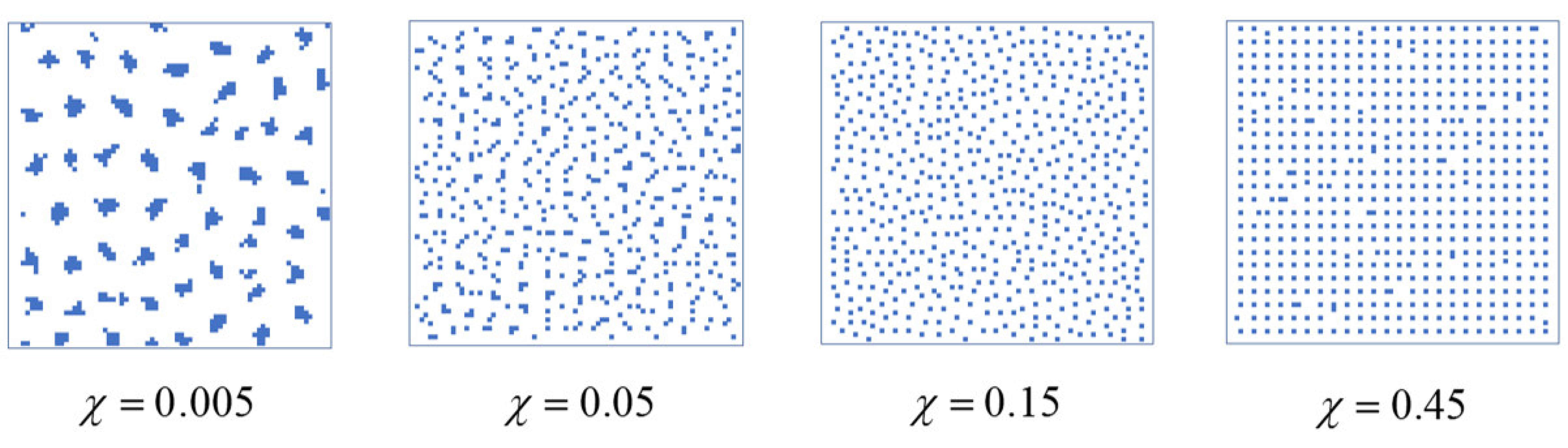}
\end{array}$
\end{center}
\caption{Constructed microstructures with $\phi = 1/9$ associated with the square exclusion regions and increasing $\chi$. The linear size of the system is $L = 72$ pixels.} \label{fig_large_circle}
\end{figure*}

\begin{figure*}[ht]
\begin{center}
$\begin{array}{c}\\
\includegraphics[width=0.75\textwidth]{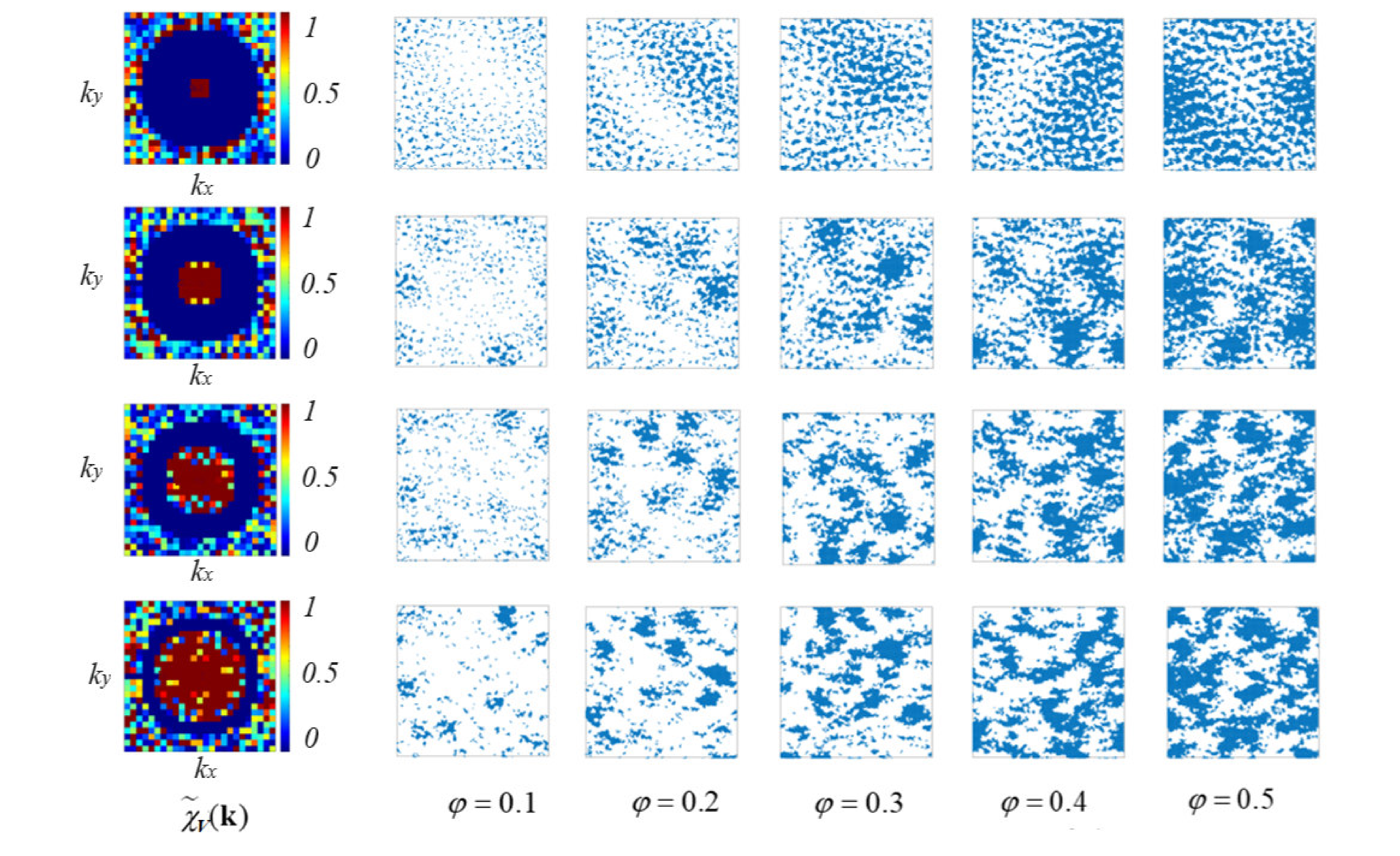}
\end{array}$
\end{center}
\caption{Non-hyperuniform stealthy composites with $\tilde \chi_{_V}({\bf k}) = 0$ in a circular-ring exclusion region defined by $K_1 \le |{\bf k}| \le K_2$ in the Fourier space with $K_2 = 10$, and varying $K_1$ and volume fraction $\phi$. The left most column shows representative $\tilde \chi_{_V}({\bf k})$: from top to bottom $K_1 = 2, 4, 6, 8$. The phase volume fractions for the realizations from left to right are $\phi = 0.1, 0.2, 0.3, 0.4, 0.5$, respectively. The linear size of the system is $L = 150$ pixels.} \label{fig_ring}
\end{figure*}

\section{Conclusions and Discussion}



In this work, we devised a Fourier-space based numerical construction procedure that explicitly incorporates the vector-dependent spectral density function ${\tilde \chi}_{_V}({\bf k})$ to design and generate anisotropic microstructures with targeted directional scattering properties. We mainly focused on anisotropic SHU composites, which possess a spectral density ${\tilde \chi}_{_V}({\bf k}) = 0$ in an exclusion region $\Omega$ around the origin in the Fourier space [see Eq. (\ref{eq_1})]. We systematically investigate the different shaped exclusion regions $\Omega$ with various discrete symmetries, including circular-disk, elliptical-disk, square, rectangular, butterfly-shaped and lemniscate-shaped exclusion regions, with the latter two instances leading to directionally hyperuniform composites. Our study allows us to understand how different discrete symmetries of $\Omega$ affect the directional scattering properties of the resulting composite, which is crucial to the engineering of novel disordered photonic and phononic two-phase media, as elaborated below.


We have found that the circular-disk $\Omega$ regions give rise to isotropic hyperuniform structures on both global and local scales, and the resulting microstructures possess phase-inversion symmetry for volume fractions $\phi$ approximately in the range [0.4, 0.6]. Increasing $\phi$ leads to enhanced phase connectedness and subsequent percolation at $\phi_c = 0.5$. The anisotropic microstructures associated with the elliptical-disk, square and rectangular exclusion regions also possess phase-inversion symmetry and directional percolation at $\phi_c \approx 0.5$. On the other hand, the directionally hyperuniform microstructures associated with the butterfly-shaped and lemniscate-shaped $\Omega$ do not possess phase-inversion symmetry and percolate along certain directions at much lower volume fractions ($\phi_c \approx 0.45$ for the butterfly-shaped $\Omega$ and $\phi \approx 0.4$ for the lemniscate-shaped $\Omega$). Increasing the size of $\Omega$ for all different shapes considered here results in finer structures, which is required to suppress local volume fraction fluctuations over a broader range of length scales associated with the larger exclusion region.

The directional hyperuniform behavior imposed by the shape asymmetry of exclusion regions is clearly manifested in the anisotropic phase morphology of the constructed media. For example, in the case of lemniscate-shaped exclusion regions, the resulting media are hyperuniform along the direction associated with the chain-like structures, which possess a much more uniform distribution of the phases along these direction on the global scale. On the other hand, the media are non-hyperuniform along the direction perpendicular to these chain-like structures, along which large gaps and voids are present and the distribution of the phases is much less uniform, both of which can lead to large local volume fraction fluctuations that destroy hyperuniformity. The exclusion regions possessing only two-fold rotational symmetry, such as the elliptical-disk and rectangular regions, impose different length scales (along the two orthogonal directions) over which the systems are required to be stealthy hyperuniform, which is also achieved by the anisotropic chain-like or stripe-like structures in the associated composite microstructures.



In addition, while the anisotropic exclusion region imposes strong constraints on the {\it global} symmetry of the resulting microstructure, the composite can still have almost isotropic {\it local} morphology. For example, in the case of square exclusion regions, the constructed microstructures contain almost isotropic ``particles'' (local clusters of pixels), arranged on distorted ``square lattices'' to realize the global four-fold rotational symmetry imposed by the exclusion region. We also found that the anisotropic phase morphologies seem to be more sensitive to the overall asymmetry but not the detailed shapes for $\Omega$ regions possessing two-fold rotational symmetry. This can be seen from the comparison of the microstructures associated with the elliptical-disk and rectangular exclusion regions: both of them possess similar chain-like or stripe-like structures along the horizontal directions. 





Although all of the illustrative construction examples we considered are hyperuniform, the general construction procedure can be readily employed to generate microstructures of composites with an arbitrary anisotropic spectral density function ${\tilde \chi}_{_V}({\bf k})$. Indeed, in Appendix B, we demonstrate this utility by generating stealthy nonhyperuniform composites, which exhibit a multi-scale structure in order to achieve the prescribed scattering behavior across scales. In addition, our procedure can be straightforwardly generalized to three dimensions. The resulting digitized microstructures can be experimentally fabricated using 3D printing techniques \cite{liu2019structural, liu20223d}.

While our focus in the present work was to engineer anisotropic scattering properties directly encoded in the spectral density function, our general construction procedure is a key initial step in the computational design of disordered hyperuniform composites with other photonic. Such designs could be achieved, for example, by leveraging recently developed predictive formulations including nonlocal theories for the effective dynamic elastic moduli and dielectric constant \cite{kim2020multifunctional, kim2020effective, torquato2021nonlocal} and the spreadability for time-dependent diffusive transport behaviors \cite{torquato2021diffusion, wang2022dynamic, spreadability}. Such theories rigorously connect effective properties of the composites to their spectral density ${\tilde \chi}_{_V}({\bf k})$, allowing us to achieve desirable composite properties by tuning a targeted spectral density. The associated microstructure can then be obtained using our construction procedure. In future work, we will explore this framework to design disordered hyperuniform composites with targeted electromagnetic and transport properties.






\begin{acknowledgments}
This work was supported by the Army Research Office under Cooperative Agreement Number W911NF-22-2-0103.
\end{acknowledgments}
\smallskip

\appendix




\section{Auto-covariance Functions for Anisotropic Composite Microstructures}

In this section, we present results on rescaled autocovariance functions (see Eq. (\ref{eq_fr})) for the anisotropic composite microstructures. In particular, Figs. \ref{fig_fr_ellipse}, \ref{fig_fr_square} and \ref{fig_fr_rectangle} respectively show $f(r)$ for the anisotropic hyperuniform microstructures associated with the elliptical-disk, square and rectangular exclusion regions. It can be seen that the rescaled autocovariance functions for the microstructures with $\phi \in [0.4, 0.6]$ are virtually identical, indicating these microstructures possess phase-inversion symmetry (see Sec. IV.G). On the other hand, Figs. \ref{fig_fr_butterfly} and \ref{fig_fr_lemniscate} respectively show $f(r)$ for the directionally hyperuniform microstructures associated with the butterfly-shaped and lemniscate-shaped exclusion regions. It can be seen that the rescaled autocovariance functions for the microstructures are distinctly different for all corresponding $\phi$ values, indicating these microstructures do not possess phase-inversion symmetry.    





\section{Effects of Increasing $\Omega(\ell)$}


We have shown that increasing $\ell$ (i.e., size of $\Omega$ region) results in finer morphologies and enhances dispersion of individual pixels in the constructed composite microstructures. As discussed in Sec. IV.A, this was due to the increasing number of constrained ${\bf k}$ vectors, which requires suppression of local volume fraction fluctuations on a broader range. Here, we further show that the system indeed behaves like ``particles'' on a lattice as $\ell$ (or equivalently $\chi$) increases. 

Previous studies have shown that for a point configuration in a two-dimensional Euclidean space, increasing $\chi$ leads to increasing order in the distribution of the points and a perfect distribution of points on the triangular-lattice can be achieved for $\chi \ge 0.5$ \cite{To18a}. Here the composite microstructures are composed of pixels arranged on a square grid (lattice). Therefore, we investigate the evolution of microstructures associated with the {\it square} exclusion regions with increasing size, whose symmetry is compatible with the underlying square lattice. In particular, we consider a system with $\phi = 1/9$, $L = 72$ pixels and $N = 576$, which possesses an order microstructure with the individual ``blue'' pixels arranged on a perfect square lattice with a lattice constant $a = 3$ pixels. The smaller system size allows fast convergence of the optimization algorithm.


Figure \ref{fig_large_circle} shows the constructed microstructures associated with the square $\Omega$ and increasing $\ell$, and thus increasing $\chi$. It can be seen that for low $\chi$ values, the system is in the ``random-medium'' regime, and the resulting microstructures contain clusters of blue phase pixels. As $\chi$ increases, the blue pixels behave more like ``particles'' with an increasing degree of repulsion, and the resulting microstructures contain distributions of individual pixels with increasing local order. An almost perfect square-lattice packing of the blue pixels is obtained at $\chi = 0.45$. These results are consistent with SHU point configurations associated with increasing $\chi$ values \cite{To18a}.
\bigskip


\section{Construction of Stealthy Non-hyperuniform Systems}


To demonstrate the utility of our general construction procedure, we employ it to render realizations of stealthy but non-hyperuniform composite systems. Without loss of generality, we consider a class of spectral density function chi $\tilde \chi_{_V}({\bf k}) = 0$ for $K_1\le|{\bf k}|\le K_2$, where $0<K_1<K_2$, i.e., the $\Omega$ region corresponds to a circular ring with inner radius $K_1$ and outer radius $K_2$. Here we use $L = 150$ and fix $K_2 = 10$ and vary $K_1$ by choosing $K_1 = 2, 4, 6, 8$.  The realizations and representative $\tilde \chi_{_V}({\bf k}) = 0$ are shown in Fig. \ref{fig_ring}. It can be seen that the realizations include ``particles'' which are grouped into clusters. As $K_1$ increases, the clusters get denser and their size decreases. Thus, the systems can be considered to possess two characteristic length scales, i.e., the particle size corresponding to $K_2$ and the cluster size corresponding to $K_1$. These structural elements (e.g., particles and particle clusters) are arranged in a way such that scatterings on multiple length scale within these bounds are completely suppressed. In future work, we will explore additional designs for the stealthy non-hyperuniformity composite systems, with a focus on engineering anisotropic scattering behaviors.


\begin{thebibliography}{140}
\expandafter\ifx\csname
natexlab\endcsname\relax\def\natexlab#1{#1}\fi
\expandafter\ifx\csname bibnamefont\endcsname\relax
  \def\bibnamefont#1{#1}\fi
\expandafter\ifx\csname bibfnamefont\endcsname\relax
  \def\bibfnamefont#1{#1}\fi
\expandafter\ifx\csname citenamefont\endcsname\relax
  \def\citenamefont#1{#1}\fi
\expandafter\ifx\csname url\endcsname\relax
  \def\url#1{\texttt{#1}}\fi
\expandafter\ifx\csname
urlprefix\endcsname\relax\def\urlprefix{URL }\fi
\providecommand{\bibinfo}[2]{#2}
\providecommand{\eprint}[2][]{\url{#2}}

\bibitem[{\citenamefont{Torquato and Stillinger}(2003)}]{To03}
\bibinfo{author}{\bibfnamefont{S.}~\bibnamefont{Torquato}} \bibnamefont{and}
  \bibinfo{author}{\bibfnamefont{F.~H.} \bibnamefont{Stillinger}},
  Local density fluctuations, hyperuniformity, and order metrics.
  \bibinfo{journal}{Phys. Rev. E} \textbf{\bibinfo{volume}{68}},
  \bibinfo{pages}{041113} (\bibinfo{year}{2003}).

\bibitem[{\citenamefont{Torquato}(2018)}]{To18a}
\bibinfo{author}{\bibfnamefont{S.}~\bibnamefont{Torquato}},
Hyperuniform states of matter.
  \bibinfo{journal}{Phys. Rep.} \textbf{\bibinfo{volume}{745}},
  \bibinfo{pages}{1} (\bibinfo{year}{2018}).

\bibitem[{\citenamefont{Zachary and Torquato}(2009)}]{Za09}
\bibinfo{author}{\bibfnamefont{C.~E.} \bibnamefont{Zachary}} \bibnamefont{and}
  \bibinfo{author}{\bibfnamefont{S.}~\bibnamefont{Torquato}},
  Hyperuniformity in point patterns and two-phase random heterogeneous media.
  \bibinfo{journal}{J. Stat. Mech. Theor. Exp.}
  \textbf{\bibinfo{volume}{2009}}, \bibinfo{pages}{P12015}
  (\bibinfo{year}{2009}).

\bibitem[{\citenamefont{Torquato}(2016{\natexlab{a}})}]{To16a}
\bibinfo{author}{\bibfnamefont{S.}~\bibnamefont{Torquato}},
Hyperuniformity and its generalizations.
  \bibinfo{journal}{Phys. Rev. E} \textbf{\bibinfo{volume}{94}},
  \bibinfo{pages}{022122} (\bibinfo{year}{2016}{\natexlab{a}}).

\bibitem[{\citenamefont{Torquato}(2002)}]{To02a}
\bibinfo{author}{\bibfnamefont{S.}~\bibnamefont{Torquato}},
  \emph{\bibinfo{title}{Random Heterogeneous Materials: Microstructure and
  Macroscopic Properties}} (\bibinfo{publisher}{Springer-Verlag, New York},
  \bibinfo{year}{2002}).

\bibitem[{\citenamefont{Chen and
  Torquato}(2018{\natexlab{a}})}]{chen2018designing}
\bibinfo{author}{\bibfnamefont{D.}~\bibnamefont{Chen}} \bibnamefont{and}
  \bibinfo{author}{\bibfnamefont{S.}~\bibnamefont{Torquato}}, Designing disordered hyperuniform two-phase materials with novel physical properties.
  \bibinfo{journal}{Acta Mater.} \textbf{\bibinfo{volume}{142}},
  \bibinfo{pages}{152} (\bibinfo{year}{2018}{\natexlab{a}}).

\bibitem[{\citenamefont{Gabrielli et~al.}(2002)\citenamefont{Gabrielli, Joyce,
  and Labini}}]{ref3}
\bibinfo{author}{\bibfnamefont{A.}~\bibnamefont{Gabrielli}},
  \bibinfo{author}{\bibfnamefont{M.}~\bibnamefont{Joyce}}, \bibnamefont{and}
  \bibinfo{author}{\bibfnamefont{F.~S.} \bibnamefont{Labini}},
  Glass-like universe: Real-space correlation properties of standard cosmological models.
  \bibinfo{journal}{Phys. Rev. D} \textbf{\bibinfo{volume}{65}},
  \bibinfo{pages}{083523} (\bibinfo{year}{2002}).

\bibitem[{\citenamefont{Donev et~al.}(2005)\citenamefont{Donev, Stillinger, and
  Torquato}}]{ref4}
\bibinfo{author}{\bibfnamefont{A.}~\bibnamefont{Donev}},
  \bibinfo{author}{\bibfnamefont{F.~H.} \bibnamefont{Stillinger}},
  \bibnamefont{and} \bibinfo{author}{\bibfnamefont{S.}~\bibnamefont{Torquato}}, Unexpected density fluctuations in jammed disordered sphere packings.
  \bibinfo{journal}{Phys. Rev. Lett.} \textbf{\bibinfo{volume}{95}},
  \bibinfo{pages}{090604} (\bibinfo{year}{2005}).

\bibitem[{\citenamefont{Zachary
  et~al.}(2011{\natexlab{a}})\citenamefont{Zachary, Jiao, and Torquato}}]{ref5}
\bibinfo{author}{\bibfnamefont{C.~E.} \bibnamefont{Zachary}},
  \bibinfo{author}{\bibfnamefont{Y.}~\bibnamefont{Jiao}}, \bibnamefont{and}
  \bibinfo{author}{\bibfnamefont{S.}~\bibnamefont{Torquato}}, Hyperuniform long-range correlations are a signature of disordered jammed hard-particle packings.
  \bibinfo{journal}{Phys. Rev. Lett.} \textbf{\bibinfo{volume}{106}},
  \bibinfo{pages}{178001} (\bibinfo{year}{2011}{\natexlab{a}}).

\bibitem[{\citenamefont{Jiao and Torquato}(2011)}]{ref6}
\bibinfo{author}{\bibfnamefont{Y.}~\bibnamefont{Jiao}} \bibnamefont{and}
  \bibinfo{author}{\bibfnamefont{S.}~\bibnamefont{Torquato}}, Maximally random jammed packings of Platonic solids: Hyperuniform long-range correlations and isostaticity.
  \bibinfo{journal}{Phys. Rev. E} \textbf{\bibinfo{volume}{84}},
  \bibinfo{pages}{041309} (\bibinfo{year}{2011}).

\bibitem[{\citenamefont{Chen et~al.}(2014)\citenamefont{Chen, Jiao, and
  Torquato}}]{ref7}
\bibinfo{author}{\bibfnamefont{D.}~\bibnamefont{Chen}},
  \bibinfo{author}{\bibfnamefont{Y.}~\bibnamefont{Jiao}}, \bibnamefont{and}
  \bibinfo{author}{\bibfnamefont{S.}~\bibnamefont{Torquato}}, Equilibrium phase behavior and maximally random jammed state of truncated tetrahedra.
  \bibinfo{journal}{J. Phys. Chem. B}
  \textbf{\bibinfo{volume}{118}}, \bibinfo{pages}{7981} (\bibinfo{year}{2014}).

\bibitem[{\citenamefont{Zachary and Torquato}(2011{\natexlab{a}})}]{ref8}
\bibinfo{author}{\bibfnamefont{C.~E.} \bibnamefont{Zachary}} \bibnamefont{and}
  \bibinfo{author}{\bibfnamefont{S.}~\bibnamefont{Torquato}}, Anomalous local coordination, density fluctuations, and void statistics in disordered hyperuniform many-particle ground states.
  \bibinfo{journal}{Phys. Rev. E} \textbf{\bibinfo{volume}{83}},
  \bibinfo{pages}{051133} (\bibinfo{year}{2011}{\natexlab{a}}).

\bibitem[{\citenamefont{Torquato
  et~al.}(2015{\natexlab{a}})\citenamefont{Torquato, Zhang, and
  Stillinger}}]{ref9}
\bibinfo{author}{\bibfnamefont{S.}~\bibnamefont{Torquato}},
  \bibinfo{author}{\bibfnamefont{G.}~\bibnamefont{Zhang}}, \bibnamefont{and}
  \bibinfo{author}{\bibfnamefont{F.}~\bibnamefont{Stillinger}}, Ensemble theory for stealthy hyperuniform disordered ground states.
  \bibinfo{journal}{Phys. Rev. X} \textbf{\bibinfo{volume}{5}},
  \bibinfo{pages}{021020} (\bibinfo{year}{2015}{\natexlab{a}}).

\bibitem[{\citenamefont{Uche et~al.}(2004)\citenamefont{Uche, Stillinger, and
  Torquato}}]{ref10}
\bibinfo{author}{\bibfnamefont{O.~U.} \bibnamefont{Uche}},
  \bibinfo{author}{\bibfnamefont{F.~H.} \bibnamefont{Stillinger}},
  \bibnamefont{and} \bibinfo{author}{\bibfnamefont{S.}~\bibnamefont{Torquato}}, Constraints on collective density variables: Two dimensions.
  \bibinfo{journal}{Phys. Rev. E} \textbf{\bibinfo{volume}{70}},
  \bibinfo{pages}{046122} (\bibinfo{year}{2004}).

\bibitem[{\citenamefont{Batten et~al.}(2008{\natexlab{a}})\citenamefont{Batten,
  Stillinger, and Torquato}}]{ref11}
\bibinfo{author}{\bibfnamefont{R.~D.} \bibnamefont{Batten}},
  \bibinfo{author}{\bibfnamefont{F.~H.} \bibnamefont{Stillinger}},
  \bibnamefont{and} \bibinfo{author}{\bibfnamefont{S.}~\bibnamefont{Torquato}}, Classical disordered ground states: Super-ideal gases and stealth and equi-luminous materials.
  \bibinfo{journal}{J. Appl. Phys.} \textbf{\bibinfo{volume}{104}},
  \bibinfo{pages}{033504} (\bibinfo{year}{2008}{\natexlab{a}}).

\bibitem[{\citenamefont{Batten et~al.}(2009)\citenamefont{Batten, Stillinger,
  and Torquato}}]{ref12}
\bibinfo{author}{\bibfnamefont{R.~D.} \bibnamefont{Batten}},
  \bibinfo{author}{\bibfnamefont{F.~H.} \bibnamefont{Stillinger}},
  \bibnamefont{and} \bibinfo{author}{\bibfnamefont{S.}~\bibnamefont{Torquato}}, Novel low-temperature behavior in classical many-particle systems.
  \bibinfo{journal}{Phys. Rev. Lett.} \textbf{\bibinfo{volume}{103}},
  \bibinfo{pages}{050602} (\bibinfo{year}{2009}).

\bibitem[{\citenamefont{Lebowitz}(1983)}]{ref13}
\bibinfo{author}{\bibfnamefont{J.~L.} \bibnamefont{Lebowitz}}, Charge fluctuations in Coulomb systems.
  \bibinfo{journal}{Phys. Rev. A} \textbf{\bibinfo{volume}{27}},
  \bibinfo{pages}{1491} (\bibinfo{year}{1983}).

\bibitem[{\citenamefont{Zhang et~al.}(2015{\natexlab{a}})\citenamefont{Zhang,
  Stillinger, and Torquato}}]{ref14}
\bibinfo{author}{\bibfnamefont{G.}~\bibnamefont{Zhang}},
  \bibinfo{author}{\bibfnamefont{F.~H.} \bibnamefont{Stillinger}},
  \bibnamefont{and} \bibinfo{author}{\bibfnamefont{S.}~\bibnamefont{Torquato}}, Ground states of stealthy hyperuniform potentials: I. Entropically favored configurations.
  \bibinfo{journal}{Phys. Rev. E} \textbf{\bibinfo{volume}{92}},
  \bibinfo{pages}{022119} (\bibinfo{year}{2015}{\natexlab{a}}).

\bibitem[{\citenamefont{Zhang et~al.}(2015{\natexlab{b}})\citenamefont{Zhang,
  Stillinger, and Torquato}}]{ref15}
\bibinfo{author}{\bibfnamefont{G.}~\bibnamefont{Zhang}},
  \bibinfo{author}{\bibfnamefont{F.~H.} \bibnamefont{Stillinger}},
  \bibnamefont{and} \bibinfo{author}{\bibfnamefont{S.}~\bibnamefont{Torquato}}, Ground states of stealthy hyperuniform potentials. II. Stacked-slider phases.
  \bibinfo{journal}{Phys. Rev. E} \textbf{\bibinfo{volume}{92}},
  \bibinfo{pages}{022120} (\bibinfo{year}{2015}{\natexlab{b}}).

\bibitem[{\citenamefont{Berthier et~al.}(2011)\citenamefont{Berthier,
  Chaudhuri, Coulais, Dauchot, and Sollich}}]{ref16}
\bibinfo{author}{\bibfnamefont{L.}~\bibnamefont{Berthier}},
  \bibinfo{author}{\bibfnamefont{P.}~\bibnamefont{Chaudhuri}},
  \bibinfo{author}{\bibfnamefont{C.}~\bibnamefont{Coulais}},
  \bibinfo{author}{\bibfnamefont{O.}~\bibnamefont{Dauchot}}, \bibnamefont{and}
  \bibinfo{author}{\bibfnamefont{P.}~\bibnamefont{Sollich}},
  Suppressed compressibility at large scale in jammed packings of size-disperse spheres.
  \bibinfo{journal}{Phys. Rev. Lett.} \textbf{\bibinfo{volume}{106}},
  \bibinfo{pages}{120601} (\bibinfo{year}{2011}).

\bibitem[{\citenamefont{Kurita and Weeks}(2011)}]{ref17}
\bibinfo{author}{\bibfnamefont{R.}~\bibnamefont{Kurita}} \bibnamefont{and}
  \bibinfo{author}{\bibfnamefont{E.~R.} \bibnamefont{Weeks}}, Incompressibility of polydisperse random-close-packed colloidal particles.
  \bibinfo{journal}{Phys. Rev. E} \textbf{\bibinfo{volume}{84}},
  \bibinfo{pages}{030401} (\bibinfo{year}{2011}).

\bibitem[{\citenamefont{Hunter and Weeks}(2012)}]{ref18}
\bibinfo{author}{\bibfnamefont{G.~L.} \bibnamefont{Hunter}} \bibnamefont{and}
  \bibinfo{author}{\bibfnamefont{E.~R.} \bibnamefont{Weeks}}, The physics of the colloidal glass transition.
  \bibinfo{journal}{Rep. Prog. Phys.}
  \textbf{\bibinfo{volume}{75}}, \bibinfo{pages}{066501}
  (\bibinfo{year}{2012}).

\bibitem[{\citenamefont{Dreyfus et~al.}(2015)\citenamefont{Dreyfus, Xu, Still,
  Hough, Yodh, and Torquato}}]{ref19}
\bibinfo{author}{\bibfnamefont{R.}~\bibnamefont{Dreyfus}},
  \bibinfo{author}{\bibfnamefont{Y.}~\bibnamefont{Xu}},
  \bibinfo{author}{\bibfnamefont{T.}~\bibnamefont{Still}},
  \bibinfo{author}{\bibfnamefont{L.~A.} \bibnamefont{Hough}},
  \bibinfo{author}{\bibfnamefont{A.}~\bibnamefont{Yodh}}, \bibnamefont{and}
  \bibinfo{author}{\bibfnamefont{S.}~\bibnamefont{Torquato}}, Diagnosing hyperuniformity in two-dimensional, disordered, jammed packings of soft spheres.
  \bibinfo{journal}{Phys. Rev. E} \textbf{\bibinfo{volume}{91}},
  \bibinfo{pages}{012302} (\bibinfo{year}{2015}).

\bibitem[{\citenamefont{Hexner and Levine}(2015{\natexlab{a}})}]{ref20}
\bibinfo{author}{\bibfnamefont{D.}~\bibnamefont{Hexner}} \bibnamefont{and}
  \bibinfo{author}{\bibfnamefont{D.}~\bibnamefont{Levine}}, Hyperuniformity of critical absorbing states.
  \bibinfo{journal}{Phys. Rev. Lett.} \textbf{\bibinfo{volume}{114}},
  \bibinfo{pages}{110602} (\bibinfo{year}{2015}{\natexlab{a}}).

\bibitem[{\citenamefont{Jack et~al.}(2015)\citenamefont{Jack, Thompson, and
  Sollich}}]{ref21}
\bibinfo{author}{\bibfnamefont{R.~L.} \bibnamefont{Jack}},
  \bibinfo{author}{\bibfnamefont{I.~R.} \bibnamefont{Thompson}},
  \bibnamefont{and} \bibinfo{author}{\bibfnamefont{P.}~\bibnamefont{Sollich}}, Hyperuniformity and phase separation in biased ensembles of trajectories for diffusive systems.
  \bibinfo{journal}{Phys. Rev. Lett.} \textbf{\bibinfo{volume}{114}},
  \bibinfo{pages}{060601} (\bibinfo{year}{2015}).

\bibitem[{\citenamefont{Weijs et~al.}(2015)\citenamefont{Weijs, Jeanneret,
  Dreyfus, and Bartolo}}]{ref22}
\bibinfo{author}{\bibfnamefont{J.~H.} \bibnamefont{Weijs}},
  \bibinfo{author}{\bibfnamefont{R.}~\bibnamefont{Jeanneret}},
  \bibinfo{author}{\bibfnamefont{R.}~\bibnamefont{Dreyfus}}, \bibnamefont{and}
  \bibinfo{author}{\bibfnamefont{D.}~\bibnamefont{Bartolo}}, Emergent hyperuniformity in periodically driven emulsions.
  \bibinfo{journal}{Phys. Rev. Lett.} \textbf{\bibinfo{volume}{115}},
  \bibinfo{pages}{108301} (\bibinfo{year}{2015}).

\bibitem[{\citenamefont{Tjhung and Berthier}(2015)}]{ref23}
\bibinfo{author}{\bibfnamefont{E.}~\bibnamefont{Tjhung}} \bibnamefont{and}
  \bibinfo{author}{\bibfnamefont{L.}~\bibnamefont{Berthier}}, Hyperuniform density fluctuations and diverging dynamic correlations in periodically driven colloidal suspensions.
  \bibinfo{journal}{Phys. Rev. Lett.} \textbf{\bibinfo{volume}{114}},
  \bibinfo{pages}{148301} (\bibinfo{year}{2015}).

\bibitem[{\citenamefont{Salvalaglio et~al.}(2020)\citenamefont{Salvalaglio,
  Bouabdellaoui, Bollani, Benali, Favre, Claude, Wenger, de~Anna, Intonti,
  Voigt et~al.}}]{salvalaglio2020hyperuniform}
\bibinfo{author}{\bibfnamefont{M.}~\bibnamefont{Salvalaglio}},
  \bibinfo{author}{\bibfnamefont{M.}~\bibnamefont{Bouabdellaoui}},
  \bibinfo{author}{\bibfnamefont{M.}~\bibnamefont{Bollani}},
  \bibinfo{author}{\bibfnamefont{A.}~\bibnamefont{Benali}},
  \bibinfo{author}{\bibfnamefont{L.}~\bibnamefont{Favre}},
  \bibinfo{author}{\bibfnamefont{J.-B.} \bibnamefont{Claude}},
  \bibinfo{author}{\bibfnamefont{J.}~\bibnamefont{Wenger}},
  \bibinfo{author}{\bibfnamefont{P.}~\bibnamefont{de~Anna}},
  \bibinfo{author}{\bibfnamefont{F.}~\bibnamefont{Intonti}},
  \bibinfo{author}{\bibfnamefont{A.}~\bibnamefont{Voigt}},
  \bibnamefont{et~al.},
  Hyperuniform monocrystalline structures by spinodal solid-state dewetting. 
  \bibinfo{journal}{Phys. Rev. Lett.}
  \textbf{\bibinfo{volume}{125}}, \bibinfo{pages}{126101}
  (\bibinfo{year}{2020}).

\bibitem[{\citenamefont{Torquato et~al.}(2008)\citenamefont{Torquato,
  Scardicchio, and Zachary}}]{ref24}
\bibinfo{author}{\bibfnamefont{S.}~\bibnamefont{Torquato}},
  \bibinfo{author}{\bibfnamefont{A.}~\bibnamefont{Scardicchio}},
  \bibnamefont{and} \bibinfo{author}{\bibfnamefont{C.~E.}
  \bibnamefont{Zachary}}, 
  Point processes in arbitrary dimension from fermionic gases, random matrix theory, and number theory.
  \bibinfo{journal}{J. Stat. Mech.:
  Theor. Exper.} \textbf{\bibinfo{volume}{2008}},
  \bibinfo{pages}{P11019} (\bibinfo{year}{2008}).

\bibitem[{\citenamefont{Feynman and Cohen}(1956)}]{ref25}
\bibinfo{author}{\bibfnamefont{R.}~\bibnamefont{Feynman}} \bibnamefont{and}
  \bibinfo{author}{\bibfnamefont{M.}~\bibnamefont{Cohen}},
  Energy spectrum of the excitations in liquid helium.
  \bibinfo{journal}{Phys. Rev.} \textbf{\bibinfo{volume}{102}},
  \bibinfo{pages}{1189} (\bibinfo{year}{1956}).

\bibitem[{\citenamefont{Jiao et~al.}(2014)\citenamefont{Jiao, Lau, Hatzikirou,
  Meyer-Hermann, Corbo, and Torquato}}]{ref26}
\bibinfo{author}{\bibfnamefont{Y.}~\bibnamefont{Jiao}},
  \bibinfo{author}{\bibfnamefont{T.}~\bibnamefont{Lau}},
  \bibinfo{author}{\bibfnamefont{H.}~\bibnamefont{Hatzikirou}},
  \bibinfo{author}{\bibfnamefont{M.}~\bibnamefont{Meyer-Hermann}},
  \bibinfo{author}{\bibfnamefont{J.~C.} \bibnamefont{Corbo}}, \bibnamefont{and}
  \bibinfo{author}{\bibfnamefont{S.}~\bibnamefont{Torquato}}, Avian photoreceptor patterns represent a disordered hyperuniform solution to a multiscale packing problem.
  \bibinfo{journal}{Phys. Rev. E} \textbf{\bibinfo{volume}{89}},
  \bibinfo{pages}{022721} (\bibinfo{year}{2014}).

\bibitem[{\citenamefont{Mayer et~al.}(2015)\citenamefont{Mayer,
  Balasubramanian, Mora, and Walczak}}]{ref27}
\bibinfo{author}{\bibfnamefont{A.}~\bibnamefont{Mayer}},
  \bibinfo{author}{\bibfnamefont{V.}~\bibnamefont{Balasubramanian}},
  \bibinfo{author}{\bibfnamefont{T.}~\bibnamefont{Mora}}, \bibnamefont{and}
  \bibinfo{author}{\bibfnamefont{A.~M.} \bibnamefont{Walczak}}, How a well-adapted immune system is organized.
  \bibinfo{journal}{Proc. Natl. Acad. Sci. U.S.A.}
  \textbf{\bibinfo{volume}{112}}, \bibinfo{pages}{5950} (\bibinfo{year}{2015}).

\bibitem[{\citenamefont{Hejna et~al.}(2013)\citenamefont{Hejna, Steinhardt, and
  Torquato}}]{ref28}
\bibinfo{author}{\bibfnamefont{M.}~\bibnamefont{Hejna}},
  \bibinfo{author}{\bibfnamefont{P.~J.} \bibnamefont{Steinhardt}},
  \bibnamefont{and} \bibinfo{author}{\bibfnamefont{S.}~\bibnamefont{Torquato}}, Nearly hyperuniform network models of amorphous silicon.
  \bibinfo{journal}{Phys. Rev. B} \textbf{\bibinfo{volume}{87}},
  \bibinfo{pages}{245204} (\bibinfo{year}{2013}).

\bibitem[{\citenamefont{Xie et~al.}(2013)\citenamefont{Xie, Long, Weigand,
  Moss, Carvalho, Roorda, Hejna, Torquato, and Steinhardt}}]{ref29}
\bibinfo{author}{\bibfnamefont{R.}~\bibnamefont{Xie}},
  \bibinfo{author}{\bibfnamefont{G.~G.} \bibnamefont{Long}},
  \bibinfo{author}{\bibfnamefont{S.~J.} \bibnamefont{Weigand}},
  \bibinfo{author}{\bibfnamefont{S.~C.} \bibnamefont{Moss}},
  \bibinfo{author}{\bibfnamefont{T.}~\bibnamefont{Carvalho}},
  \bibinfo{author}{\bibfnamefont{S.}~\bibnamefont{Roorda}},
  \bibinfo{author}{\bibfnamefont{M.}~\bibnamefont{Hejna}},
  \bibinfo{author}{\bibfnamefont{S.}~\bibnamefont{Torquato}}, \bibnamefont{and}
  \bibinfo{author}{\bibfnamefont{P.~J.} \bibnamefont{Steinhardt}}, Hyperuniformity in amorphous silicon based on the measurement of the infinite-wavelength limit of the structure factor.
  \bibinfo{journal}{Proc. Natl. Acad. Sci. U.S.A.}
  \textbf{\bibinfo{volume}{110}}, \bibinfo{pages}{13250}
  (\bibinfo{year}{2013}).

\bibitem[{\citenamefont{Klatt et~al.}(2019)\citenamefont{Klatt, Lovri{\'c},
  Chen, Kapfer, Schaller, Sch{\"o}nh{\"o}fer, Gardiner, Smith,
  Schr{\"o}der-Turk, and Torquato}}]{ref30}
\bibinfo{author}{\bibfnamefont{M.~A.} \bibnamefont{Klatt}},
  \bibinfo{author}{\bibfnamefont{J.}~\bibnamefont{Lovri{\'c}}},
  \bibinfo{author}{\bibfnamefont{D.}~\bibnamefont{Chen}},
  \bibinfo{author}{\bibfnamefont{S.~C.} \bibnamefont{Kapfer}},
  \bibinfo{author}{\bibfnamefont{F.~M.} \bibnamefont{Schaller}},
  \bibinfo{author}{\bibfnamefont{P.~W.} \bibnamefont{Sch{\"o}nh{\"o}fer}},
  \bibinfo{author}{\bibfnamefont{B.~S.} \bibnamefont{Gardiner}},
  \bibinfo{author}{\bibfnamefont{A.-S.} \bibnamefont{Smith}},
  \bibinfo{author}{\bibfnamefont{G.~E.} \bibnamefont{Schr{\"o}der-Turk}},
  \bibnamefont{and} \bibinfo{author}{\bibfnamefont{S.}~\bibnamefont{Torquato}}, Universal hidden order in amorphous cellular geometries.
  \bibinfo{journal}{Nat. Comm.} \textbf{\bibinfo{volume}{10}},
  \bibinfo{pages}{811} (\bibinfo{year}{2019}).

\bibitem[{\citenamefont{Hexner and Levine}(2017)}]{hexner2017noise}
\bibinfo{author}{\bibfnamefont{D.}~\bibnamefont{Hexner}} \bibnamefont{and}
  \bibinfo{author}{\bibfnamefont{D.}~\bibnamefont{Levine}},
  Noise, diffusion, and hyperuniformity.
  \bibinfo{journal}{Phys. Rev. Lett.} \textbf{\bibinfo{volume}{118}},
  \bibinfo{pages}{020601} (\bibinfo{year}{2017}).

\bibitem[{\citenamefont{Hexner et~al.}(2017)\citenamefont{Hexner, Chaikin, and
  Levine}}]{hexner2017enhanced}
\bibinfo{author}{\bibfnamefont{D.}~\bibnamefont{Hexner}},
  \bibinfo{author}{\bibfnamefont{P.~M.} \bibnamefont{Chaikin}},
  \bibnamefont{and} \bibinfo{author}{\bibfnamefont{D.}~\bibnamefont{Levine}},
  Enhanced hyperuniformity from random reorganization.
  \bibinfo{journal}{Proc. Natl. Acad. Sci. U.S.A.}
  \textbf{\bibinfo{volume}{114}}, \bibinfo{pages}{4294} (\bibinfo{year}{2017}).

\bibitem[{\citenamefont{Weijs and Bartolo}(2017)}]{weijs2017mixing}
\bibinfo{author}{\bibfnamefont{J.~H.} \bibnamefont{Weijs}} \bibnamefont{and}
  \bibinfo{author}{\bibfnamefont{D.}~\bibnamefont{Bartolo}},
  Mixing by unstirring: Hyperuniform dispersion of interacting particles upon chaotic advection.
  \bibinfo{journal}{Phys. Rev. Lett.} \textbf{\bibinfo{volume}{119}},
  \bibinfo{pages}{048002} (\bibinfo{year}{2017}).

\bibitem[{\citenamefont{Lei et~al.}(2019)\citenamefont{Lei, Ciamarra, and
  Ni}}]{lei2019nonequilibrium}
\bibinfo{author}{\bibfnamefont{Q.-L.} \bibnamefont{Lei}},
  \bibinfo{author}{\bibfnamefont{M.~P.} \bibnamefont{Ciamarra}},
  \bibnamefont{and} \bibinfo{author}{\bibfnamefont{R.}~\bibnamefont{Ni}}, Nonequilibrium strongly hyperuniform fluids of circle active particles with large local density fluctuations.
  \bibinfo{journal}{Sci. Adv.} \textbf{\bibinfo{volume}{5}},
  \bibinfo{pages}{eaau7423} (\bibinfo{year}{2019}).

\bibitem[{\citenamefont{Lei and Ni}(2019)}]{lei2019random}
\bibinfo{author}{\bibfnamefont{Q.}~\bibnamefont{Lei}} \bibnamefont{and}
  \bibinfo{author}{\bibfnamefont{R.}~\bibnamefont{Ni}}, 
  Hydrodynamics of random-organizing hyperuniform fluids.
  \bibinfo{journal}{arXiv
  preprint arXiv:1904.07514}  (\bibinfo{year}{2019}).

\bibitem[{\citenamefont{Gerasimenko et~al.}(2019)\citenamefont{Gerasimenko,
  Vaskivskyi, Litskevich, Ravnik, Vodeb, Diego, Kabanov, and
  Mihailovic}}]{Ge19}
\bibinfo{author}{\bibfnamefont{Y.~A.} \bibnamefont{Gerasimenko}},
  \bibinfo{author}{\bibfnamefont{I.}~\bibnamefont{Vaskivskyi}},
  \bibinfo{author}{\bibfnamefont{M.}~\bibnamefont{Litskevich}},
  \bibinfo{author}{\bibfnamefont{J.}~\bibnamefont{Ravnik}},
  \bibinfo{author}{\bibfnamefont{J.}~\bibnamefont{Vodeb}},
  \bibinfo{author}{\bibfnamefont{M.}~\bibnamefont{Diego}},
  \bibinfo{author}{\bibfnamefont{V.}~\bibnamefont{Kabanov}}, \bibnamefont{and}
  \bibinfo{author}{\bibfnamefont{D.}~\bibnamefont{Mihailovic}}, Quantum jamming transition to a correlated electron glass in 1T-TaS2.
  \bibinfo{journal}{Nat. Mater.} \textbf{\bibinfo{volume}{18}},
  \bibinfo{pages}{1078} (\bibinfo{year}{2019}).

\bibitem[{\citenamefont{Sakai et~al.}(2022)\citenamefont{Sakai, Arita, and
  Ohtsuki}}]{sakai2022quantum}
\bibinfo{author}{\bibfnamefont{S.}~\bibnamefont{Sakai}},
  \bibinfo{author}{\bibfnamefont{R.}~\bibnamefont{Arita}}, \bibnamefont{and}
  \bibinfo{author}{\bibfnamefont{T.}~\bibnamefont{Ohtsuki}},
  Quantum phase transition between hyperuniform density distributions.
  \bibinfo{journal}{arXiv preprint arXiv:2207.09698}  (\bibinfo{year}{2022}).

\bibitem[{\citenamefont{Rumi et~al.}(2019)\citenamefont{Rumi, S{\'a}nchez,
  El{\'\i}as, Maldonado, Puig, Bolecek, Nieva, Konczykowski, Fasano, and
  Kolton}}]{Ru19}
\bibinfo{author}{\bibfnamefont{G.}~\bibnamefont{Rumi}},
  \bibinfo{author}{\bibfnamefont{J.~A.} \bibnamefont{S{\'a}nchez}},
  \bibinfo{author}{\bibfnamefont{F.}~\bibnamefont{El{\'\i}as}},
  \bibinfo{author}{\bibfnamefont{R.~C.} \bibnamefont{Maldonado}},
  \bibinfo{author}{\bibfnamefont{J.}~\bibnamefont{Puig}},
  \bibinfo{author}{\bibfnamefont{N.~R.~C.} \bibnamefont{Bolecek}},
  \bibinfo{author}{\bibfnamefont{G.}~\bibnamefont{Nieva}},
  \bibinfo{author}{\bibfnamefont{M.}~\bibnamefont{Konczykowski}},
  \bibinfo{author}{\bibfnamefont{Y.}~\bibnamefont{Fasano}}, \bibnamefont{and}
  \bibinfo{author}{\bibfnamefont{A.~B.} \bibnamefont{Kolton}},
  \bibinfo{journal}{Phys. Rev. Res.} \textbf{\bibinfo{volume}{1}},
  \bibinfo{pages}{033057} (\bibinfo{year}{2019}).

\bibitem[{\citenamefont{S{\'a}nchez et~al.}(2019)\citenamefont{S{\'a}nchez,
  Maldonado, Bolecek, Rumi, Pedrazzini, Dolz, Nieva, van~der Beek,
  Konczykowski, Dewhurst et~al.}}]{Sa19}
\bibinfo{author}{\bibfnamefont{J.~A.} \bibnamefont{S{\'a}nchez}},
  \bibinfo{author}{\bibfnamefont{R.~C.} \bibnamefont{Maldonado}},
  \bibinfo{author}{\bibfnamefont{N.~R.~C.} \bibnamefont{Bolecek}},
  \bibinfo{author}{\bibfnamefont{G.}~\bibnamefont{Rumi}},
  \bibinfo{author}{\bibfnamefont{P.}~\bibnamefont{Pedrazzini}},
  \bibinfo{author}{\bibfnamefont{M.~I.} \bibnamefont{Dolz}},
  \bibinfo{author}{\bibfnamefont{G.}~\bibnamefont{Nieva}},
  \bibinfo{author}{\bibfnamefont{C.~J.} \bibnamefont{van~der Beek}},
  \bibinfo{author}{\bibfnamefont{M.}~\bibnamefont{Konczykowski}},
  \bibinfo{author}{\bibfnamefont{C.~D.} \bibnamefont{Dewhurst}},
  \bibnamefont{et~al.}, \bibinfo{journal}{Commun. Phys.}
  \textbf{\bibinfo{volume}{2}}, \bibinfo{pages}{1} (\bibinfo{year}{2019}).

\bibitem[{\citenamefont{Chen et~al.}(2023)\citenamefont{Chen, Jiang, Wang,
  Zhuang, and Jiao}}]{chen2021multihyperuniform}
\bibinfo{author}{\bibfnamefont{D.}~\bibnamefont{Chen}},
  \bibinfo{author}{\bibfnamefont{X.}~\bibnamefont{Jiang}},
  \bibinfo{author}{\bibfnamefont{D.}~\bibnamefont{Wang}},
  \bibinfo{author}{\bibfnamefont{H.}~\bibnamefont{Zhuang}}, \bibnamefont{and}
  \bibinfo{author}{\bibfnamefont{Y.}~\bibnamefont{Jiao}},
  Multihyperuniform long-range order in medium-entropy alloys.
  \bibinfo{journal}{Acta Mater.} \textbf{\bibinfo{volume}{246}},
  \bibinfo{pages}{118678} (\bibinfo{year}{2023}).

\bibitem[{\citenamefont{Zheng et~al.}(2020)\citenamefont{Zheng, Liu, Nan, Shen,
  Zhang, Chen, He, Xu, Chen, Jiao et~al.}}]{Zh20}
\bibinfo{author}{\bibfnamefont{Y.}~\bibnamefont{Zheng}},
  \bibinfo{author}{\bibfnamefont{L.}~\bibnamefont{Liu}},
  \bibinfo{author}{\bibfnamefont{H.}~\bibnamefont{Nan}},
  \bibinfo{author}{\bibfnamefont{Z.-X.} \bibnamefont{Shen}},
  \bibinfo{author}{\bibfnamefont{G.}~\bibnamefont{Zhang}},
  \bibinfo{author}{\bibfnamefont{D.}~\bibnamefont{Chen}},
  \bibinfo{author}{\bibfnamefont{L.}~\bibnamefont{He}},
  \bibinfo{author}{\bibfnamefont{W.}~\bibnamefont{Xu}},
  \bibinfo{author}{\bibfnamefont{M.}~\bibnamefont{Chen}},
  \bibinfo{author}{\bibfnamefont{Y.}~\bibnamefont{Jiao}}, \bibnamefont{et~al.}, Disordered hyperuniformity in two-dimensional amorphous silica.
  \bibinfo{journal}{Sci. Adv.} \textbf{\bibinfo{volume}{6}},
  \bibinfo{pages}{eaba0826} (\bibinfo{year}{2020}).

\bibitem[{\citenamefont{Chen et~al.}(2021{\natexlab{a}})\citenamefont{Chen,
  Zheng, Liu, Zhang, Chen, Jiao, and Zhuang}}]{Ch21}
\bibinfo{author}{\bibfnamefont{D.}~\bibnamefont{Chen}},
  \bibinfo{author}{\bibfnamefont{Y.}~\bibnamefont{Zheng}},
  \bibinfo{author}{\bibfnamefont{L.}~\bibnamefont{Liu}},
  \bibinfo{author}{\bibfnamefont{G.}~\bibnamefont{Zhang}},
  \bibinfo{author}{\bibfnamefont{M.}~\bibnamefont{Chen}},
  \bibinfo{author}{\bibfnamefont{Y.}~\bibnamefont{Jiao}}, \bibnamefont{and}
  \bibinfo{author}{\bibfnamefont{H.}~\bibnamefont{Zhuang}},
  Stone–wales defects preserve hyperuniformity in amorphous two-dimensional networks. 
  \bibinfo{journal}{Proc. Natl. Acad. Sci. U.S.A.}
  \textbf{\bibinfo{volume}{118}}, \bibinfo{pages}{e2016862118}
  (\bibinfo{year}{2021}{\natexlab{a}}).

\bibitem[{\citenamefont{Chen et~al.}(2021{\natexlab{b}})\citenamefont{Chen,
  Zheng, Lee, Kang, Zhu, Zhuang, Huang, and Jiao}}]{PhysRevB.103.224102}
\bibinfo{author}{\bibfnamefont{D.}~\bibnamefont{Chen}},
  \bibinfo{author}{\bibfnamefont{Y.}~\bibnamefont{Zheng}},
  \bibinfo{author}{\bibfnamefont{C.-H.} \bibnamefont{Lee}},
  \bibinfo{author}{\bibfnamefont{S.}~\bibnamefont{Kang}},
  \bibinfo{author}{\bibfnamefont{W.}~\bibnamefont{Zhu}},
  \bibinfo{author}{\bibfnamefont{H.}~\bibnamefont{Zhuang}},
  \bibinfo{author}{\bibfnamefont{P.~Y.} \bibnamefont{Huang}}, \bibnamefont{and}
  \bibinfo{author}{\bibfnamefont{Y.}~\bibnamefont{Jiao}}, Nearly hyperuniform, nonhyperuniform, and antihyperuniform density fluctuations in two-dimensional transition metal dichalcogenides with defects.
  \bibinfo{journal}{Phys. Rev. B} \textbf{\bibinfo{volume}{103}},
  \bibinfo{pages}{224102} (\bibinfo{year}{2021}{\natexlab{b}}).

\bibitem[{\citenamefont{Zheng et~al.}(2021)\citenamefont{Zheng, Chen, Liu, Liu,
  Chen, Zhuang, and Jiao}}]{Zh21}
\bibinfo{author}{\bibfnamefont{Y.}~\bibnamefont{Zheng}},
  \bibinfo{author}{\bibfnamefont{D.}~\bibnamefont{Chen}},
  \bibinfo{author}{\bibfnamefont{L.}~\bibnamefont{Liu}},
  \bibinfo{author}{\bibfnamefont{Y.}~\bibnamefont{Liu}},
  \bibinfo{author}{\bibfnamefont{M.}~\bibnamefont{Chen}},
  \bibinfo{author}{\bibfnamefont{H.}~\bibnamefont{Zhuang}}, \bibnamefont{and}
  \bibinfo{author}{\bibfnamefont{Y.}~\bibnamefont{Jiao}},
  Topological transformations in hyperuniform pentagonal two-dimensional materials induced by Stone-Wales defects.
  \bibinfo{journal}{Phys. Rev. B} \textbf{\bibinfo{volume}{103}},
  \bibinfo{pages}{245413} (\bibinfo{year}{2021}).

\bibitem{apr_dhu}
D. Chen, H. Zhuang, M. Chen, P. Huang, V. Vleck, and Y. Jiao, Disordered hyperuniform solid state materials. Appl. Phys. Rev. {\bf 10}, 021310 (2023).

\bibitem[{\citenamefont{Chen et~al.}(2022{\natexlab{a}})\citenamefont{Chen,
  Liu, Zhuang, Chen, and Jiao}}]{nanotube}
\bibinfo{author}{\bibfnamefont{D.}~\bibnamefont{Chen}},
  \bibinfo{author}{\bibfnamefont{Y.}~\bibnamefont{Liu}},
  \bibinfo{author}{\bibfnamefont{H.}~\bibnamefont{Zhuang}},
  \bibinfo{author}{\bibfnamefont{M.}~\bibnamefont{Chen}}, \bibnamefont{and}
  \bibinfo{author}{\bibfnamefont{Y.}~\bibnamefont{Jiao}},
  Disordered hyperuniform quasi-one-dimensional materials.
  \bibinfo{journal}{Phys. Rev. B} \textbf{\bibinfo{volume}{106}},
  \bibinfo{pages}{235427} (\bibinfo{year}{2022}{\natexlab{a}}).

\bibitem[{\citenamefont{Zhang et~al.}(2023)\citenamefont{Zhang, Wang, Zhang,
  Yu, and Douglas}}]{zhang2023approach}
\bibinfo{author}{\bibfnamefont{H.}~\bibnamefont{Zhang}},
  \bibinfo{author}{\bibfnamefont{X.}~\bibnamefont{Wang}},
  \bibinfo{author}{\bibfnamefont{J.}~\bibnamefont{Zhang}},
  \bibinfo{author}{\bibfnamefont{H.-B.} \bibnamefont{Yu}}, \bibnamefont{and}
  \bibinfo{author}{\bibfnamefont{J.~F.} \bibnamefont{Douglas}}, 
  Approach to Hyperuniformity in a Metallic Glass-Forming Material Exhibiting a Fragile to Strong Glass Transition.
  \bibinfo{journal}{arXiv preprint arXiv:2302.01429}  (\bibinfo{year}{2023}).

\bibitem[{\citenamefont{Florescu et~al.}(2009)\citenamefont{Florescu, Torquato,
  and Steinhardt}}]{ref31}
\bibinfo{author}{\bibfnamefont{M.}~\bibnamefont{Florescu}},
  \bibinfo{author}{\bibfnamefont{S.}~\bibnamefont{Torquato}}, \bibnamefont{and}
  \bibinfo{author}{\bibfnamefont{P.~J.} \bibnamefont{Steinhardt}}, Designer disordered materials with large, complete photonic band gaps.
  \bibinfo{journal}{Proc. Natl. Acad. Sci.}
  \textbf{\bibinfo{volume}{106}}, \bibinfo{pages}{20658}
  (\bibinfo{year}{2009}).

\bibitem[{\citenamefont{Man et~al.}(2013{\natexlab{a}})\citenamefont{Man,
  Florescu, Matsuyama, Yadak, Nahal, Hashemizad, Williamson, Steinhardt,
  Torquato, and Chaikin}}]{ref32}
\bibinfo{author}{\bibfnamefont{W.}~\bibnamefont{Man}},
  \bibinfo{author}{\bibfnamefont{M.}~\bibnamefont{Florescu}},
  \bibinfo{author}{\bibfnamefont{K.}~\bibnamefont{Matsuyama}},
  \bibinfo{author}{\bibfnamefont{P.}~\bibnamefont{Yadak}},
  \bibinfo{author}{\bibfnamefont{G.}~\bibnamefont{Nahal}},
  \bibinfo{author}{\bibfnamefont{S.}~\bibnamefont{Hashemizad}},
  \bibinfo{author}{\bibfnamefont{E.}~\bibnamefont{Williamson}},
  \bibinfo{author}{\bibfnamefont{P.}~\bibnamefont{Steinhardt}},
  \bibinfo{author}{\bibfnamefont{S.}~\bibnamefont{Torquato}}, \bibnamefont{and}
  \bibinfo{author}{\bibfnamefont{P.}~\bibnamefont{Chaikin}},
  Photonic band gap in isotropic hyperuniform disordered solids with low dielectric contrast. 
  \bibinfo{journal}{Opt. Exp.} \textbf{\bibinfo{volume}{21}},
  \bibinfo{pages}{19972} (\bibinfo{year}{2013}{\natexlab{a}}).

\bibitem[{\citenamefont{Man et~al.}(2013{\natexlab{b}})\citenamefont{Man,
  Florescu, Williamson, He, Hashemizad, Leung, Liner, Torquato, Chaikin, and
  Steinhardt}}]{ref33}
\bibinfo{author}{\bibfnamefont{W.}~\bibnamefont{Man}},
  \bibinfo{author}{\bibfnamefont{M.}~\bibnamefont{Florescu}},
  \bibinfo{author}{\bibfnamefont{E.~P.} \bibnamefont{Williamson}},
  \bibinfo{author}{\bibfnamefont{Y.}~\bibnamefont{He}},
  \bibinfo{author}{\bibfnamefont{S.~R.} \bibnamefont{Hashemizad}},
  \bibinfo{author}{\bibfnamefont{B.~Y.} \bibnamefont{Leung}},
  \bibinfo{author}{\bibfnamefont{D.~R.} \bibnamefont{Liner}},
  \bibinfo{author}{\bibfnamefont{S.}~\bibnamefont{Torquato}},
  \bibinfo{author}{\bibfnamefont{P.~M.} \bibnamefont{Chaikin}},
  \bibnamefont{and} \bibinfo{author}{\bibfnamefont{P.~J.}
  \bibnamefont{Steinhardt}}, 
  Isotropic band gaps and freeform waveguides observed in hyperuniform disordered photonic solids. \bibinfo{journal}{Proc. Natl.
  Acad. Sci.} \textbf{\bibinfo{volume}{110}}, \bibinfo{pages}{15886}
  (\bibinfo{year}{2013}{\natexlab{b}}).

\bibitem[{\citenamefont{Yu}(2023)}]{scattering}
\bibinfo{author}{\bibfnamefont{S.}~\bibnamefont{Yu}},  Evolving scattering networks for engineering disorder.
\bibinfo{journal}{Nat.
  Comput. Sci.} \textbf{\bibinfo{volume}{1}}, \bibinfo{pages}{1}
  (\bibinfo{year}{2023}).

\bibitem[{\citenamefont{Granchi et~al.}(2022)\citenamefont{Granchi, Spalding,
  Lodde, Petruzzella, Otten, Fiore, Intonti, Sapienza, Florescu, and
  Gurioli}}]{granchi2022near}
\bibinfo{author}{\bibfnamefont{N.}~\bibnamefont{Granchi}},
  \bibinfo{author}{\bibfnamefont{R.}~\bibnamefont{Spalding}},
  \bibinfo{author}{\bibfnamefont{M.}~\bibnamefont{Lodde}},
  \bibinfo{author}{\bibfnamefont{M.}~\bibnamefont{Petruzzella}},
  \bibinfo{author}{\bibfnamefont{F.~W.} \bibnamefont{Otten}},
  \bibinfo{author}{\bibfnamefont{A.}~\bibnamefont{Fiore}},
  \bibinfo{author}{\bibfnamefont{F.}~\bibnamefont{Intonti}},
  \bibinfo{author}{\bibfnamefont{R.}~\bibnamefont{Sapienza}},
  \bibinfo{author}{\bibfnamefont{M.}~\bibnamefont{Florescu}}, \bibnamefont{and}
  \bibinfo{author}{\bibfnamefont{M.}~\bibnamefont{Gurioli}},
  Near‐Field Investigation of Luminescent Hyperuniform Disordered Materials. 
  \bibinfo{journal}{Adv. Opt. Mater.} \textbf{\bibinfo{volume}{10}},
  \bibinfo{pages}{2102565} (\bibinfo{year}{2022}).

\bibitem[{\citenamefont{Park et~al.}(2021)\citenamefont{Park, Lee, Kim, Park,
  and Yu}}]{park2021hearing}
\bibinfo{author}{\bibfnamefont{S.}~\bibnamefont{Park}},
  \bibinfo{author}{\bibfnamefont{I.}~\bibnamefont{Lee}},
  \bibinfo{author}{\bibfnamefont{J.}~\bibnamefont{Kim}},
  \bibinfo{author}{\bibfnamefont{N.}~\bibnamefont{Park}}, \bibnamefont{and}
  \bibinfo{author}{\bibfnamefont{S.}~\bibnamefont{Yu}}, Hearing the shape of a drum for light: isospectrality in photonics.
  \bibinfo{journal}{Nanophotonics} \textbf{\bibinfo{volume}{11}},
  \bibinfo{pages}{2763} (\bibinfo{year}{2021}).

\bibitem[{\citenamefont{Klatt et~al.}(2022)\citenamefont{Klatt, Steinhardt, and
  Torquato}}]{klatt2022wave}
\bibinfo{author}{\bibfnamefont{M.~A.} \bibnamefont{Klatt}},
  \bibinfo{author}{\bibfnamefont{P.~J.} \bibnamefont{Steinhardt}},
  \bibnamefont{and} \bibinfo{author}{\bibfnamefont{S.}~\bibnamefont{Torquato}}, Wave propagation and band tails of two-dimensional disordered systems in the thermodynamic limit.
  \bibinfo{journal}{Proc. Natl. Acad. Sci.}
  \textbf{\bibinfo{volume}{119}}, \bibinfo{pages}{e2213633119}
  (\bibinfo{year}{2022}).

\bibitem[{\citenamefont{Tavakoli et~al.}(2022)\citenamefont{Tavakoli, Spalding,
  Lambertz, Koppejan, Gkantzounis, Wan, Rohrich, Kontoleta, Koenderink,
  Sapienza et~al.}}]{tavakoli2022over}
\bibinfo{author}{\bibfnamefont{N.}~\bibnamefont{Tavakoli}},
  \bibinfo{author}{\bibfnamefont{R.}~\bibnamefont{Spalding}},
  \bibinfo{author}{\bibfnamefont{A.}~\bibnamefont{Lambertz}},
  \bibinfo{author}{\bibfnamefont{P.}~\bibnamefont{Koppejan}},
  \bibinfo{author}{\bibfnamefont{G.}~\bibnamefont{Gkantzounis}},
  \bibinfo{author}{\bibfnamefont{C.}~\bibnamefont{Wan}},
  \bibinfo{author}{\bibfnamefont{R.}~\bibnamefont{Rohrich}},
  \bibinfo{author}{\bibfnamefont{E.}~\bibnamefont{Kontoleta}},
  \bibinfo{author}{\bibfnamefont{A.~F.} \bibnamefont{Koenderink}},
  \bibinfo{author}{\bibfnamefont{R.}~\bibnamefont{Sapienza}},
  \bibnamefont{et~al.}, Over 65 percent sunlight absorption in a 1 micron Si slab with hyperuniform texture.
  \bibinfo{journal}{ACS Photo.}
  \textbf{\bibinfo{volume}{9}}, \bibinfo{pages}{1206} (\bibinfo{year}{2022}).

\bibitem[{\citenamefont{Ch{\'e}ron et~al.}(2022)\citenamefont{Ch{\'e}ron,
  F{\'e}lix, Groby, Pagneux, and Romero-Garc{\'\i}a}}]{cheron2022wave}
\bibinfo{author}{\bibfnamefont{{\'E}.}~\bibnamefont{Ch{\'e}ron}},
  \bibinfo{author}{\bibfnamefont{S.}~\bibnamefont{F{\'e}lix}},
  \bibinfo{author}{\bibfnamefont{J.-P.} \bibnamefont{Groby}},
  \bibinfo{author}{\bibfnamefont{V.}~\bibnamefont{Pagneux}}, \bibnamefont{and}
  \bibinfo{author}{\bibfnamefont{V.}~\bibnamefont{Romero-Garc{\'\i}a}}, Wave transport in stealth hyperuniform materials: The diffusive regime and beyond.
  \bibinfo{journal}{Appl. Phys. Lett.} \textbf{\bibinfo{volume}{121}},
  \bibinfo{pages}{061702} (\bibinfo{year}{2022}).

\bibitem[{\citenamefont{Zhang et~al.}(2016)\citenamefont{Zhang, Stillinger, and
  Torquato}}]{ref34}
\bibinfo{author}{\bibfnamefont{G.}~\bibnamefont{Zhang}},
  \bibinfo{author}{\bibfnamefont{F.}~\bibnamefont{Stillinger}},
  \bibnamefont{and} \bibinfo{author}{\bibfnamefont{S.}~\bibnamefont{Torquato}}, Transport, geometrical, and topological properties of stealthy disordered hyperuniform two-phase systems.
  \bibinfo{journal}{J. Chem. Phys.}
  \textbf{\bibinfo{volume}{145}}, \bibinfo{pages}{244109}
  (\bibinfo{year}{2016}).

\bibitem[{\citenamefont{Torquato}(2021)}]{torquato2021diffusion}
\bibinfo{author}{\bibfnamefont{S.}~\bibnamefont{Torquato}}, Diffusion spreadability as a probe of the microstructure of complex media across length scales.
  \bibinfo{journal}{Phys. Rev. E} \textbf{\bibinfo{volume}{104}},
  \bibinfo{pages}{054102} (\bibinfo{year}{2021}).

\bibitem[{\citenamefont{Maher et~al.}(2022)\citenamefont{Maher, Stillinger, and
  Torquato}}]{maher2022characterization}
\bibinfo{author}{\bibfnamefont{C.~E.} \bibnamefont{Maher}},
  \bibinfo{author}{\bibfnamefont{F.~H.} \bibnamefont{Stillinger}},
  \bibnamefont{and} \bibinfo{author}{\bibfnamefont{S.}~\bibnamefont{Torquato}}, Characterization of void space, large-scale structure, and transport properties of maximally random jammed packings of superballs.
  \bibinfo{journal}{Phys. Rev. Mater.} \textbf{\bibinfo{volume}{6}},
  \bibinfo{pages}{025603} (\bibinfo{year}{2022}).

\bibitem[{\citenamefont{Xu et~al.}(2017)\citenamefont{Xu, Chen, Chen, Xu, and
  Jiao}}]{ref35}
\bibinfo{author}{\bibfnamefont{Y.}~\bibnamefont{Xu}},
  \bibinfo{author}{\bibfnamefont{S.}~\bibnamefont{Chen}},
  \bibinfo{author}{\bibfnamefont{P.-E.} \bibnamefont{Chen}},
  \bibinfo{author}{\bibfnamefont{W.}~\bibnamefont{Xu}}, \bibnamefont{and}
  \bibinfo{author}{\bibfnamefont{Y.}~\bibnamefont{Jiao}}, Microstructure and mechanical properties of hyperuniform heterogeneous materials.
  \bibinfo{journal}{Phys. Rev. E} \textbf{\bibinfo{volume}{96}},
  \bibinfo{pages}{043301} (\bibinfo{year}{2017}).

\bibitem[{\citenamefont{Puig et~al.}(2022)\citenamefont{Puig, El{\'\i}as,
  Arag{\'o}n~S{\'a}nchez, Cort{\'e}s~Maldonado, Rumi, Nieva, Pedrazzini,
  Kolton, and Fasano}}]{puig2022anisotropic}
\bibinfo{author}{\bibfnamefont{J.}~\bibnamefont{Puig}},
  \bibinfo{author}{\bibfnamefont{F.}~\bibnamefont{El{\'\i}as}},
  \bibinfo{author}{\bibfnamefont{J.}~\bibnamefont{Arag{\'o}n~S{\'a}nchez}},
  \bibinfo{author}{\bibfnamefont{R.}~\bibnamefont{Cort{\'e}s~Maldonado}},
  \bibinfo{author}{\bibfnamefont{G.}~\bibnamefont{Rumi}},
  \bibinfo{author}{\bibfnamefont{G.}~\bibnamefont{Nieva}},
  \bibinfo{author}{\bibfnamefont{P.}~\bibnamefont{Pedrazzini}},
  \bibinfo{author}{\bibfnamefont{A.~B.} \bibnamefont{Kolton}},
  \bibnamefont{and} \bibinfo{author}{\bibfnamefont{Y.}~\bibnamefont{Fasano}},
  \bibinfo{journal}{Comm. Mater.} \textbf{\bibinfo{volume}{3}},
  \bibinfo{pages}{32} (\bibinfo{year}{2022}).

\bibitem[{\citenamefont{Torquato and Chen}(2018)}]{ref36}
\bibinfo{author}{\bibfnamefont{S.}~\bibnamefont{Torquato}} \bibnamefont{and}
  \bibinfo{author}{\bibfnamefont{D.}~\bibnamefont{Chen}}, Multifunctional hyperuniform cellular networks: Optimality, anisotropy and disorder.
  \bibinfo{journal}{Multifun. Mater.} \textbf{\bibinfo{volume}{1}},
  \bibinfo{pages}{015001} (\bibinfo{year}{2018}).

\bibitem[{\citenamefont{Kim and
  Torquato}(2020{\natexlab{a}})}]{kim2020multifunctional}
\bibinfo{author}{\bibfnamefont{J.}~\bibnamefont{Kim}} \bibnamefont{and}
  \bibinfo{author}{\bibfnamefont{S.}~\bibnamefont{Torquato}}, Multifunctional composites for elastic and electromagnetic wave propagation.
  \bibinfo{journal}{Proc. Natl. Acad. Sci.}
  \textbf{\bibinfo{volume}{117}}, \bibinfo{pages}{8764}
  (\bibinfo{year}{2020}{\natexlab{a}}).

\bibitem[{\citenamefont{Torquato}(2022)}]{torquato2022extraordinary}
\bibinfo{author}{\bibfnamefont{S.}~\bibnamefont{Torquato}}, Extraordinary disordered hyperuniform multifunctional composites.
  \bibinfo{journal}{J. Compos. Mater.}
  \textbf{\bibinfo{volume}{56}}, \bibinfo{pages}{3635} (\bibinfo{year}{2022}).

\bibitem[{\citenamefont{Florescu et~al.}(2013)\citenamefont{Florescu,
  Steinhardt, and Torquato}}]{Fl13}
\bibinfo{author}{\bibfnamefont{M.}~\bibnamefont{Florescu}},
  \bibinfo{author}{\bibfnamefont{P.~J.} \bibnamefont{Steinhardt}},
  \bibnamefont{and} \bibinfo{author}{\bibfnamefont{S.}~\bibnamefont{Torquato}}, Optical cavities and waveguides in hyperuniform disordered photonic solids.
  \bibinfo{journal}{Phys. Rev. B} \textbf{\bibinfo{volume}{87}},
  \bibinfo{pages}{165116} (\bibinfo{year}{2013}).

\bibitem[{\citenamefont{Tang et~al.}(2022)\citenamefont{Tang, Hao, Liu, Tian,
  Niu, and Zang}}]{tang2022soft}
\bibinfo{author}{\bibfnamefont{H.}~\bibnamefont{Tang}},
  \bibinfo{author}{\bibfnamefont{Z.}~\bibnamefont{Hao}},
  \bibinfo{author}{\bibfnamefont{Y.}~\bibnamefont{Liu}},
  \bibinfo{author}{\bibfnamefont{Y.}~\bibnamefont{Tian}},
  \bibinfo{author}{\bibfnamefont{H.}~\bibnamefont{Niu}}, \bibnamefont{and}
  \bibinfo{author}{\bibfnamefont{J.}~\bibnamefont{Zang}}, Soft and disordered hyperuniform elastic metamaterials for highly efficient vibration concentration.
  \bibinfo{journal}{Nat. Sci. Rev.} \textbf{\bibinfo{volume}{9}},
  \bibinfo{pages}{nwab133} (\bibinfo{year}{2022}).

\bibitem{new1}
O. Leseur, R. Pierrat, and R. Carminati, High-Density Hyperuniform Materials Can Be Transparent. Optica {\bf 3}, 763 (2016);

\bibitem[{\citenamefont{Torquato and Kim}(2021)}]{torquato2021nonlocal}
\bibinfo{author}{\bibfnamefont{S.}~\bibnamefont{Torquato}} \bibnamefont{and}
  \bibinfo{author}{\bibfnamefont{J.}~\bibnamefont{Kim}}, Nonlocal effective electromagnetic wave characteristics of composite media: Beyond the quasistatic regime.
  \bibinfo{journal}{Phys. Rev. X} \textbf{\bibinfo{volume}{11}},
  \bibinfo{pages}{021002} (\bibinfo{year}{2021}).

\bibitem{new2}
Kim and S. Torquato, Effective electromagnetic wave properties of disordered stealthy hyperuniform layered media beyond the quasistatic regime. Optica {\bf 10}, 965 (2023).


\bibitem[{\citenamefont{Wu et~al.}(2017)\citenamefont{Wu, Sheng, and
  Hao}}]{wu2017effective}
\bibinfo{author}{\bibfnamefont{B.-Y.} \bibnamefont{Wu}},
  \bibinfo{author}{\bibfnamefont{X.-Q.} \bibnamefont{Sheng}}, \bibnamefont{and}
  \bibinfo{author}{\bibfnamefont{Y.}~\bibnamefont{Hao}}, 
  Effective media properties of hyperuniform disordered composite materials. 
  \bibinfo{journal}{PloS
  One} \textbf{\bibinfo{volume}{12}}, \bibinfo{pages}{e0185921}
  (\bibinfo{year}{2017}).

\bibitem[{\citenamefont{Zhang and Hao}(2019)}]{zhang2019metasurface}
\bibinfo{author}{\bibfnamefont{H.}~\bibnamefont{Zhang}} \bibnamefont{and}
  \bibinfo{author}{\bibfnamefont{Y.}~\bibnamefont{Hao}}, in
  \emph{\bibinfo{booktitle}{2019 IEEE International Symposium on Antennas and Propagation and USNC-URSI Radio Science Meeting}}
  (\bibinfo{organization}{IEEE}, \bibinfo{year}{2019}), pp.
  \bibinfo{pages}{1205--1206}.

\bibitem[{\citenamefont{Zhang et~al.}(2019)\citenamefont{Zhang, Chu, Giddens,
  Wu, and Hao}}]{zhang2019experimental}
\bibinfo{author}{\bibfnamefont{H.}~\bibnamefont{Zhang}},
  \bibinfo{author}{\bibfnamefont{H.}~\bibnamefont{Chu}},
  \bibinfo{author}{\bibfnamefont{H.}~\bibnamefont{Giddens}},
  \bibinfo{author}{\bibfnamefont{W.}~\bibnamefont{Wu}}, \bibnamefont{and}
  \bibinfo{author}{\bibfnamefont{Y.}~\bibnamefont{Hao}},
  Experimental demonstration of Luneburg lens based on hyperuniform disordered media.
  \bibinfo{journal}{Appl. Phys. Lett.} \textbf{\bibinfo{volume}{114}},
  \bibinfo{pages}{053507} (\bibinfo{year}{2019}).

\bibitem[{\citenamefont{Zhang et~al.}(2021)\citenamefont{Zhang, Cheng, Chu,
  Christogeorgos, Wu, and Hao}}]{zhang2021hyperuniform}
\bibinfo{author}{\bibfnamefont{H.}~\bibnamefont{Zhang}},
  \bibinfo{author}{\bibfnamefont{Q.}~\bibnamefont{Cheng}},
  \bibinfo{author}{\bibfnamefont{H.}~\bibnamefont{Chu}},
  \bibinfo{author}{\bibfnamefont{O.}~\bibnamefont{Christogeorgos}},
  \bibinfo{author}{\bibfnamefont{W.}~\bibnamefont{Wu}}, \bibnamefont{and}
  \bibinfo{author}{\bibfnamefont{Y.}~\bibnamefont{Hao}},
  Hyperuniform disordered distribution metasurface for scattering reduction.
  \bibinfo{journal}{Appl. Phys. Lett.} \textbf{\bibinfo{volume}{118}},
  \bibinfo{pages}{101601} (\bibinfo{year}{2021}).

\bibitem[{\citenamefont{Zhang et~al.}(2022)\citenamefont{Zhang, Wu, Cheng,
  Chen, Yu, and Fang}}]{zhang2022reconfigurable}
\bibinfo{author}{\bibfnamefont{H.}~\bibnamefont{Zhang}},
  \bibinfo{author}{\bibfnamefont{W.}~\bibnamefont{Wu}},
  \bibinfo{author}{\bibfnamefont{Q.}~\bibnamefont{Cheng}},
  \bibinfo{author}{\bibfnamefont{Q.}~\bibnamefont{Chen}},
  \bibinfo{author}{\bibfnamefont{Y.-H.} \bibnamefont{Yu}}, \bibnamefont{and}
  \bibinfo{author}{\bibfnamefont{D.-G.} \bibnamefont{Fang}},
  Reconfigurable Reflectarray Antenna Based on Hyperuniform Disordered Distribution.
  \bibinfo{journal}{IEEE Trans. Anten. Prop.}
  (\bibinfo{year}{2022}).

\bibitem[{\citenamefont{Haberko et~al.}(2013)\citenamefont{Haberko, Muller, and
  Scheffold}}]{ref37}
\bibinfo{author}{\bibfnamefont{J.}~\bibnamefont{Haberko}},
  \bibinfo{author}{\bibfnamefont{N.}~\bibnamefont{Muller}}, \bibnamefont{and}
  \bibinfo{author}{\bibfnamefont{F.}~\bibnamefont{Scheffold}}, Direct laser writing of three-dimensional network structures as templates for disordered photonic materials.
  \bibinfo{journal}{Phys. Rev. A} \textbf{\bibinfo{volume}{88}},
  \bibinfo{pages}{043822} (\bibinfo{year}{2013}).

\bibitem[{\citenamefont{Zito et~al.}(2015)\citenamefont{Zito, Rusciano, Pesce,
  Malafronte, Di~Girolamo, Ausanio, Vecchione, and Sasso}}]{ref38}
\bibinfo{author}{\bibfnamefont{G.}~\bibnamefont{Zito}},
  \bibinfo{author}{\bibfnamefont{G.}~\bibnamefont{Rusciano}},
  \bibinfo{author}{\bibfnamefont{G.}~\bibnamefont{Pesce}},
  \bibinfo{author}{\bibfnamefont{A.}~\bibnamefont{Malafronte}},
  \bibinfo{author}{\bibfnamefont{R.}~\bibnamefont{Di~Girolamo}},
  \bibinfo{author}{\bibfnamefont{G.}~\bibnamefont{Ausanio}},
  \bibinfo{author}{\bibfnamefont{A.}~\bibnamefont{Vecchione}},
  \bibnamefont{and} \bibinfo{author}{\bibfnamefont{A.}~\bibnamefont{Sasso}},
  Nanoscale engineering of two-dimensional disordered hyperuniform block-copolymer assemblies. 
  \bibinfo{journal}{Phys. Rev. E} \textbf{\bibinfo{volume}{92}},
  \bibinfo{pages}{050601} (\bibinfo{year}{2015}).

\bibitem[{\citenamefont{Yu et~al.}(2021)\citenamefont{Yu, Qiu, Chong, Torquato,
  and Park}}]{yu2021engineered}
\bibinfo{author}{\bibfnamefont{S.}~\bibnamefont{Yu}},
  \bibinfo{author}{\bibfnamefont{C.-W.} \bibnamefont{Qiu}},
  \bibinfo{author}{\bibfnamefont{Y.}~\bibnamefont{Chong}},
  \bibinfo{author}{\bibfnamefont{S.}~\bibnamefont{Torquato}}, \bibnamefont{and}
  \bibinfo{author}{\bibfnamefont{N.}~\bibnamefont{Park}}, Engineered disorder in photonics.
  \bibinfo{journal}{Nat. Rev. Mater.} \textbf{\bibinfo{volume}{6}},
  \bibinfo{pages}{226} (\bibinfo{year}{2021}).

\bibitem[{\citenamefont{Yeong and Torquato}(1998{\natexlab{a}})}]{Ye98a}
\bibinfo{author}{\bibfnamefont{C.~L.~Y.} \bibnamefont{Yeong}} \bibnamefont{and}
  \bibinfo{author}{\bibfnamefont{S.}~\bibnamefont{Torquato}}, Reconstructing random media.
  \bibinfo{journal}{Phys. Rev. E} \textbf{\bibinfo{volume}{57}},
  \bibinfo{pages}{495} (\bibinfo{year}{1998}{\natexlab{a}}).

\bibitem[{\citenamefont{Yeong and Torquato}(1998{\natexlab{b}})}]{Ye98b}
\bibinfo{author}{\bibfnamefont{C.~L.~Y.} \bibnamefont{Yeong}} \bibnamefont{and}
  \bibinfo{author}{\bibfnamefont{S.}~\bibnamefont{Torquato}}, Reconstructing random media. II. 
  \bibinfo{journal}{Phys. Rev. E} \textbf{\bibinfo{volume}{58}},
  \bibinfo{pages}{224} (\bibinfo{year}{1998}{\natexlab{b}}).

\bibitem[{\citenamefont{Roberts}(1997)}]{roberts1997statistical}
\bibinfo{author}{\bibfnamefont{A.~P.} \bibnamefont{Roberts}}, Statistical reconstruction of three-dimensional porous media from two-dimensional images.
  \bibinfo{journal}{Phys. Rev. E} \textbf{\bibinfo{volume}{56}},
  \bibinfo{pages}{3203} (\bibinfo{year}{1997}).

\bibitem{dhu_rand_field}
Y. Gao, Y. Jiao, and Y. Liu, Ultraefficient reconstruction of effectively hyperuniform disordered biphase materials via non-Gaussian random fields. Phys. Rev. E {\bf 105}, 045305 (2022).

\bibitem[{\citenamefont{Niezgoda et~al.}(2008)\citenamefont{Niezgoda, Fullwood,
  and Kalidindi}}]{niezgoda2008delineation}
\bibinfo{author}{\bibfnamefont{S.}~\bibnamefont{Niezgoda}},
  \bibinfo{author}{\bibfnamefont{D.}~\bibnamefont{Fullwood}}, \bibnamefont{and}
  \bibinfo{author}{\bibfnamefont{S.}~\bibnamefont{Kalidindi}}, Delineation of the space of 2-point correlations in a composite material system.
  \bibinfo{journal}{Acta Mater.} \textbf{\bibinfo{volume}{56}},
  \bibinfo{pages}{5285} (\bibinfo{year}{2008}).

\bibitem[{\citenamefont{Fullwood et~al.}(2008)\citenamefont{Fullwood, Niezgoda,
  and Kalidindi}}]{fullwood2008microstructure}
\bibinfo{author}{\bibfnamefont{D.~T.} \bibnamefont{Fullwood}},
  \bibinfo{author}{\bibfnamefont{S.~R.} \bibnamefont{Niezgoda}},
  \bibnamefont{and} \bibinfo{author}{\bibfnamefont{S.~R.}
  \bibnamefont{Kalidindi}}, Microstructure reconstructions from 2-point statistics using phase-recovery algorithms. \bibinfo{journal}{Acta Mater.}
  \textbf{\bibinfo{volume}{56}}, \bibinfo{pages}{942} (\bibinfo{year}{2008}).

\bibitem[{\citenamefont{Cherkasov et~al.}(2021)\citenamefont{Cherkasov, Ananev,
  Karsanina, Khlyupin, and Gerke}}]{cherkasov2021adaptive}
\bibinfo{author}{\bibfnamefont{A.}~\bibnamefont{Cherkasov}},
  \bibinfo{author}{\bibfnamefont{A.}~\bibnamefont{Ananev}},
  \bibinfo{author}{\bibfnamefont{M.}~\bibnamefont{Karsanina}},
  \bibinfo{author}{\bibfnamefont{A.}~\bibnamefont{Khlyupin}}, \bibnamefont{and}
  \bibinfo{author}{\bibfnamefont{K.}~\bibnamefont{Gerke}},
  Adaptive phase-retrieval stochastic reconstruction with correlation functions: Three-dimensional images from two-dimensional cuts.
  \bibinfo{journal}{Phys. Rev. E} \textbf{\bibinfo{volume}{104}},
  \bibinfo{pages}{035304} (\bibinfo{year}{2021}).

\bibitem[{\citenamefont{Okabe and Blunt}(2005)}]{okabe2005pore}
\bibinfo{author}{\bibfnamefont{H.}~\bibnamefont{Okabe}} \bibnamefont{and}
  \bibinfo{author}{\bibfnamefont{M.~J.} \bibnamefont{Blunt}}, Pore space reconstruction using multiple-point statistics.
  \bibinfo{journal}{J. Petro. Sci. Eng.}
  \textbf{\bibinfo{volume}{46}}, \bibinfo{pages}{121} (\bibinfo{year}{2005}).

\bibitem[{\citenamefont{Hajizadeh et~al.}(2011)\citenamefont{Hajizadeh,
  Safekordi, and Farhadpour}}]{hajizadeh2011multiple}
\bibinfo{author}{\bibfnamefont{A.}~\bibnamefont{Hajizadeh}},
  \bibinfo{author}{\bibfnamefont{A.}~\bibnamefont{Safekordi}},
  \bibnamefont{and} \bibinfo{author}{\bibfnamefont{F.~A.}
  \bibnamefont{Farhadpour}}, A multiple-point statistics algorithm for 3D pore space reconstruction from 2D images. \bibinfo{journal}{Adv. Water Res.}
  \textbf{\bibinfo{volume}{34}}, \bibinfo{pages}{1256} (\bibinfo{year}{2011}).

\bibitem[{\citenamefont{Tahmasebi and Sahimi}(2013)}]{tahmasebi2013cross}
\bibinfo{author}{\bibfnamefont{P.}~\bibnamefont{Tahmasebi}} \bibnamefont{and}
  \bibinfo{author}{\bibfnamefont{M.}~\bibnamefont{Sahimi}}, Cross-correlation function for accurate reconstruction of heterogeneous media.
  \bibinfo{journal}{Phys. Rev. Lett.} \textbf{\bibinfo{volume}{110}},
  \bibinfo{pages}{078002} (\bibinfo{year}{2013}).

\bibitem[{\citenamefont{Tahmasebi et~al.}(2012)\citenamefont{Tahmasebi,
  Hezarkhani, and Sahimi}}]{tahmasebi2012multiple}
\bibinfo{author}{\bibfnamefont{P.}~\bibnamefont{Tahmasebi}},
  \bibinfo{author}{\bibfnamefont{A.}~\bibnamefont{Hezarkhani}},
  \bibnamefont{and} \bibinfo{author}{\bibfnamefont{M.}~\bibnamefont{Sahimi}},
  Multiple-point geostatistical modeling based on the cross-correlation functions.
  \bibinfo{journal}{Comput. Geosci.} \textbf{\bibinfo{volume}{16}},
  \bibinfo{pages}{779} (\bibinfo{year}{2012}).

\bibitem[{\citenamefont{Xu et~al.}(2014)\citenamefont{Xu, Li, Brinson, and
  Chen}}]{xu2014descriptor}
\bibinfo{author}{\bibfnamefont{H.}~\bibnamefont{Xu}},
  \bibinfo{author}{\bibfnamefont{Y.}~\bibnamefont{Li}},
  \bibinfo{author}{\bibfnamefont{C.}~\bibnamefont{Brinson}}, \bibnamefont{and}
  \bibinfo{author}{\bibfnamefont{W.}~\bibnamefont{Chen}}, 
  \bibinfo{journal}{J. Mech. Design}
  \textbf{\bibinfo{volume}{136}} (\bibinfo{year}{2014}).

\bibitem[{\citenamefont{Cang et~al.}(2017)\citenamefont{Cang, Xu, Chen, Liu,
  Jiao, and Yi~Ren}}]{cang2017microstructure}
\bibinfo{author}{\bibfnamefont{R.}~\bibnamefont{Cang}},
  \bibinfo{author}{\bibfnamefont{Y.}~\bibnamefont{Xu}},
  \bibinfo{author}{\bibfnamefont{S.}~\bibnamefont{Chen}},
  \bibinfo{author}{\bibfnamefont{Y.}~\bibnamefont{Liu}},
  \bibinfo{author}{\bibfnamefont{Y.}~\bibnamefont{Jiao}}, \bibnamefont{and}
  \bibinfo{author}{\bibfnamefont{M.}~\bibnamefont{Yi~Ren}}, Microstructure representation and reconstruction of heterogeneous materials via deep belief network for computational material design.
  \bibinfo{journal}{J. Mech. Design}
  \textbf{\bibinfo{volume}{139}} (\bibinfo{year}{2017}).

\bibitem[{\citenamefont{Cang et~al.}(2018)\citenamefont{Cang, Li, Yao, Jiao,
  and Ren}}]{cang2018improving}
\bibinfo{author}{\bibfnamefont{R.}~\bibnamefont{Cang}},
  \bibinfo{author}{\bibfnamefont{H.}~\bibnamefont{Li}},
  \bibinfo{author}{\bibfnamefont{H.}~\bibnamefont{Yao}},
  \bibinfo{author}{\bibfnamefont{Y.}~\bibnamefont{Jiao}}, \bibnamefont{and}
  \bibinfo{author}{\bibfnamefont{Y.}~\bibnamefont{Ren}}, Improving direct physical properties prediction of heterogeneous materials from imaging data via convolutional neural network and a morphology-aware generative models.
  \bibinfo{journal}{Comput. Mater. Sci.}
  \textbf{\bibinfo{volume}{150}}, \bibinfo{pages}{212} (\bibinfo{year}{2018}).

\bibitem[{\citenamefont{Yang et~al.}(2018)\citenamefont{Yang, Li,
  Catherine~Brinson, Choudhary, Chen, and Agrawal}}]{yang2018microstructural}
\bibinfo{author}{\bibfnamefont{Z.}~\bibnamefont{Yang}},
  \bibinfo{author}{\bibfnamefont{X.}~\bibnamefont{Li}},
  \bibinfo{author}{\bibfnamefont{L.}~\bibnamefont{Catherine~Brinson}},
  \bibinfo{author}{\bibfnamefont{A.~N.} \bibnamefont{Choudhary}},
  \bibinfo{author}{\bibfnamefont{W.}~\bibnamefont{Chen}}, \bibnamefont{and}
  \bibinfo{author}{\bibfnamefont{A.}~\bibnamefont{Agrawal}}, Microstructural Materials Design Via Deep Adversarial Learning Methodology
  \bibinfo{journal}{J. Mech. Design}
  \textbf{\bibinfo{volume}{140}}, 111416 (\bibinfo{year}{2018}).

\bibitem[{\citenamefont{Li et~al.}(2018)\citenamefont{Li, Zhang, Zhao,
  Burkhart, Brinson, and Chen}}]{li2018transfer}
\bibinfo{author}{\bibfnamefont{X.}~\bibnamefont{Li}},
  \bibinfo{author}{\bibfnamefont{Y.}~\bibnamefont{Zhang}},
  \bibinfo{author}{\bibfnamefont{H.}~\bibnamefont{Zhao}},
  \bibinfo{author}{\bibfnamefont{C.}~\bibnamefont{Burkhart}},
  \bibinfo{author}{\bibfnamefont{L.~C.} \bibnamefont{Brinson}},
  \bibnamefont{and} \bibinfo{author}{\bibfnamefont{W.}~\bibnamefont{Chen}},
  \bibinfo{journal}{Sci. Rep.} \textbf{\bibinfo{volume}{8}},
  \bibinfo{pages}{1} (\bibinfo{year}{2018}).

\bibitem[{\citenamefont{Cheng et~al.}(2022)\citenamefont{Cheng, Jiao, and
  Ren}}]{cheng2022data}
\bibinfo{author}{\bibfnamefont{S.}~\bibnamefont{Cheng}},
  \bibinfo{author}{\bibfnamefont{Y.}~\bibnamefont{Jiao}}, \bibnamefont{and}
  \bibinfo{author}{\bibfnamefont{Y.}~\bibnamefont{Ren}}, Data-driven learning of 3-point correlation functions as microstructure representations. \bibinfo{journal}{Acta
  Mater.} \textbf{\bibinfo{volume}{229}}, \bibinfo{pages}{117800}
  (\bibinfo{year}{2022}).

\bibitem[{\citenamefont{Kirkpatrick et~al.}(1983)\citenamefont{Kirkpatrick,
  Gelatt~Jr, and Vecchi}}]{kirkpatrick1983optimization}
\bibinfo{author}{\bibfnamefont{S.}~\bibnamefont{Kirkpatrick}},
  \bibinfo{author}{\bibfnamefont{C.~D.} \bibnamefont{Gelatt~Jr}},
  \bibnamefont{and} \bibinfo{author}{\bibfnamefont{M.~P.}
  \bibnamefont{Vecchi}}, Optimization by simulated annealing. 
  \bibinfo{journal}{Science}
  \textbf{\bibinfo{volume}{220}}, \bibinfo{pages}{671} (\bibinfo{year}{1983}).

\bibitem[{\citenamefont{Torquato}(2006)}]{torquato2006necessary}
\bibinfo{author}{\bibfnamefont{S.}~\bibnamefont{Torquato}}, Necessary conditions on realizable two-point correlation functions of random media.
  \bibinfo{journal}{Ind. Eng. Chem. Res.}
  \textbf{\bibinfo{volume}{45}}, \bibinfo{pages}{6923} (\bibinfo{year}{2006}).

\bibitem[{\citenamefont{Jiao et~al.}(2007)\citenamefont{Jiao, Stillinger, and
  Torquato}}]{jiao2007modeling}
\bibinfo{author}{\bibfnamefont{Y.}~\bibnamefont{Jiao}},
  \bibinfo{author}{\bibfnamefont{F.}~\bibnamefont{Stillinger}},
  \bibnamefont{and} \bibinfo{author}{\bibfnamefont{S.}~\bibnamefont{Torquato}}, Modeling heterogeneous materials via two-point correlation functions: Basic principles.
  \bibinfo{journal}{Phys. Rev. E} \textbf{\bibinfo{volume}{76}},
  \bibinfo{pages}{031110} (\bibinfo{year}{2007}).

\bibitem[{\citenamefont{Jiao et~al.}(2008{\natexlab{a}})\citenamefont{Jiao,
  Stillinger, and Torquato}}]{jiao2008modeling}
\bibinfo{author}{\bibfnamefont{Y.}~\bibnamefont{Jiao}},
  \bibinfo{author}{\bibfnamefont{F.}~\bibnamefont{Stillinger}},
  \bibnamefont{and} \bibinfo{author}{\bibfnamefont{S.}~\bibnamefont{Torquato}}, Modeling heterogeneous materials via two-point correlation functions. II. Algorithmic details and applications.
  \bibinfo{journal}{Phys. Rev. E} \textbf{\bibinfo{volume}{77}},
  \bibinfo{pages}{031135} (\bibinfo{year}{2008}{\natexlab{a}}).

\bibitem[{\citenamefont{Jiao et~al.}(2009)\citenamefont{Jiao, Stillinger, and
  Torquato}}]{jiao2009superior}
\bibinfo{author}{\bibfnamefont{Y.}~\bibnamefont{Jiao}},
  \bibinfo{author}{\bibfnamefont{F.}~\bibnamefont{Stillinger}},
  \bibnamefont{and} \bibinfo{author}{\bibfnamefont{S.}~\bibnamefont{Torquato}}, A superior descriptor of random textures and its predictive capacity.
  \bibinfo{journal}{Proc. Natl. Acad. Sci. USA}
  \textbf{\bibinfo{volume}{106}}, \bibinfo{pages}{17634}
  (\bibinfo{year}{2009}).

\bibitem[{\citenamefont{Chen et~al.}(2019)\citenamefont{Chen, Xu, Chawla, Ren,
  and Jiao}}]{chen2019hierarchical}
\bibinfo{author}{\bibfnamefont{P.-E.} \bibnamefont{Chen}},
  \bibinfo{author}{\bibfnamefont{W.}~\bibnamefont{Xu}},
  \bibinfo{author}{\bibfnamefont{N.}~\bibnamefont{Chawla}},
  \bibinfo{author}{\bibfnamefont{Y.}~\bibnamefont{Ren}}, \bibnamefont{and}
  \bibinfo{author}{\bibfnamefont{Y.}~\bibnamefont{Jiao}}, Hierarchical n-point polytope functions for quantitative representation of complex heterogeneous materials and microstructural evolution.
  \bibinfo{journal}{Acta Mater.} \textbf{\bibinfo{volume}{179}},
  \bibinfo{pages}{317} (\bibinfo{year}{2019}).

\bibitem[{\citenamefont{Chen et~al.}(2020)\citenamefont{Chen, Xu, Ren, and
  Jiao}}]{chen2020probing}
\bibinfo{author}{\bibfnamefont{P.-E.} \bibnamefont{Chen}},
  \bibinfo{author}{\bibfnamefont{W.}~\bibnamefont{Xu}},
  \bibinfo{author}{\bibfnamefont{Y.}~\bibnamefont{Ren}}, \bibnamefont{and}
  \bibinfo{author}{\bibfnamefont{Y.}~\bibnamefont{Jiao}}, Probing information content of hierarchical -point polytope functions for quantifying and reconstructing disordered systems.
  \bibinfo{journal}{Phys. Rev. E} \textbf{\bibinfo{volume}{102}},
  \bibinfo{pages}{013305} (\bibinfo{year}{2020}).

\bibitem[{\citenamefont{Gerke and Karsanina}(2015)}]{gerke2015improving}
\bibinfo{author}{\bibfnamefont{K.~M.} \bibnamefont{Gerke}} \bibnamefont{and}
  \bibinfo{author}{\bibfnamefont{M.~V.} \bibnamefont{Karsanina}}, Improving stochastic reconstructions by weighting correlation functions in an objective function.
  \bibinfo{journal}{EPL} \textbf{\bibinfo{volume}{111}},
  \bibinfo{pages}{56002} (\bibinfo{year}{2015}).

\bibitem[{\citenamefont{Karsanina and Gerke}(2018)}]{karsanina2018hierarchical}
\bibinfo{author}{\bibfnamefont{M.~V.} \bibnamefont{Karsanina}}
  \bibnamefont{and} \bibinfo{author}{\bibfnamefont{K.~M.} \bibnamefont{Gerke}}, Hierarchical optimization: Fast and robust multiscale stochastic reconstructions with rescaled correlation functions.
  \bibinfo{journal}{Phys. Rev. Lett.} \textbf{\bibinfo{volume}{121}},
  \bibinfo{pages}{265501} (\bibinfo{year}{2018}).

\bibitem[{\citenamefont{Feng et~al.}(2018)\citenamefont{Feng, Teng, He, and
  Wu}}]{feng2018accelerating}
\bibinfo{author}{\bibfnamefont{J.}~\bibnamefont{Feng}},
  \bibinfo{author}{\bibfnamefont{Q.}~\bibnamefont{Teng}},
  \bibinfo{author}{\bibfnamefont{X.}~\bibnamefont{He}}, \bibnamefont{and}
  \bibinfo{author}{\bibfnamefont{X.}~\bibnamefont{Wu}},  Accelerating multi-point statistics reconstruction method for porous media via deep learning. 
  \bibinfo{journal}{Acta
  Mater.} \textbf{\bibinfo{volume}{159}}, \bibinfo{pages}{296}
  (\bibinfo{year}{2018}).

\bibitem[{\citenamefont{Jiao et~al.}(2013)\citenamefont{Jiao, Padilla, and
  Chawla}}]{jiao2013modeling}
\bibinfo{author}{\bibfnamefont{Y.}~\bibnamefont{Jiao}},
  \bibinfo{author}{\bibfnamefont{E.}~\bibnamefont{Padilla}}, \bibnamefont{and}
  \bibinfo{author}{\bibfnamefont{N.}~\bibnamefont{Chawla}}, Modeling and predicting microstructure evolution in lead/tin alloy via correlation functions and stochastic material reconstruction.
  \bibinfo{journal}{Acta Mater.} \textbf{\bibinfo{volume}{61}},
  \bibinfo{pages}{3370} (\bibinfo{year}{2013}).

\bibitem[{\citenamefont{Chen et~al.}(2015)\citenamefont{Chen, Li, and
  Jiao}}]{chen2015dynamic}
\bibinfo{author}{\bibfnamefont{S.}~\bibnamefont{Chen}},
  \bibinfo{author}{\bibfnamefont{H.}~\bibnamefont{Li}}, \bibnamefont{and}
  \bibinfo{author}{\bibfnamefont{Y.}~\bibnamefont{Jiao}}, Dynamic reconstruction of heterogeneous materials and microstructure evolution.
  \bibinfo{journal}{Phys. Rev. E} \textbf{\bibinfo{volume}{92}},
  \bibinfo{pages}{023301} (\bibinfo{year}{2015}).

\bibitem[{\citenamefont{Jiao and Chawla}(2014)}]{jiao2014modeling}
\bibinfo{author}{\bibfnamefont{Y.}~\bibnamefont{Jiao}} \bibnamefont{and}
  \bibinfo{author}{\bibfnamefont{N.}~\bibnamefont{Chawla}}, Modeling and characterizing anisotropic inclusion orientation in heterogeneous material via directional cluster functions and stochastic microstructure reconstruction.
  \bibinfo{journal}{J. Appl. Phys.} \textbf{\bibinfo{volume}{115}},
  \bibinfo{pages}{093511} (\bibinfo{year}{2014}).

\bibitem[{\citenamefont{Guo et~al.}(2014)\citenamefont{Guo, Chawla, Jing,
  Torquato, and Jiao}}]{guo2014accurate}
\bibinfo{author}{\bibfnamefont{E.-Y.} \bibnamefont{Guo}},
  \bibinfo{author}{\bibfnamefont{N.}~\bibnamefont{Chawla}},
  \bibinfo{author}{\bibfnamefont{T.}~\bibnamefont{Jing}},
  \bibinfo{author}{\bibfnamefont{S.}~\bibnamefont{Torquato}}, \bibnamefont{and}
  \bibinfo{author}{\bibfnamefont{Y.}~\bibnamefont{Jiao}}, Accurate modeling and reconstruction of three-dimensional percolating filamentary microstructures from two-dimensional micrographs via dilation-erosion method.
  \bibinfo{journal}{Mater. Charac.} \textbf{\bibinfo{volume}{89}},
  \bibinfo{pages}{33} (\bibinfo{year}{2014}).

\bibitem[{\citenamefont{Chen et~al.}(2016)\citenamefont{Chen, Kirubanandham,
  Chawla, and Jiao}}]{chen2016stochastic}
\bibinfo{author}{\bibfnamefont{S.}~\bibnamefont{Chen}},
  \bibinfo{author}{\bibfnamefont{A.}~\bibnamefont{Kirubanandham}},
  \bibinfo{author}{\bibfnamefont{N.}~\bibnamefont{Chawla}}, \bibnamefont{and}
  \bibinfo{author}{\bibfnamefont{Y.}~\bibnamefont{Jiao}}, Stochastic multi-scale reconstruction of 3D microstructure consisting of polycrystalline grains and second-phase particles from 2D micrographs.
  \bibinfo{journal}{Metal. Mater. Trans. A}
  \textbf{\bibinfo{volume}{47}}, \bibinfo{pages}{1440} (\bibinfo{year}{2016}).

\bibitem[{\citenamefont{Gerke et~al.}(2019)\citenamefont{Gerke, Karsanina, and
  Katsman}}]{gerke2019calculation}
\bibinfo{author}{\bibfnamefont{K.~M.} \bibnamefont{Gerke}},
  \bibinfo{author}{\bibfnamefont{M.~V.} \bibnamefont{Karsanina}},
  \bibnamefont{and} \bibinfo{author}{\bibfnamefont{R.}~\bibnamefont{Katsman}}, Calculation of tensorial flow properties on pore level: Exploring the influence of boundary conditions on the permeability of three-dimensional stochastic reconstructions.
  \bibinfo{journal}{Phys. Rev. E} \textbf{\bibinfo{volume}{100}},
  \bibinfo{pages}{053312} (\bibinfo{year}{2019}).

\bibitem[{\citenamefont{Karsanina and Gerke}(2022)}]{karsanina2022stochastic}
\bibinfo{author}{\bibfnamefont{M.~V.} \bibnamefont{Karsanina}}
  \bibnamefont{and} \bibinfo{author}{\bibfnamefont{K.~M.} \bibnamefont{Gerke}}, Stochastic (re) constructions of non-stationary material structures: Using ensemble averaged correlation functions and non-uniform phase distributions.
  \bibinfo{journal}{Physica A: Stat. Mech. Appl.} p.
  \bibinfo{pages}{128417} (\bibinfo{year}{2022}).

\bibitem[{\citenamefont{Chen et~al.}(2022{\natexlab{b}})\citenamefont{Chen,
  Raghavan, Zheng, Li, Ankit, and Jiao}}]{chen2022quantifying}
\bibinfo{author}{\bibfnamefont{P.-E.} \bibnamefont{Chen}},
  \bibinfo{author}{\bibfnamefont{R.}~\bibnamefont{Raghavan}},
  \bibinfo{author}{\bibfnamefont{Y.}~\bibnamefont{Zheng}},
  \bibinfo{author}{\bibfnamefont{H.}~\bibnamefont{Li}},
  \bibinfo{author}{\bibfnamefont{K.}~\bibnamefont{Ankit}}, \bibnamefont{and}
  \bibinfo{author}{\bibfnamefont{Y.}~\bibnamefont{Jiao}}, Quantifying microstructural evolution via time-dependent reduced-dimension metrics based on hierarchical -point polytope functions.
  \bibinfo{journal}{Phys. Rev. E} \textbf{\bibinfo{volume}{105}},
  \bibinfo{pages}{025306} (\bibinfo{year}{2022}{\natexlab{b}}).

\bibitem[{\citenamefont{Chen and Torquato}(2018{\natexlab{b}})}]{Ch18a}
\bibinfo{author}{\bibfnamefont{D.}~\bibnamefont{Chen}} \bibnamefont{and}
  \bibinfo{author}{\bibfnamefont{S.}~\bibnamefont{Torquato}}, Designing disordered hyperuniform two-phase materials with novel physical properties.
  \bibinfo{journal}{Acta Mater.} \textbf{\bibinfo{volume}{142}},
  \bibinfo{pages}{152} (\bibinfo{year}{2018}{\natexlab{b}}).

\bibitem[{\citenamefont{Martis et~al.}(2013)\citenamefont{Martis, Marcotte,
  Stillinger, and Torquato}}]{martis2013exotic}
\bibinfo{author}{\bibfnamefont{S.}~\bibnamefont{Martis}},
  \bibinfo{author}{\bibfnamefont{{\'E}.}~\bibnamefont{Marcotte}},
  \bibinfo{author}{\bibfnamefont{F.~H.} \bibnamefont{Stillinger}},
  \bibnamefont{and} \bibinfo{author}{\bibfnamefont{S.}~\bibnamefont{Torquato}}, Exotic ground states of directional pair potentials via collective-density variables.
  \bibinfo{journal}{J. Stat. Phys.}
  \textbf{\bibinfo{volume}{150}}, \bibinfo{pages}{414} (\bibinfo{year}{2013}).

\bibitem[{\citenamefont{Podolskiy and Narimanov}(2005)}]{waveguide}
\bibinfo{author}{\bibfnamefont{V.~A.} \bibnamefont{Podolskiy}}
  \bibnamefont{and} \bibinfo{author}{\bibfnamefont{E.~E.}
  \bibnamefont{Narimanov}}, Strongly anisotropic waveguide as a nonmagnetic left-handed system. \bibinfo{journal}{Phys. Rev. B}
  \textbf{\bibinfo{volume}{71}}, \bibinfo{pages}{201101}
  (\bibinfo{year}{2005}).




\bibitem[{\citenamefont{Damaskos et~al.}(1982)\citenamefont{Damaskos, Maffett,
  and Uslenghi}}]{damaskos1982dispersion}
\bibinfo{author}{\bibfnamefont{N.}~\bibnamefont{Damaskos}},
  \bibinfo{author}{\bibfnamefont{A.}~\bibnamefont{Maffett}}, \bibnamefont{and}
  \bibinfo{author}{\bibfnamefont{P.}~\bibnamefont{Uslenghi}},
  \bibinfo{journal}{IEEE Tran. Anten. Prop.}
  \textbf{\bibinfo{volume}{30}}, \bibinfo{pages}{991} (\bibinfo{year}{1982}).

\bibitem[{\citenamefont{Itin}(2010)}]{itin2010dispersion}
\bibinfo{author}{\bibfnamefont{Y.}~\bibnamefont{Itin}}, Dispersion relation for electromagnetic waves in anisotropic media.
  \bibinfo{journal}{Phys. Lett. A} \textbf{\bibinfo{volume}{374}},
  \bibinfo{pages}{1113} (\bibinfo{year}{2010}).

\bibitem[{\citenamefont{Torquato}(2016{\natexlab{b}})}]{To16b}
\bibinfo{author}{\bibfnamefont{S.}~\bibnamefont{Torquato}}, Disordered hyperuniform heterogeneous materials.
  \bibinfo{journal}{J. Phys. Condens. Matter} \textbf{\bibinfo{volume}{28}},
  \bibinfo{pages}{414012} (\bibinfo{year}{2016}{\natexlab{b}}).

\bibitem[{\citenamefont{Torquato
  et~al.}(2015{\natexlab{b}})\citenamefont{Torquato, Zhang, and
  Stillinger}}]{To15}
\bibinfo{author}{\bibfnamefont{S.}~\bibnamefont{Torquato}},
  \bibinfo{author}{\bibfnamefont{G.}~\bibnamefont{Zhang}}, \bibnamefont{and}
  \bibinfo{author}{\bibfnamefont{F.~H.} \bibnamefont{Stillinger}}, Ensemble theory for stealthy hyperuniform disordered ground states.
  \bibinfo{journal}{Phys. Rev. X} \textbf{\bibinfo{volume}{5}},
  \bibinfo{pages}{021020} (\bibinfo{year}{2015}{\natexlab{b}}).


\bibitem[{\citenamefont{Kim and
  Torquato}(2020{\natexlab{b}})}]{kim2020effective}
\bibinfo{author}{\bibfnamefont{J.}~\bibnamefont{Kim}} \bibnamefont{and}
  \bibinfo{author}{\bibfnamefont{S.}~\bibnamefont{Torquato}}, Effective elastic wave characteristics of composite media.
  \bibinfo{journal}{New J. of Phys.} \textbf{\bibinfo{volume}{22}},
  \bibinfo{pages}{123050} (\bibinfo{year}{2020}{\natexlab{b}}).



\bibitem[{\citenamefont{Wang and Torquato}(2022)}]{wang2022dynamic}
\bibinfo{author}{\bibfnamefont{H.}~\bibnamefont{Wang}} \bibnamefont{and}
  \bibinfo{author}{\bibfnamefont{S.}~\bibnamefont{Torquato}}, Dynamic measure of hyperuniformity and nonhyperuniformity in heterogeneous media via the diffusion spreadability.
  \bibinfo{journal}{Phys. Rev. Appl.} \textbf{\bibinfo{volume}{17}},
  \bibinfo{pages}{034022} (\bibinfo{year}{2022}).

\bibitem[{\citenamefont{Xu et~al.}(2022)\citenamefont{Xu, Chen, Li, Xu, Ren,
  Shan, and Jiao}}]{xu2022correlation}
\bibinfo{author}{\bibfnamefont{Y.}~\bibnamefont{Xu}},
  \bibinfo{author}{\bibfnamefont{P.-E.} \bibnamefont{Chen}},
  \bibinfo{author}{\bibfnamefont{H.}~\bibnamefont{Li}},
  \bibinfo{author}{\bibfnamefont{W.}~\bibnamefont{Xu}},
  \bibinfo{author}{\bibfnamefont{Y.}~\bibnamefont{Ren}},
  \bibinfo{author}{\bibfnamefont{W.}~\bibnamefont{Shan}}, \bibnamefont{and}
  \bibinfo{author}{\bibfnamefont{Y.}~\bibnamefont{Jiao}},  Correlation-function-based microstructure design of alloy-polymer composites for dynamic dry adhesion tuning in soft gripping.
  \bibinfo{journal}{J. Appl. Phys.} \textbf{\bibinfo{volume}{131}},
  \bibinfo{pages}{115104} (\bibinfo{year}{2022}).

\bibitem[{\citenamefont{Iyer et~al.}(2020)\citenamefont{Iyer, Dulal, Zhang,
  Ghumman, Chien, Balasubramanian, and Chen}}]{iyer2020designing}
\bibinfo{author}{\bibfnamefont{A.}~\bibnamefont{Iyer}},
  \bibinfo{author}{\bibfnamefont{R.}~\bibnamefont{Dulal}},
  \bibinfo{author}{\bibfnamefont{Y.}~\bibnamefont{Zhang}},
  \bibinfo{author}{\bibfnamefont{U.~F.} \bibnamefont{Ghumman}},
  \bibinfo{author}{\bibfnamefont{T.}~\bibnamefont{Chien}},
  \bibinfo{author}{\bibfnamefont{G.}~\bibnamefont{Balasubramanian}},
  \bibnamefont{and} \bibinfo{author}{\bibfnamefont{W.}~\bibnamefont{Chen}}, Designing anisotropic microstructures with spectral density function.
  \bibinfo{journal}{Comput. Mater. Sci.}
  \textbf{\bibinfo{volume}{179}}, \bibinfo{pages}{109559}
  (\bibinfo{year}{2020}).

\bibitem[{\citenamefont{Farooq~Ghumman
  et~al.}(2018)\citenamefont{Farooq~Ghumman, Iyer, Dulal, Munshi, Wang, Chien,
  Balasubramanian, and Chen}}]{farooq2018spectral}
\bibinfo{author}{\bibfnamefont{U.}~\bibnamefont{Farooq~Ghumman}},
  \bibinfo{author}{\bibfnamefont{A.}~\bibnamefont{Iyer}},
  \bibinfo{author}{\bibfnamefont{R.}~\bibnamefont{Dulal}},
  \bibinfo{author}{\bibfnamefont{J.}~\bibnamefont{Munshi}},
  \bibinfo{author}{\bibfnamefont{A.}~\bibnamefont{Wang}},
  \bibinfo{author}{\bibfnamefont{T.}~\bibnamefont{Chien}},
  \bibinfo{author}{\bibfnamefont{G.}~\bibnamefont{Balasubramanian}},
  \bibnamefont{and} \bibinfo{author}{\bibfnamefont{W.}~\bibnamefont{Chen}}, A spectral density function approach for active layer design of organic photovoltaic cells.
  \bibinfo{journal}{J. Mech. Design}
  \textbf{\bibinfo{volume}{140}} (\bibinfo{year}{2018}).

\bibitem[{\citenamefont{Lu and Torquato}(1990)}]{lu1990local}
\bibinfo{author}{\bibfnamefont{B.}~\bibnamefont{Lu}} \bibnamefont{and}
  \bibinfo{author}{\bibfnamefont{S.}~\bibnamefont{Torquato}}, Local volume fraction fluctuations in heterogeneous media.
  \bibinfo{journal}{J. Chem. Phys.}
  \textbf{\bibinfo{volume}{93}}, \bibinfo{pages}{3452} (\bibinfo{year}{1990}).

\bibitem{debye}
P. Debye and A. M. Bueche, Scattering by an inhomogeneous solid. J. Appl. Phys. {\bf 20}, 518 (1949).

\bibitem[{\citenamefont{Ishimaru}(1978)}]{ishimaru1978wave}
\bibinfo{author}{\bibfnamefont{A.}~\bibnamefont{Ishimaru}},
  \emph{\bibinfo{title}{Wave propagation and scattering in random media}},
  vol.~\bibinfo{volume}{2} (\bibinfo{publisher}{Academic press New York},
  \bibinfo{year}{1978}).

\bibitem[{\citenamefont{{O{\u g}uz} et~al.}(2017)\citenamefont{{O{\u g}uz},
  {Socolar}, {Steinhardt}, and {Torquato}}}]{Og17}
\bibinfo{author}{\bibfnamefont{E.~C.} \bibnamefont{{O{\u g}uz}}},
  \bibinfo{author}{\bibfnamefont{J.~E.~S.} \bibnamefont{{Socolar}}},
  \bibinfo{author}{\bibfnamefont{P.~J.} \bibnamefont{{Steinhardt}}},
  \bibnamefont{and}
  \bibinfo{author}{\bibfnamefont{S.}~\bibnamefont{{Torquato}}}, Hyperuniformity of quasicrystals.
  \bibinfo{journal}{Phys. Rev. B} \textbf{\bibinfo{volume}{95}},
  \bibinfo{pages}{054119} (\bibinfo{year}{2017}).

\bibitem[{\citenamefont{Zhang et~al.}(2017)\citenamefont{Zhang, Stillinger, and
  Torquato}}]{zhang2017classical}
\bibinfo{author}{\bibfnamefont{G.}~\bibnamefont{Zhang}},
  \bibinfo{author}{\bibfnamefont{F.~H.} \bibnamefont{Stillinger}},
  \bibnamefont{and} \bibinfo{author}{\bibfnamefont{S.}~\bibnamefont{Torquato}}, Classical many-particle systems with unique disordered ground states.
  \bibinfo{journal}{Phys. Rev. E} \textbf{\bibinfo{volume}{96}},
  \bibinfo{pages}{042146} (\bibinfo{year}{2017}).

\bibitem[{\citenamefont{Zachary
  et~al.}(2011{\natexlab{b}})\citenamefont{Zachary, Jiao, and
  Torquato}}]{Za11c}
\bibinfo{author}{\bibfnamefont{C.~E.} \bibnamefont{Zachary}},
  \bibinfo{author}{\bibfnamefont{Y.}~\bibnamefont{Jiao}}, \bibnamefont{and}
  \bibinfo{author}{\bibfnamefont{S.}~\bibnamefont{Torquato}}, Hyperuniformity, quasi-long-range correlations, and void-space constraints in maximally random jammed particle packings. I. Polydisperse spheres.
  \bibinfo{journal}{Phys. Rev. E} \textbf{\bibinfo{volume}{83}},
  \bibinfo{pages}{051308} (\bibinfo{year}{2011}{\natexlab{b}}).

\bibitem[{\citenamefont{Zachary
  et~al.}(2011{\natexlab{c}})\citenamefont{Zachary, Jiao, and
  Torquato}}]{Za11d}
\bibinfo{author}{\bibfnamefont{C.~E.} \bibnamefont{Zachary}},
  \bibinfo{author}{\bibfnamefont{Y.}~\bibnamefont{Jiao}}, \bibnamefont{and}
  \bibinfo{author}{\bibfnamefont{S.}~\bibnamefont{Torquato}}, Hyperuniformity, quasi-long-range correlations, and void-space constraints in maximally random jammed particle packings. II. Anisotropy in particle shape.
  \bibinfo{journal}{Phys. Rev. E} \textbf{\bibinfo{volume}{83}},
  \bibinfo{pages}{051309} (\bibinfo{year}{2011}{\natexlab{c}}).

\bibitem[{\citenamefont{Zachary and Torquato}(2011{\natexlab{b}})}]{Za11b}
\bibinfo{author}{\bibfnamefont{C.~E.} \bibnamefont{Zachary}} \bibnamefont{and}
  \bibinfo{author}{\bibfnamefont{S.}~\bibnamefont{Torquato}}, Anomalous local coordination, density fluctuations, and void statistics in disordered hyperuniform many-particle ground states.
  \bibinfo{journal}{Phys. Rev. E} \textbf{\bibinfo{volume}{83}},
  \bibinfo{pages}{051133} (\bibinfo{year}{2011}{\natexlab{b}}).


\bibitem[{\citenamefont{Kim and Torquato}(2018)}]{Ki18a}
\bibinfo{author}{\bibfnamefont{J.}~\bibnamefont{Kim}} \bibnamefont{and}
  \bibinfo{author}{\bibfnamefont{S.}~\bibnamefont{Torquato}}, Effect of imperfections on the hyperuniformity of many-body systems.
  \bibinfo{journal}{Phys. Rev. B} \textbf{\bibinfo{volume}{97}},
  \bibinfo{pages}{054105} (\bibinfo{year}{2018}).

\bibitem[{\citenamefont{Batten et~al.}(2008{\natexlab{b}})\citenamefont{Batten,
  Stillinger, and Torquato}}]{Ba08}
\bibinfo{author}{\bibfnamefont{R.~D.} \bibnamefont{Batten}},
  \bibinfo{author}{\bibfnamefont{F.~H.} \bibnamefont{Stillinger}},
  \bibnamefont{and} \bibinfo{author}{\bibfnamefont{S.}~\bibnamefont{Torquato}}, Classical disordered ground states: Super-ideal gases and stealth and equi-luminous materials.
  \bibinfo{journal}{J. Appl. Phys.} \textbf{\bibinfo{volume}{104}},
  \bibinfo{pages}{033504} (\bibinfo{year}{2008}{\natexlab{b}}).

\bibitem[{\citenamefont{Jiao et~al.}(2008{\natexlab{b}})\citenamefont{Jiao,
  Stillinger, and Torquato}}]{jiao2008optimal}
\bibinfo{author}{\bibfnamefont{Y.}~\bibnamefont{Jiao}},
  \bibinfo{author}{\bibfnamefont{F.}~\bibnamefont{Stillinger}},
  \bibnamefont{and} \bibinfo{author}{\bibfnamefont{S.}~\bibnamefont{Torquato}}, Optimal packings of superdisks and the role of symmetry.
  \bibinfo{journal}{Phys. Rev. Lett.} \textbf{\bibinfo{volume}{100}},
  \bibinfo{pages}{245504} (\bibinfo{year}{2008}{\natexlab{b}}).

\bibitem[{\citenamefont{Atkinson et~al.}(2012)\citenamefont{Atkinson, Jiao, and
  Torquato}}]{atkinson2012maximally}
\bibinfo{author}{\bibfnamefont{S.}~\bibnamefont{Atkinson}},
  \bibinfo{author}{\bibfnamefont{Y.}~\bibnamefont{Jiao}}, \bibnamefont{and}
  \bibinfo{author}{\bibfnamefont{S.}~\bibnamefont{Torquato}}, Maximally dense packings of two-dimensional convex and concave noncircular particles.
  \bibinfo{journal}{Phys. Rev. E} \textbf{\bibinfo{volume}{86}},
  \bibinfo{pages}{031302} (\bibinfo{year}{2012}).

\bibitem[{\citenamefont{Cox and Hyde}(2014)}]{cox2014galois}
\bibinfo{author}{\bibfnamefont{D.~A.} \bibnamefont{Cox}} \bibnamefont{and}
  \bibinfo{author}{\bibfnamefont{T.}~\bibnamefont{Hyde}},
  The Galois theory of the lemniscate.
  \bibinfo{journal}{J. Num. Theor.} \textbf{\bibinfo{volume}{135}},
  \bibinfo{pages}{43} (\bibinfo{year}{2014}).

\bibitem{perco}
W Xu, X Su, Y Jiao, Continuum Percolation of Congruent Overlapping Spherocylinders.
Phys. Rev. E {\bf 94}, 032122 (2016).

\bibitem[{\citenamefont{Liu et~al.}(2019)\citenamefont{Liu, Wang, Ho, Ng, Ng,
  Hall-Chen, Koay, Dong, Liu, Qiu et~al.}}]{liu2019structural}
\bibinfo{author}{\bibfnamefont{Y.}~\bibnamefont{Liu}},
  \bibinfo{author}{\bibfnamefont{H.}~\bibnamefont{Wang}},
  \bibinfo{author}{\bibfnamefont{J.}~\bibnamefont{Ho}},
  \bibinfo{author}{\bibfnamefont{R.~C.} \bibnamefont{Ng}},
  \bibinfo{author}{\bibfnamefont{R.~J.} \bibnamefont{Ng}},
  \bibinfo{author}{\bibfnamefont{V.~H.} \bibnamefont{Hall-Chen}},
  \bibinfo{author}{\bibfnamefont{E.~H.} \bibnamefont{Koay}},
  \bibinfo{author}{\bibfnamefont{Z.}~\bibnamefont{Dong}},
  \bibinfo{author}{\bibfnamefont{H.}~\bibnamefont{Liu}},
  \bibinfo{author}{\bibfnamefont{C.-W.} \bibnamefont{Qiu}},
  \bibnamefont{et~al.}, Structural color three-dimensional printing by shrinking photonic crystals. \bibinfo{journal}{Nat. Comm.}
  \textbf{\bibinfo{volume}{10}}, \bibinfo{pages}{4340} (\bibinfo{year}{2019}).

\bibitem[{\citenamefont{Liu et~al.}(2022)\citenamefont{Liu, Ding, Li, Niu,
  Zeng, Zhang, Du, and Gu}}]{liu20223d}
\bibinfo{author}{\bibfnamefont{K.}~\bibnamefont{Liu}},
  \bibinfo{author}{\bibfnamefont{H.}~\bibnamefont{Ding}},
  \bibinfo{author}{\bibfnamefont{S.}~\bibnamefont{Li}},
  \bibinfo{author}{\bibfnamefont{Y.}~\bibnamefont{Niu}},
  \bibinfo{author}{\bibfnamefont{Y.}~\bibnamefont{Zeng}},
  \bibinfo{author}{\bibfnamefont{J.}~\bibnamefont{Zhang}},
  \bibinfo{author}{\bibfnamefont{X.}~\bibnamefont{Du}}, \bibnamefont{and}
  \bibinfo{author}{\bibfnamefont{Z.}~\bibnamefont{Gu}},
  3D printing colloidal crystal microstructures via sacrificial-scaffold-mediated two-photon lithography.
  \bibinfo{journal}{Nat. Comm.} \textbf{\bibinfo{volume}{13}},
  \bibinfo{pages}{4563} (\bibinfo{year}{2022}).

\bibitem{spreadability}
M. Skolnick and S. Torquato, Simulated diffusion spreadability for characterizing the structure and transport properties of two-phase materials. Acta Mater. {\bf 250} 118857 (2023). 




\end{thebibliography}
\end{document}